\documentclass[useAMS,usenatbib]{mn2e}
\usepackage{epsfig}
\title[Optical Studies of BL Lacertae]{Nature of Intra-night Optical Variability of BL Lacertae}

\author[Gaur et al.]
{Haritma Gaur$^{1}$\thanks{E-mail: haritma@shao.ac.cn}, Alok C.\ Gupta$^{2,1}$, R.\ Bachev$^{3}$, 
A.\ Strigachev$^{3}$, E.\ Semkov$^{3}$, M.\ B{\"o}ttcher$^{4,5}$, 
\newauthor Paul J.\ Wiita$^{6}$, J. A. de Diego$^{7}$, Minfeng Gu$^{1}$, H. Guo$^{1}$, R. Joshi$^{8}$,
B. Mihov$^{3}$, N. Palma$^{5}$, \newauthor S. Peneva$^{3}$,  A. Rajasingam$^{5}$,
L. Slavcheva-Mihova$^{3}$  \\
\\
$^{1}$Key Laboratory for Research in Galaxies and Cosmology, Shanghai Astronomical Observatory, Chinese Academy of Sciences,\\
 80 Nandan Road, Shanghai 200030, China; haritma@shao.ac.cn \\
$^{2}$Aryabhatta Research Institute of Observational Sciences (ARIES), Manora Peak, Nainital -- 263002, India \\
$^{3}$Institute of Astronomy and National Astronomical Observatory,Bulgarian Academy of Sciences, \\
72 Tsarigradsko Shosse Blvd., 1784 Sofia, Bulgaria  \\
$^{4}$Centre for Space Research, North-West University, Potchefstroom 2520, South Africa \\
$^{5}$Astrophysical Institute, Department of Physics and Astronomy, Ohio University, Athens, OH 45701, USA  \\
$^{6}$Department of Physics, The College of New Jersey, P.O.\ Box 7718, Ewing, NJ 08628-0718, USA \\
$^{7}$Instituto de Astronom\'ia, Universidad Nacional  Aut\'onoma de M\'exico, 04310, M\'exico D.F., M\'exico \\
$^{8}$Inter-University Centre of Astronomy and Astrophysics (IUCAA), Pune - 411007, India \\
}

\begin{document}

\date{Accepted ....... Received  ......; in original form ......}

\pagerange{\pageref{firstpage}--\pageref{lastpage}} \pubyear{2010}

\maketitle

\label{firstpage}
\begin{abstract}

We present the results of extensive multi-band intra-night optical monitoring of BL Lacertae during  2010--2012. 
BL Lacertae was very active in this period and showed intense variability in almost all wavelengths.
We extensively observed it for a total for 38 nights; on 26 of them observations were done quasi-simultaneously
 in B, V, R and I bands (totaling 113 light curves), with an average sampling interval of around 8 minutes.
BL Lacertae showed significant variations on hour-like
timescales in a total of 19 nights in different optical bands. We did not find any evidence for periodicities
or characteristic variability time-scales in the light curves.
 The intranight variability amplitude is generally greater at higher frequencies and decreases as the source flux increases.
 We found spectral variations in BL Lacertae in the sense that the optical spectrum becomes flatter as the flux 
increases but in several flaring states deviates from the linear trend suggesting different jet components contributing to the
emission at different times.

\end{abstract}


\section{Introduction}

BL Lacertae is a well known source which has been used to define a class of active
galactic nuclei (AGNs) that, together with flat spectrum radio quasars, make up the highly
variable objects called blazars.  The BL Lacertae class is characterized by the absence or extreme
weakness of  emission lines (with equivalent width in the rest frame of the host galaxy of $< 5$ \AA),
intense flux and spectral variability across the
complete electromagnetic spectrum  on a wide variety of time-scales,
and highly variable optical and radio polarization  (e.g., Wagner \& Witzel 1995).  
A relativistic plasma jet pointing close to our line of sight can account for the
observed properties of these objects.
BL Lacertae is an optically bright blazar located at $z=0.0688 \pm 0.0002$ (Miller \&
Hawley 1977) hosted by a giant elliptical galaxy with R$=15.5$ (Scarpa et al.\ 2000).
BL Lacertae is an LBL (low frequency peaked blazar) as its low energy spectral component  peaks at
millimeter to micron wavelengths while the high energy spectral component peaks in the MeV-GeV  range.
On some occasions, BL Lacertae has shown broad $H \alpha$ and $H  \beta$ emission lines
in its spectrum, raising the issue of its membership in its eponymous class (Vermeulen et al. 1995). 

BL Lacertae was observed by several multi-wavelength campaigns carried out by the Whole Earth Blazar Telescope 
(WEBT/GASP; B\"ottcher et al. 2003; Villata et al.\ 2009; Raiteri et al. 2009, 2010, 2013 and references therein).
BL Lacertae is well known for its intense optical variability on short and intra-day time-scales
(e.g.\ Massaro et al.\ 1998; Tosti et al.\ 1999; Clements \& Carini 2001; Hagen-Thorn et al.\ 2004)
and strong polarization variability (Marscher et al.\ 2008; Gaur et al.\ 2014 and references therein).
Extensive light curves for BL Lacertae have been presented by many authors and hence a number
of investigations have been carried out to search for the flux variations, spectral changes 
and any possible periodicities in the light curves (e.g.\ Racine 1970; Speziali \& Natali 1998; 
B{\"o}ttcher et al.\ 2003; Fan et al.\ 2001; Hu et al.\ 2006).  Nesci et al.\ (1998) found the 
source to be variable with the amplitudes of flux
variations larger at shorter wavelengths. 
Papadakis et al.\ (2003) studied the rise and decay time-scales of the source during the course of a single 
night and found them to increase with decreasing frequency. They also studied the time-lags between the light
curves in different optical bands and found the B band to lead the I band by $\sim$0.4 hours. Villata et al.\ (2002) 
carried out a campaign in 2000--2001 with exceptionally dense
temporal sampling, which was able to measure
intra-night flux variations of this blazar. They found the optical spectrum to 
be only weakly sensitive to the long term brightness trend and argued that this achromatic modulation of the flux base level 
on long time-scales is due to variations of the jet Doppler factor. However, the short-term
 flux variations and especially the bluer-when-brighter trend indicate the importance of intrinsic  processes related to the jet 
 emission mechanism (Raiteri et al. 2013; Agarwal \& Gupta 2015).

A key motivation of this study is to look for intra-day flux and spectral variations
in optical bands during the active state of BL Lacertae in 2010--2012. We also studied interband BVRI time delays
on intra-day time-scales of BL Lacertae. As BL Lacertae is a very well known LBL and the optical bands are located
above the first peak of the spectral energy distribution, the fast intra-day variability (IDV) properties
can yield rather direct implications for the nature of  the acceleration and cooling mechanisms of the   
relativistic electron populations.
Over the course of 3 years, we performed quasi-simultaneous optical multi-band photometric monitoring of this source
from various telescopes in Bulgaria, Greece, India and the USA on intra-day time-scales.

The paper is organized as follows. In Section 2 we briefly describe the observations and data reductions. Section 3 discusses
the methods of quantifying variability. We present our results in Section 4.  Sections 5 \& 6 contain a discussion and our conclusions,
respectively.

\section{Observations and Data Reduction}

Our observations of BL Lacertae started on 10 June 2010 and ran through 26 October 2012.
 The entire observation log is presented in Table 1. The observations were carried out at 
seven telescopes in Bulgaria,  Greece, India and the USA.
The telescopes in Bulgaria, Greece and India are described in detail in Gaur et al.\ (2012, Table 1) 
and the standard data reduction methods we used at each telescope are given in Section 3 of that paper,
so we will not repeat them here. During our observations, typical seeing vary
between 1--3 arcsec.
In our  observations of BL Lacertae, comparison stars are observed in the same field as the blazar 
and their magnitudes are taken from Villata et al.\ (1998).
We used star C  for calibration as it has both magnitude and colour close to those of BL Lacertae 
during our observations.

At the MDM Observatory on the south-west ridge of Kitt Peak, Arizona, USA,
data were taken for limited periods during the nights of 3, 4, 5, 6, 7 and 8
July 2010 with the 1.3 m McGraw-Hill Telescope, using the Templeton CCD with
B, V, R, and I filters. CCD parameters are described in Table 2. The standard data reduction was performed using \texttt{IRAF}
\footnote{IRAF is distributed by the National Optical Astronomy Observatories, which are operated
by the Association of Universities for Research in Astronomy, Inc., under cooperative agreement with the
National Science Foundation.},
including bias subtraction and flat-field division. Instrumental magnitudes of BL Lacertae
 plus four comparison stars in the field (Villata et al.\ 1998) were extracted
using the \texttt{IRAF} package \texttt{DAOPHOT}\footnote{Dominion Astrophysical
Observatory Photometry software} with an aperture radius of 6 arcsec and
a sky annulus between 7.5 and 10 arcsec. 

The host galaxy of BL Lacertae is relatively bright, so, in order to remove its contribution from the
observed magnitudes, we first dereddened the magnitudes using the Galactic extinction coefficient of Schlegel
et al.\ (1998) and converted them into fluxes. We then subtracted the host galaxy contribution
from the observed fluxes in R band  by considering different aperture radii used by different observatories for the
extraction of BL Lacertae magnitudes, using Nilsson et al.\ (2007). We inferred the host 
galaxy contribution in B, V and I bands
by adopting the elliptical galaxy colours of V$-$R $ = 0.61$, B$-$V $= 0.96$ and R$-$I $= 0.70$ 
from Fukugita et al.\ (1995).
Finally, we subtracted the host galaxy contribution in the B, V and I bands in order to avoid  host
contamination in the extraction of colour indices.

\section{Variability Detection Criterion}

\subsection{Power Enhanced F-test}

The F-test, as described by de Diego  (2010), provides a standard criterion for testing for the presence of intra-night
variability. The F-statistic is defined as the ratio of two given sample variances such as s$_Q ^{2}$ for the blazar instrumental light 
curve measurements and s$_* ^{2}$ for that of the standard star, i.e.,
\begin{equation}
F=\frac {s_Q ^{2}}{s_* ^{2}} .
\end{equation}
Usually, two comparison stars in the blazar field  are used to calculate $F_{1}$ and $F_{2}$, and evidence of variability is 
claimed if both the F-tests simultaneously reject the null hypothesis at a specific significance level (usually 0.01 or 0.001) ( 
Gaur et al. 2012 and references therein).
In this case, the number of degrees of freedom for each sample, $\nu_Q$ and $\nu_*$ will be the same, and equal to the number of
measurements, $N$ minus 1 ($\nu = N - 1$).

Recently, de Diego (2014) called this procedure the ``Double Positive Test'' (DPT) and pointed out a problem with this 
procedure that DPT has very low power and in practice its significance level can not be calculated. Also,
  a large brightness difference or some variability in one of the stars may lead to an underestimation of the 
source's variability with respect to the dimmer or less variable star.
To avoid this issue we employed the power-enhanced F-test using the approach of de Diego (2014) and de Diego et al. (2015). 
It consists of increasing the number of degrees of freedom in the denominator of the F-distribution by stacking 
all the light curves of the standard stars and it consists
in transforming the comparison star differential light curves to have the same photometric noise as if their magnitudes
 matched exactly the mean magnitude of the target object. The mean brightness of both the comparison star and the target
object are matched to ensure that the photometric errors are equal. Including 
multiple standard stars reduces the possibility of false detections of intra-night variability that can be produced by 
one single peculiar comparison star light curve. More details are provided in de Diego et al. 2015. In the analysis,
 we used three standard stars B, C and H in the field of BL Lacertae whose brightnesses are very close to 
the brightness of BL Lacertae. Thus, for the $i^{th}$ observation of the 
light curve of standard star $j$, for which we have $N_j$ data points, we calculate the square deviation as: 
\begin{equation}
    s_{j,i}^2 = (m_{j,i} - \overline{m}_{j})^2.
\end{equation}

By stacking the results of all observations on the total of $k$ comparison stars, we can calculate the combined variance as:
\begin{equation}
    s_c^2 = \frac{1}{(\sum_{j=1}^k N_j) - k} \sum_{j=1}^k \sum_{i=1}^{N_j} s_{j,i}^2.
\end{equation}
Then, we compare this combined variance with the blazar light curve variance to obtain the $F$ value
with $\nu_Q = N - 1$ degrees of freedom in the numerator and $\nu_* = k (N - 1)$  degrees of freedom in the denominator. 
This value is then compared with the $F^{(\alpha)}_{\nu_Q,\nu_*}$ critical value, where $\alpha$ is the significance 
level set for the test.
The smaller the $\alpha$ value, the more improbable it is that the result is produced by chance. If $F$ is larger
than the critical value, the null hypothesis (no variability) is discarded. We have performed the
$F$-test at the $\alpha = 0.001$ level. 

\subsection{Discrete Correlation Function (DCF)}

To estimate the variability time-scales in the observed light curves of BL Lacertae and to determine the cross-correlations
between different optical bands, we used the Discrete Correlation Function (Edelson \& Krolik 1988; Hovatta et al. 2007).

The first step is to calculate the unbinned correlation (UDCF) using the given time series by:
\begin{equation}
UDCF_{ij} = {\frac{(a(i) - \bar{a})(b(j) - \bar{b})}{\sqrt{\sigma_a^2 \sigma_b^2}}}.
\end{equation}
Here $a(i)$ and $b(j)$ are the individual points in two time series $a$ and $b$, respectively, $\bar{a}$
and $\bar{b}$ are respectively the means of the time series, and $\sigma_a^2$ and $\sigma_b^2$ are their variances.
The correlation function is binned after calculation of the UDCF. The DCF can be calculated by averaging the 
UDCF values ($M$ in number) for each time delay
\( \Delta t_{ij} = (t_{yj}-t_{xi}) \) lying in the range  \( \tau - \frac{\Delta\tau}{2} \leq t_{ij} 
\leq \tau+ \frac{\Delta\tau}{2} \) via

\begin{equation}
DCF(\tau) = {\frac{1}{n}} \sum ~UDCF_{ij}(\tau) .
\end{equation}
where $\tau$ is the centre of a time bin and $n$ are the number of points in each bin.
 DCF analysis is frequently used for finding the correlation and possible lags between multi-frequency
AGN data. When the same data train is used, so $a$=$b$, there is obviously a peak at zero lag and is called Auto 
Correlation Function (ACF), indicating that there 
is no time lag between the two but any other strong peaks in the ACF give indications of variability timescales.

\section{Results}

\subsection{Flux variations}

We extensively observed the source for a  total of 38 nights during 2010--2012.   During  
26 of those nights, we observed the source quasi-simultaneously in the B, V, R and I bands, providing a total of 113 light 
curves in B, V, R and I. These light curves are displayed in Figs.\ 1 and 2.
The  lengths of the individual observations were usually between  2 and 6 hours.
It is clear from the figures that BL Lacertae was variable from day to day and also on hourly time-scales.
We searched for genuine flux variations on IDV time-scales and found 50 light curves to be variable in the whole set of filters
 during a total of 19 nights using the enhanced F-test. 
The intranight variability amplitudes (in per cent) are given by Heidt \& Wagner (1996):
\begin{eqnarray}
Amp = 100\times \sqrt{{(A_{max}-A_{min}})^2 - 2\sigma^2},
\end{eqnarray}
where $A_{max}$ and $A_{min}$ are the maximum and minimum values in the calibrated light curves of the blazar, and $\sigma$
is the average measurement error. 
The enhanced F-test values 
are presented in Table 1. 

The duty cycle (in per cent) (DC) of BL Lacertae is computed following the definition of Romero et al.\ (1999) that 
later was used by many authors (e.g., Stalin et al.\ 2004; Goyal et al.\ 2012, and references therein),

\begin{equation} 
DC = 100\frac{\sum_{i=1}^n N_i(1/\Delta t_i)}{\sum_{i=1}^n (1/\Delta t_i)}  ,
\label{eqno1} 
\end{equation}
where $\Delta t_i = \Delta t_{i,obs}(1+z)^{-1}$ is the duration of the
monitoring session of a source on the $i^{th}$ night, corrected for
its cosmological redshift, $z$. Since for a given source the monitoring durations on different nights are not always equal, 
the computation of the DC is weighted by the actual monitoring duration
$\Delta t_i$ on the $i^{th}$ night. $N_i$ is set equal  to 1 if intra-day variability 
was detected by the F-test (given in Table 1),  otherwise $N_i$ = 0.
We found the duty cycle of BL Lacertae to be around 44 per cent. 

\subsection{Amplitude of Variability}

When the variability is substantial the amplitude of variability usually is greater in higher energy bands (B) and smaller 
in lower energy bands (R) in most of the observations.
It has been found by many investigators that amplitude of variability is greater at higher frequencies for
BL Lacertaes (Ghisellini et al. 1997; Massaro et al. 1998; Bonning et al. 2012) however, the amplitude
of variability is not systemaically larger at higher frequencies is also found on some occasions (Ghosh et al. 2000;
Ramirez et al. 2004). In few cases, we found the amplitude of variability in lower energy bands 
to be comparable to or greater than the amplitude of variability in higher energy bands. Still, because 
the errors in the B band are higher in these cases, while they are often lower in the lower frequency band,  there are 
some nights during which the statistical significance of the variations falls below our thresholds in 
B (also sometimes V), when  clear variability is detected in R and I bands.  Also, differences 
in the amplitude of variability in various optical bands change from one observation to another. 
The highest fractional amplitude of variability was found to be $\sim$38 per cent on 6 July 2010, where the  
source showed a $\sim$0.3 mag change in less than 3.5 hours in V band. \\

We searched for possible correlations between the variability amplitude and the duration of the observation, as displayed
in Fig.\ 3 (left panel). We found a significant positive correlation ($\rho=0.3306$ with $p=0.0190$, where $\rho$ and $p$ are
 the Spearman Correlation coefficient and its p-value, respectively) between them, i.e., the variability amplitude 
increases with duration of the observation.  We conclude that there is an enhanced likelihood of 
seeing higher amplitudes of variability in longer duration light curves, which is in agreement with Gupta \& Joshi (2005) report
for a larger sample of blazars.

Next we examined the possible correlation between the amplitude of variability with 
the source flux (Fig.\ 3, right panel). It can be seen that amplitude of variability decreases as the source flux increases 
($\rho$=-0.4990 with $p$=0.0002). 
This might be explained as, if the source attains a high flux state, the irregularities in the jet flow decreases 
particularly if fewer 
non-axisymmetric bubbles were carried outward in the relativistic magnetized jets (Gupta et al.\ 2008).
For instance, the blazar 3C 279 was observed in 2006 during its outburst/high state but did not show any genuine micro-variability, 
while other
sources which were in their pre/post outburst state showed significant micro-variability  (Gupta et al.\ 2008). 
This anti-correlation between variability amplitude and flux could also be explained by a two component
model where in more active phases or in an outburst state, the slowly varying jet component rises and dominates on the emission
from the more variable regions, i.e., shocks/knots, and therefore reduces the fractional amplitude. However, there is 
 scatter in the plot, some of which could be due to an observational bias, as some observations were longer than others. 
Due to this, the observed variability amplitude at similar luminosities could be quite different as the
 amplitude of variability increases with the duration of the observation.

\subsection{R-band Autocorrelations}

We searched for the presence of a characteristic time-scale of variability in all of the nights by auto-correlating the
 R-band measurements, for which we had the most data. In most of the nights, the
 auto-correlation function (ACF) shows a sharp maximum at nearly zero lag, as expected, and it stays positively 
auto-correlated with itself for time lags of somewhat less than 1 hour followed by dropping to negative values. 
 Hence, we conclude that we did not find any evidence for  characteristic variability time-scales from this approach.
One example of R-band ACF is shown in the left panel of figure 4.
\begin{figure*}
   \centering
\includegraphics[width=4cm , angle=0]{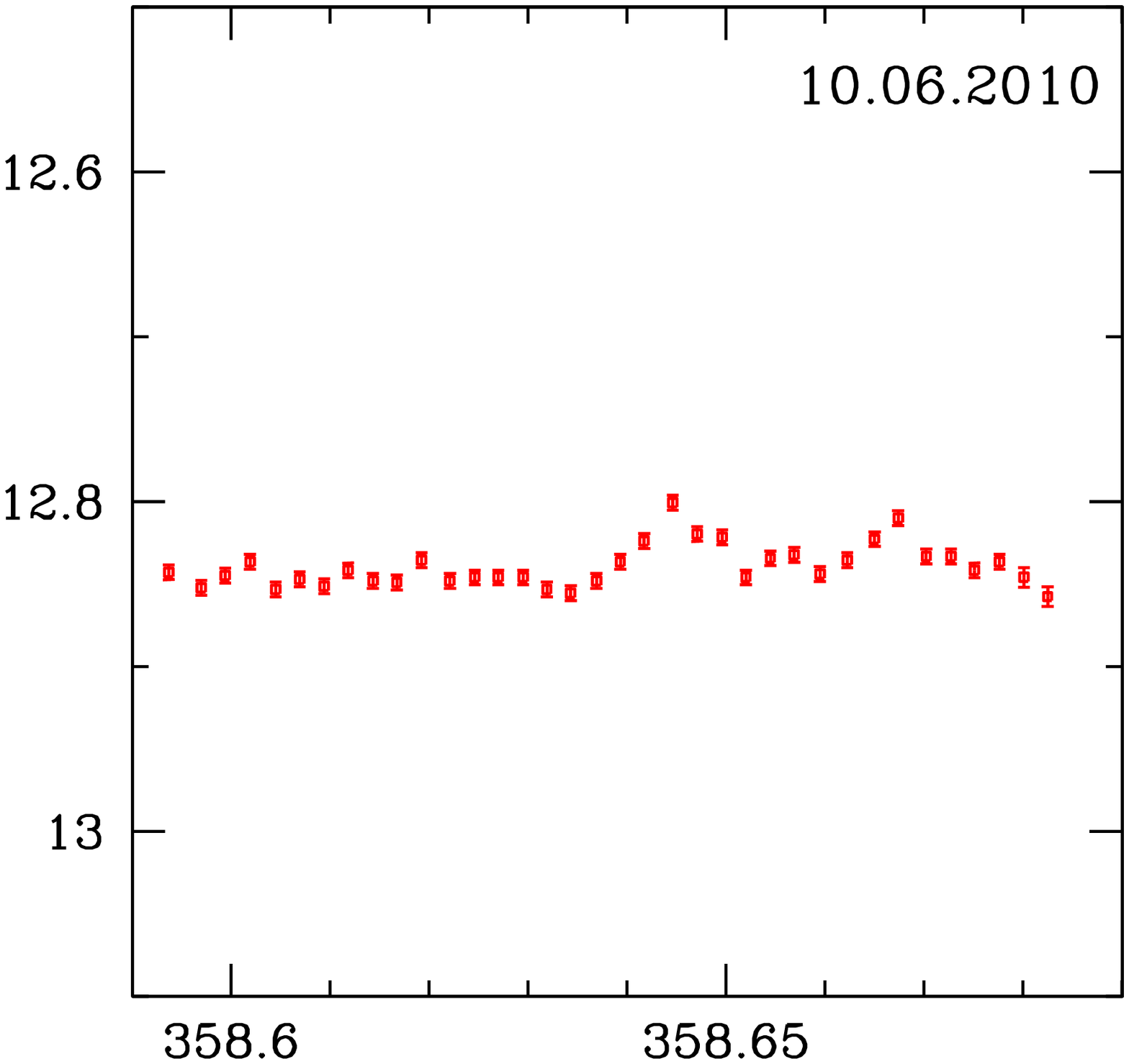}
\includegraphics[width=4cm , angle=0]{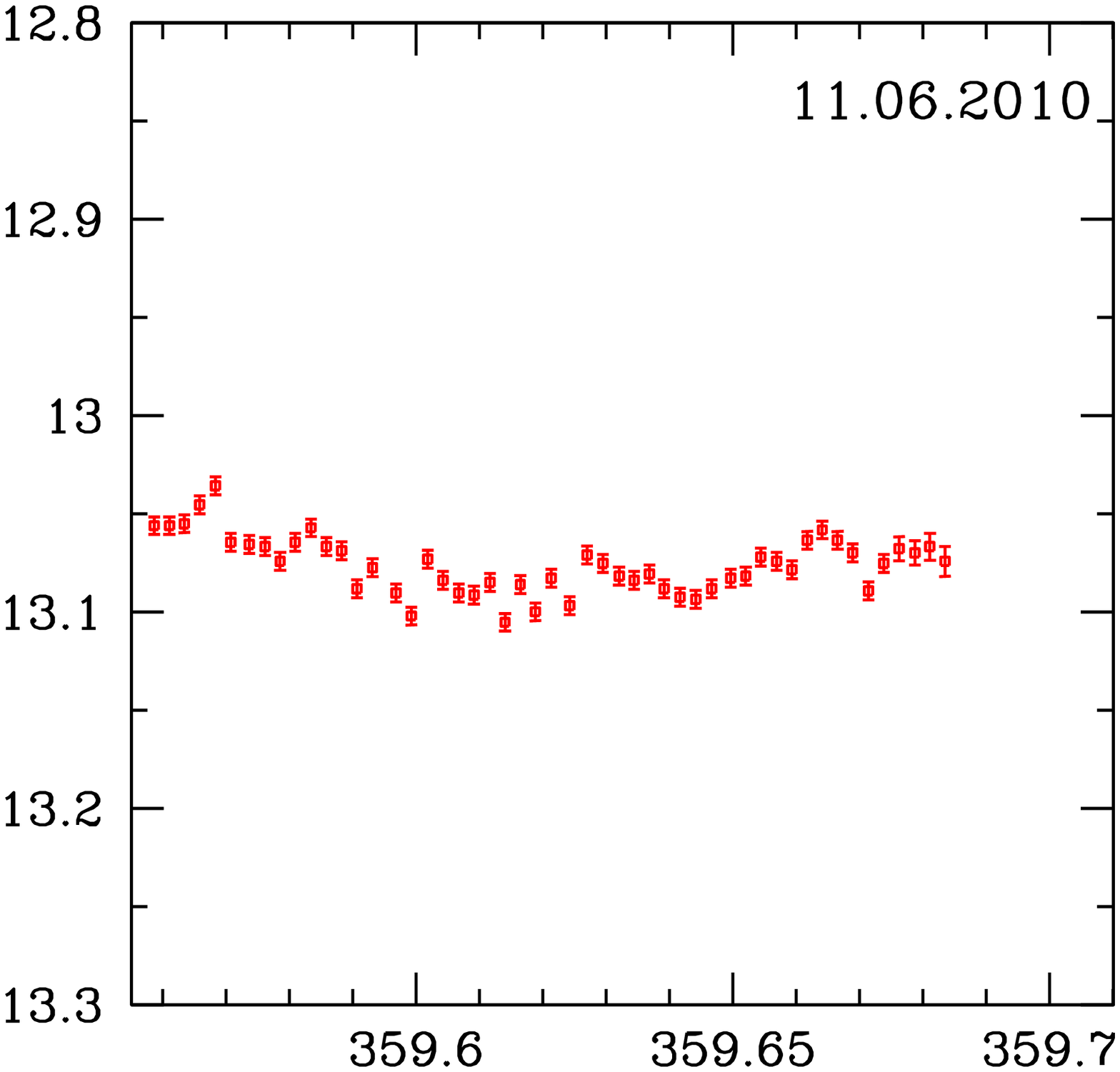}
\includegraphics[width=4cm , angle=0]{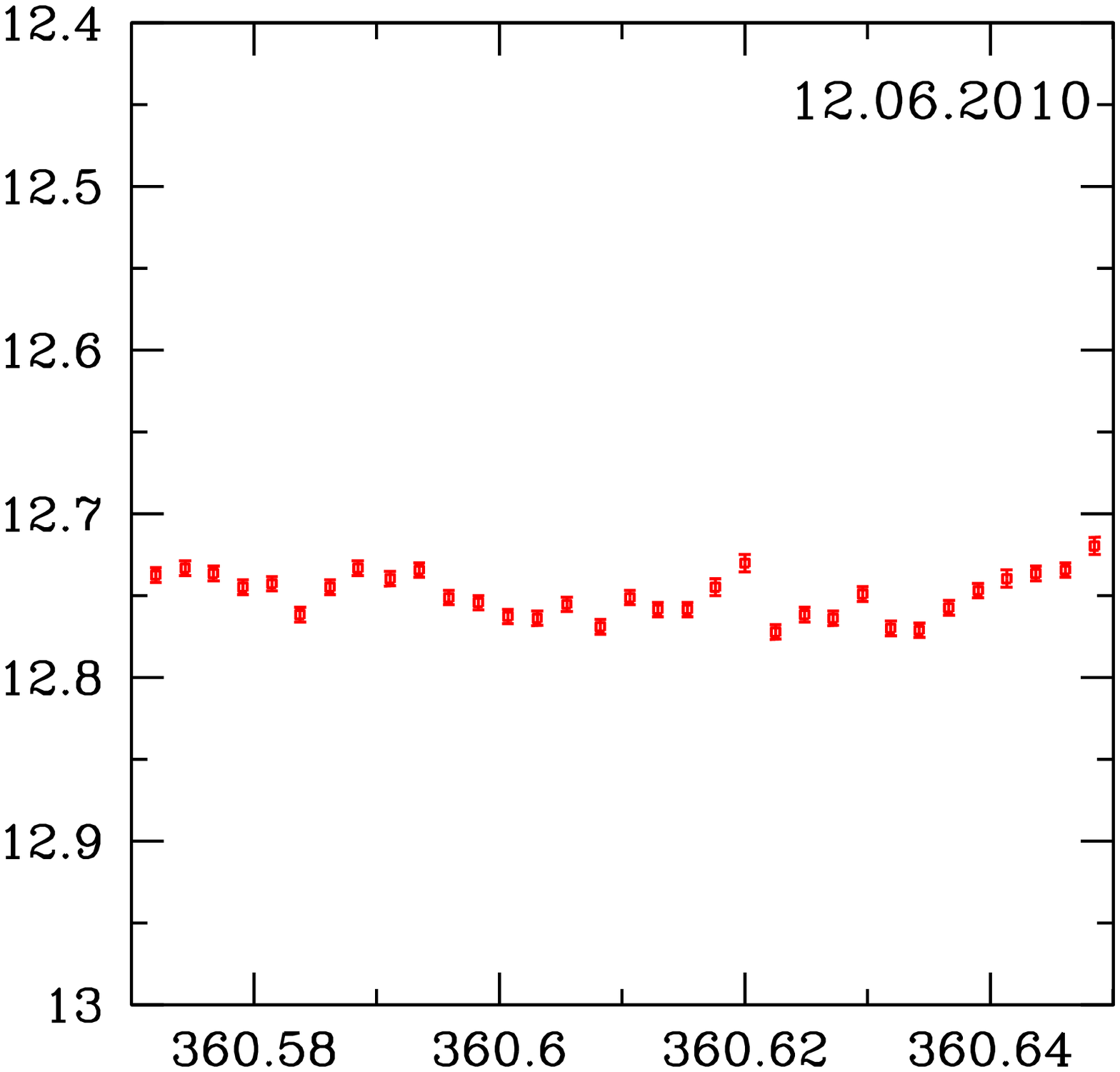}
\includegraphics[width=4cm , angle=0]{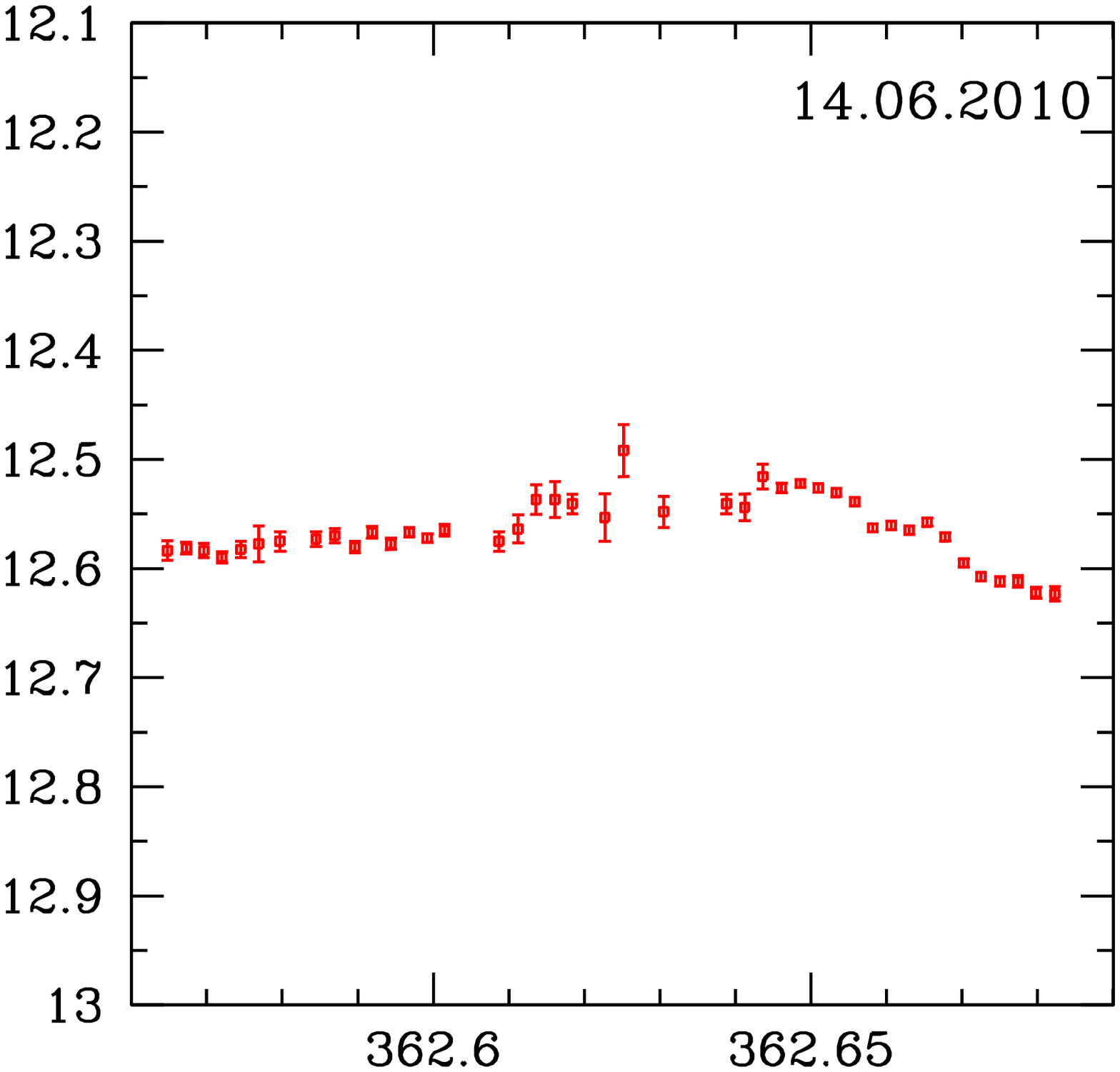}
\includegraphics[width=4cm , angle=0]{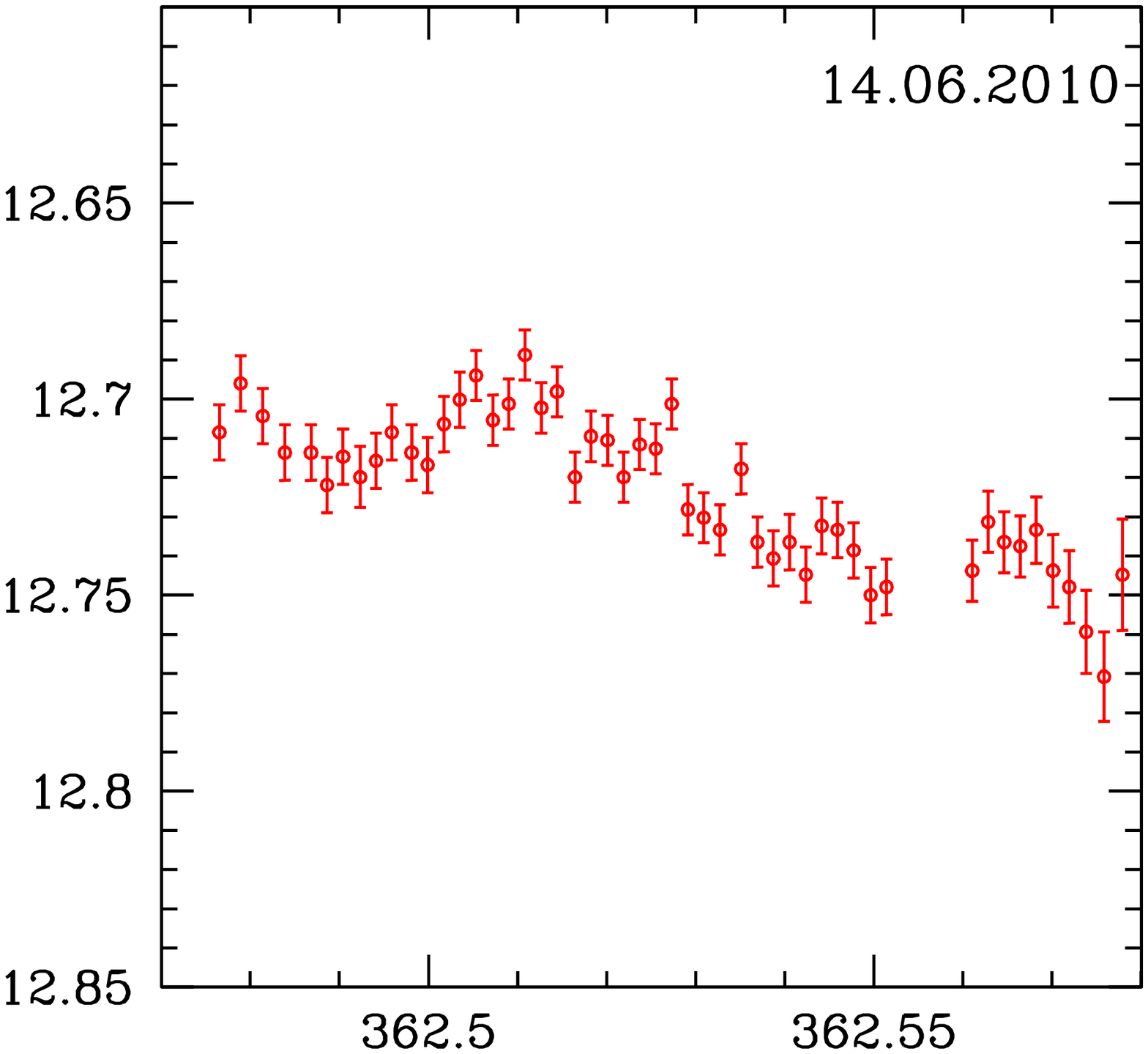}
\includegraphics[width=4cm , angle=0]{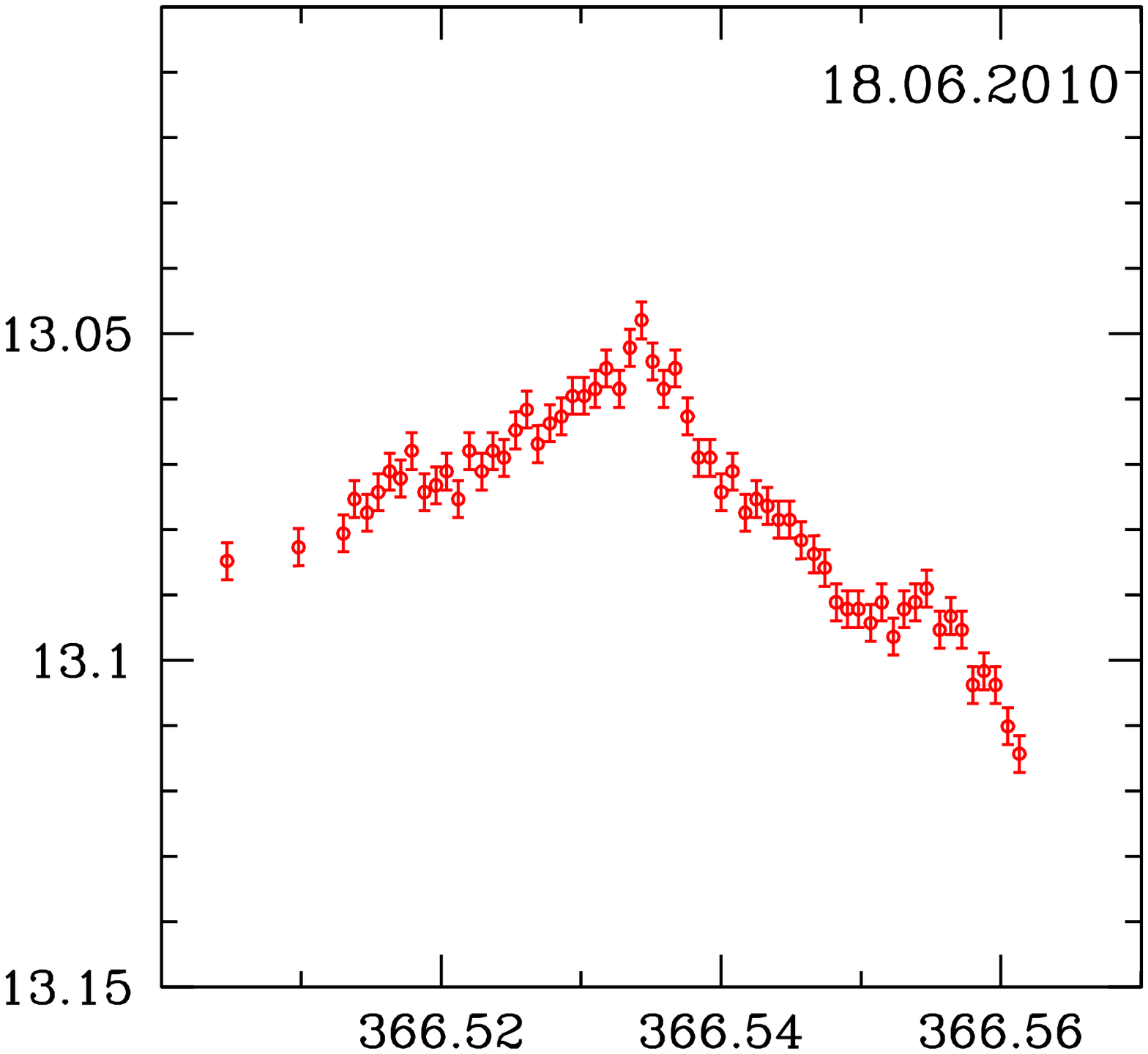}
\includegraphics[width=4cm , angle=0]{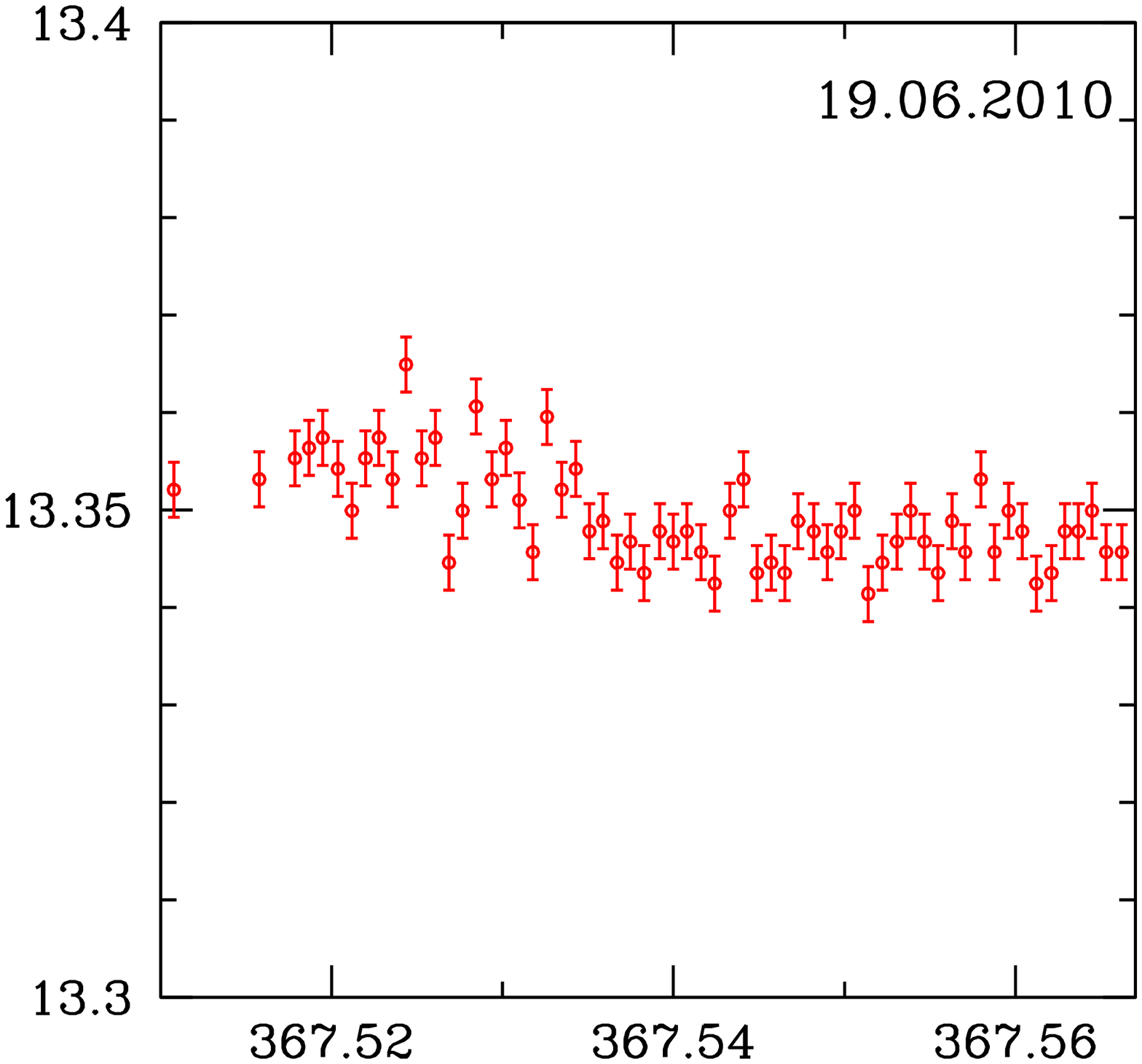}
\includegraphics[width=4cm , angle=0]{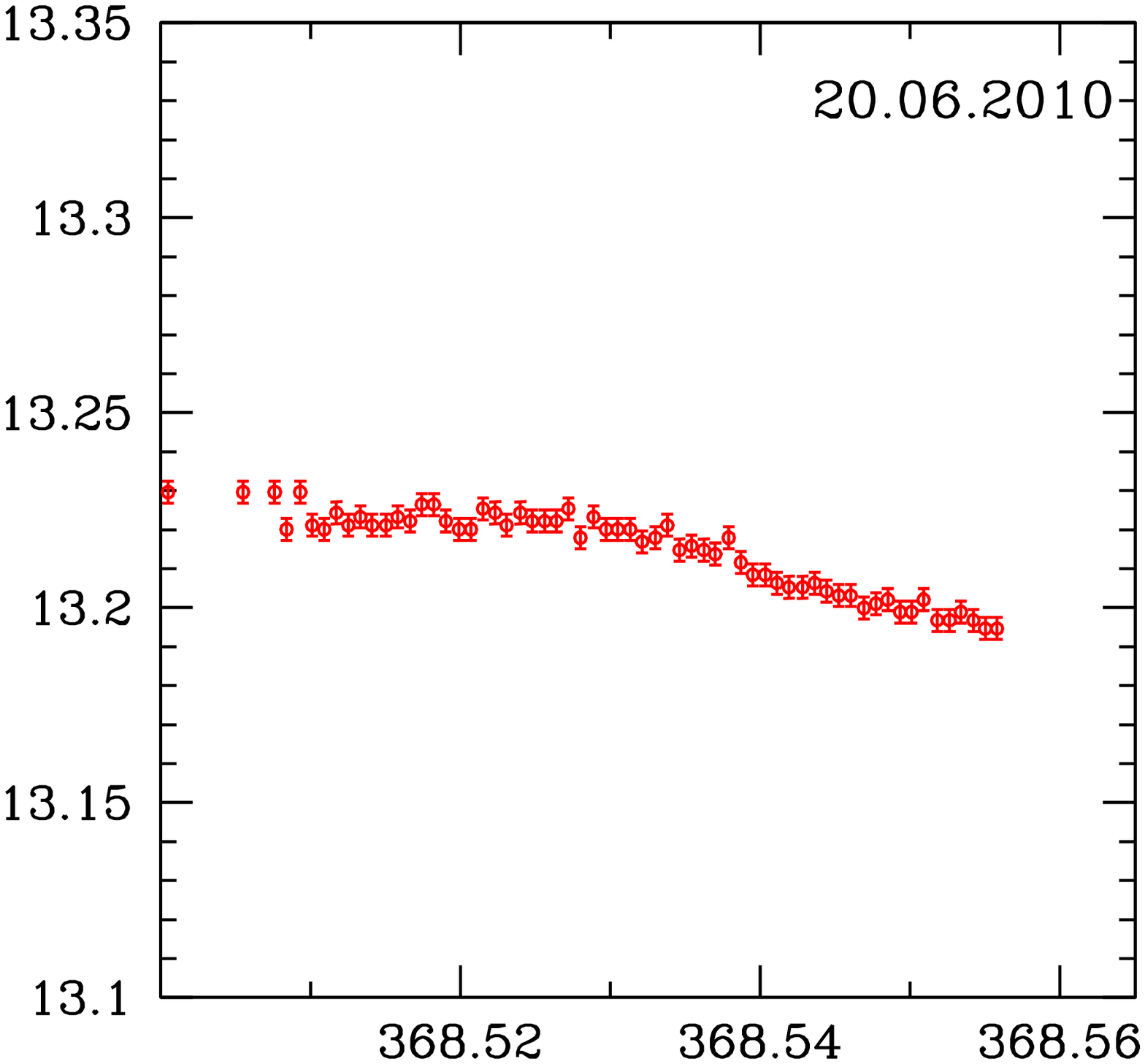}
\includegraphics[width=4cm , angle=0]{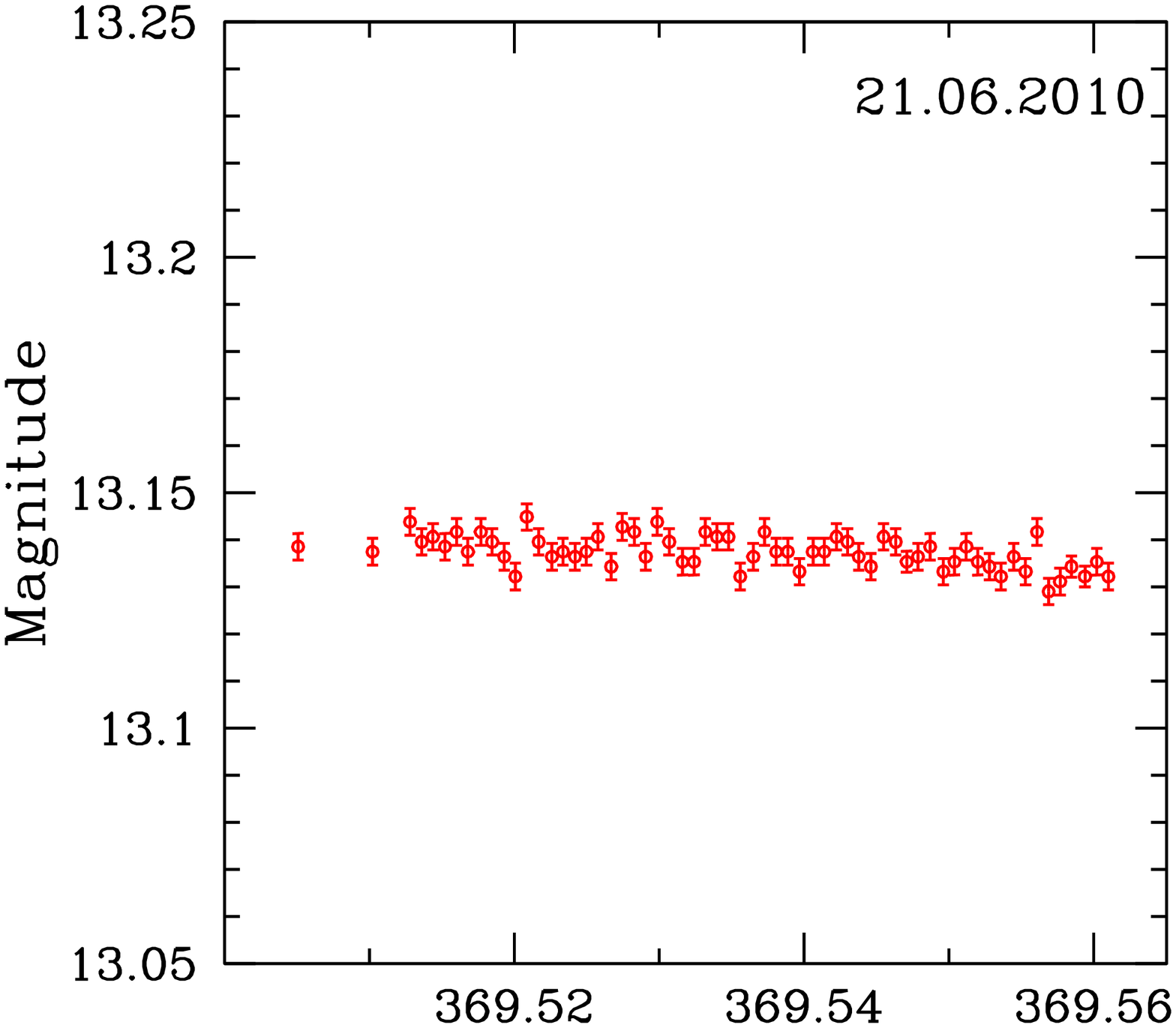}
\includegraphics[width=4cm , angle=0]{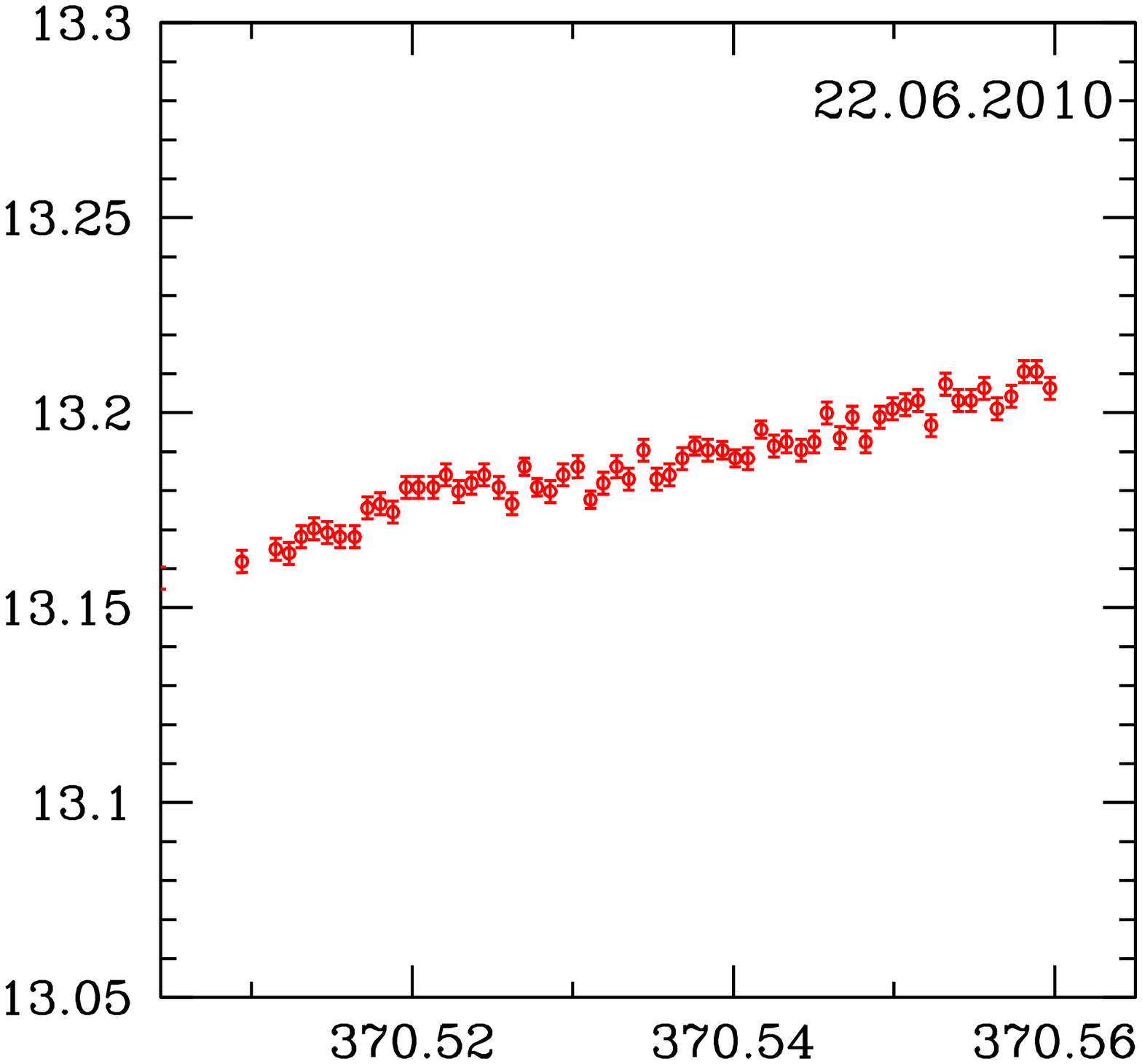}
\includegraphics[width=4cm , angle=0]{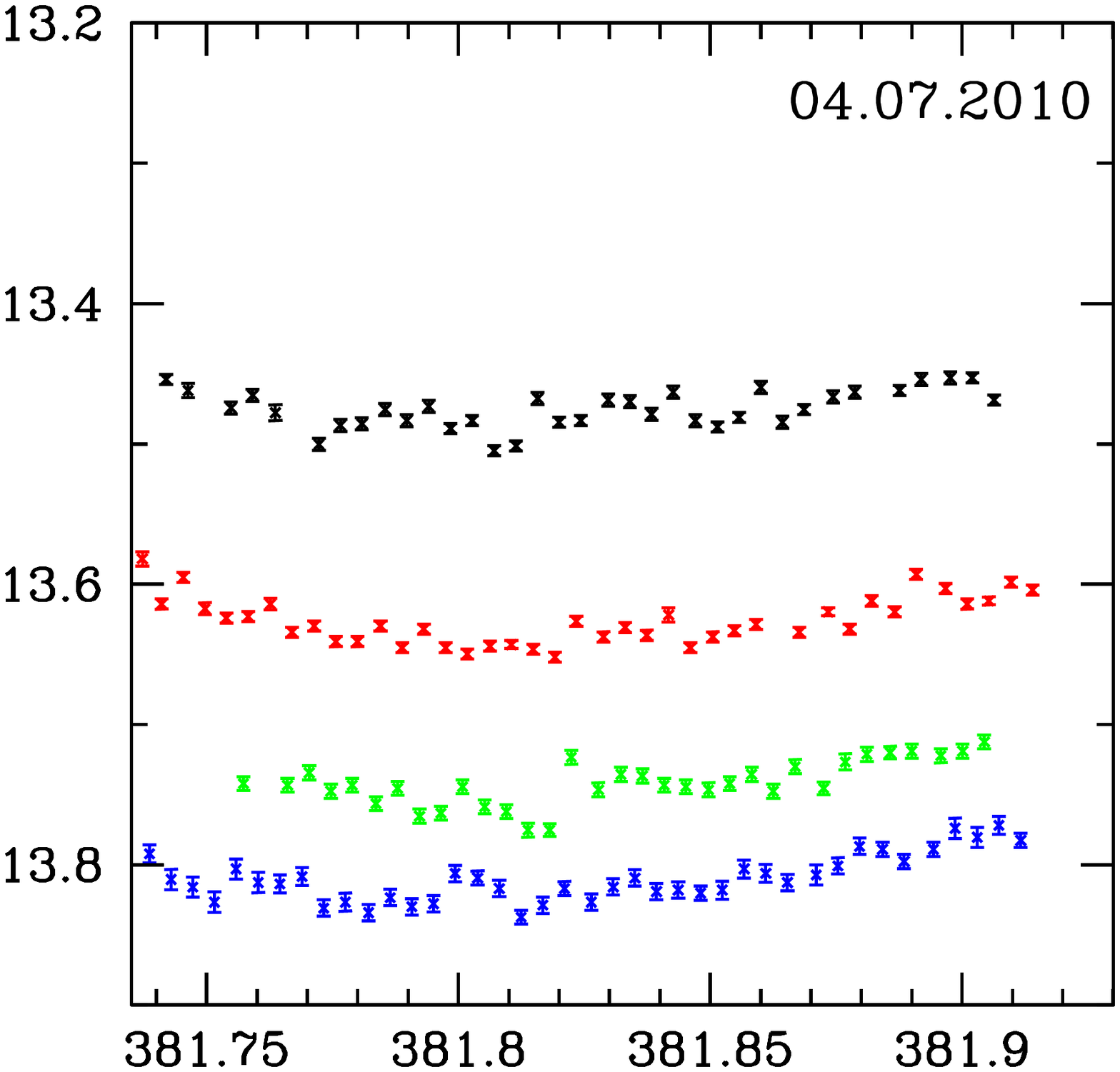}
\includegraphics[width=4cm , angle=0]{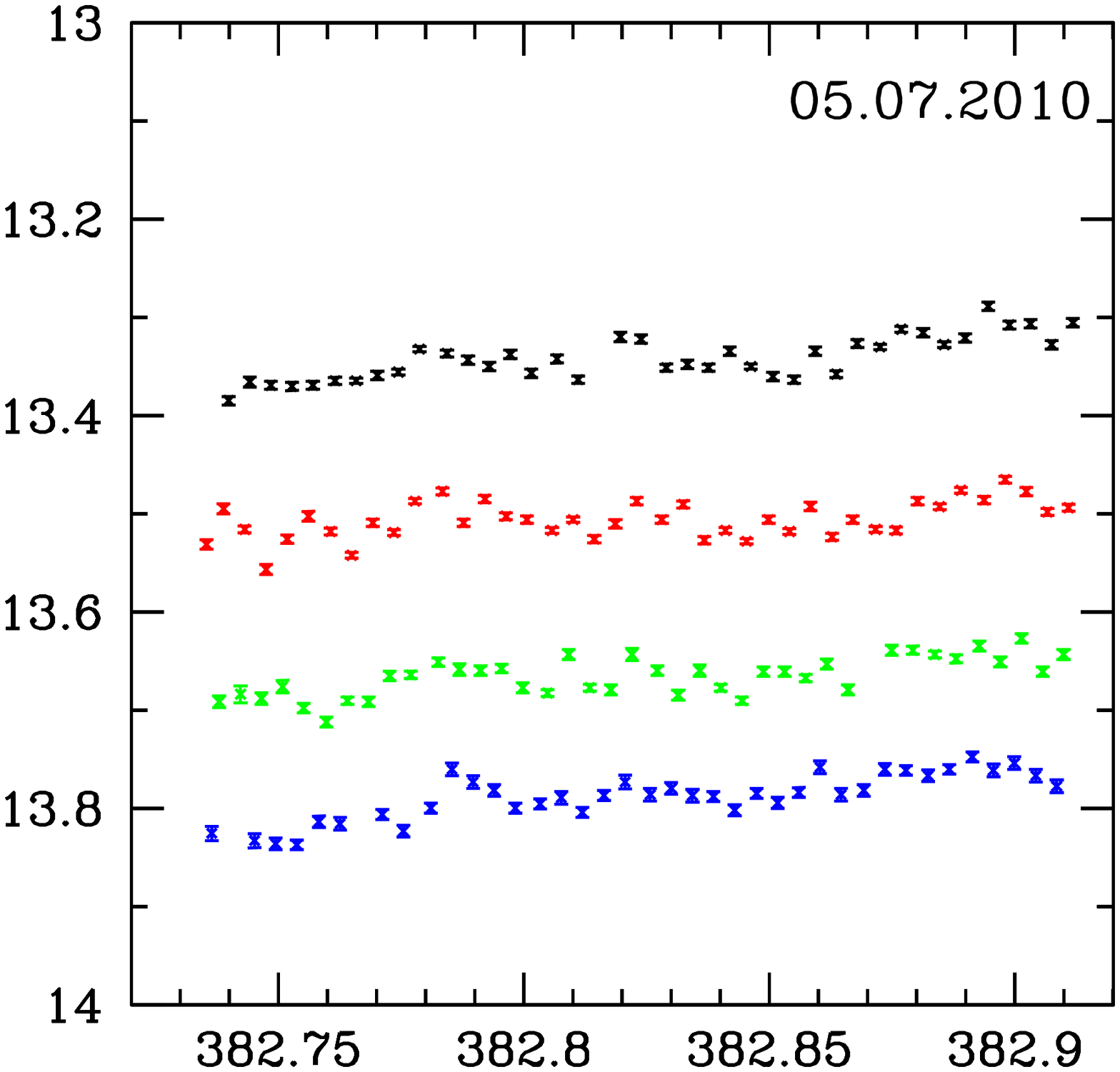}
\includegraphics[width=4cm , angle=0]{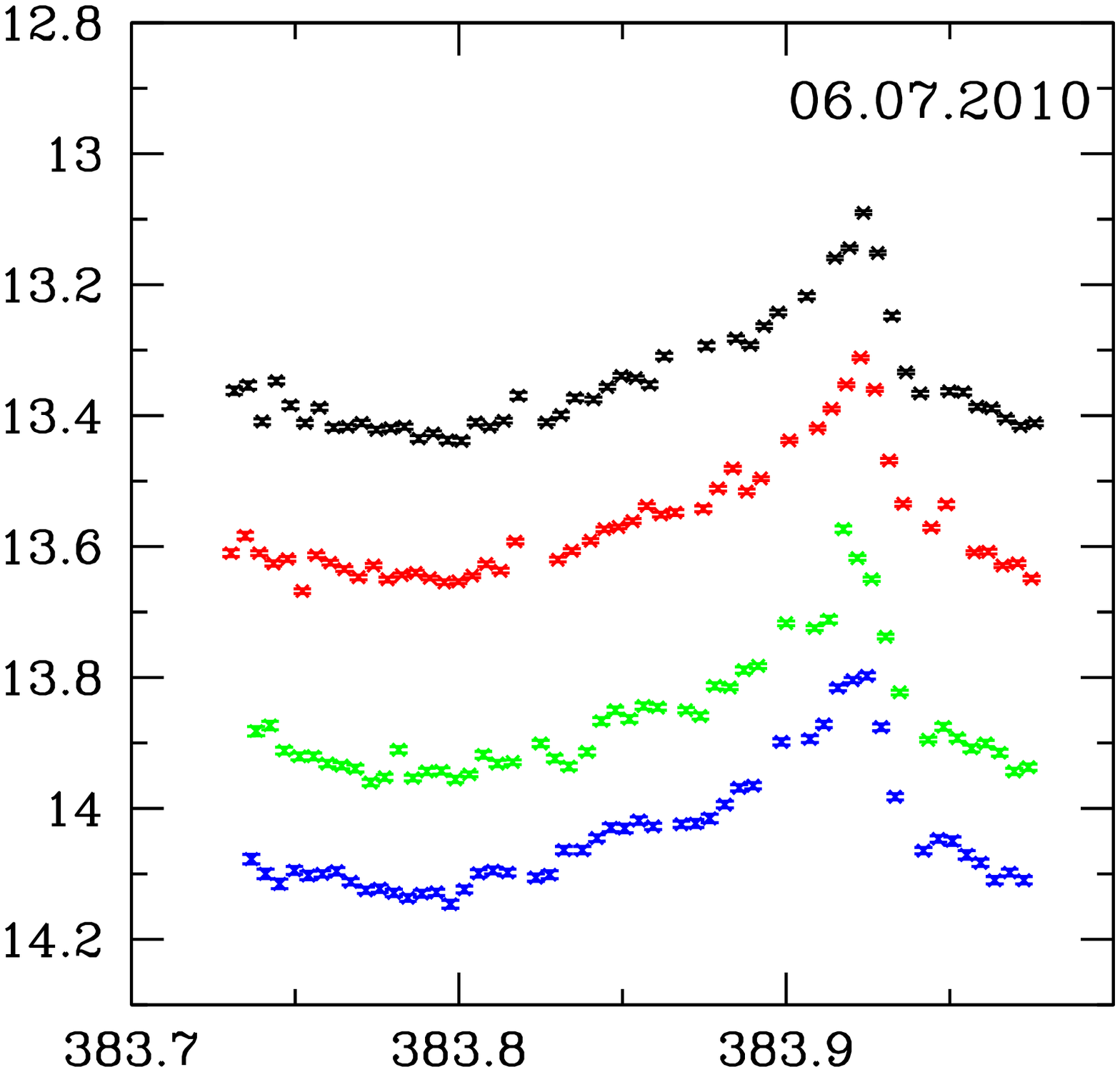}
\includegraphics[width=4cm , angle=0]{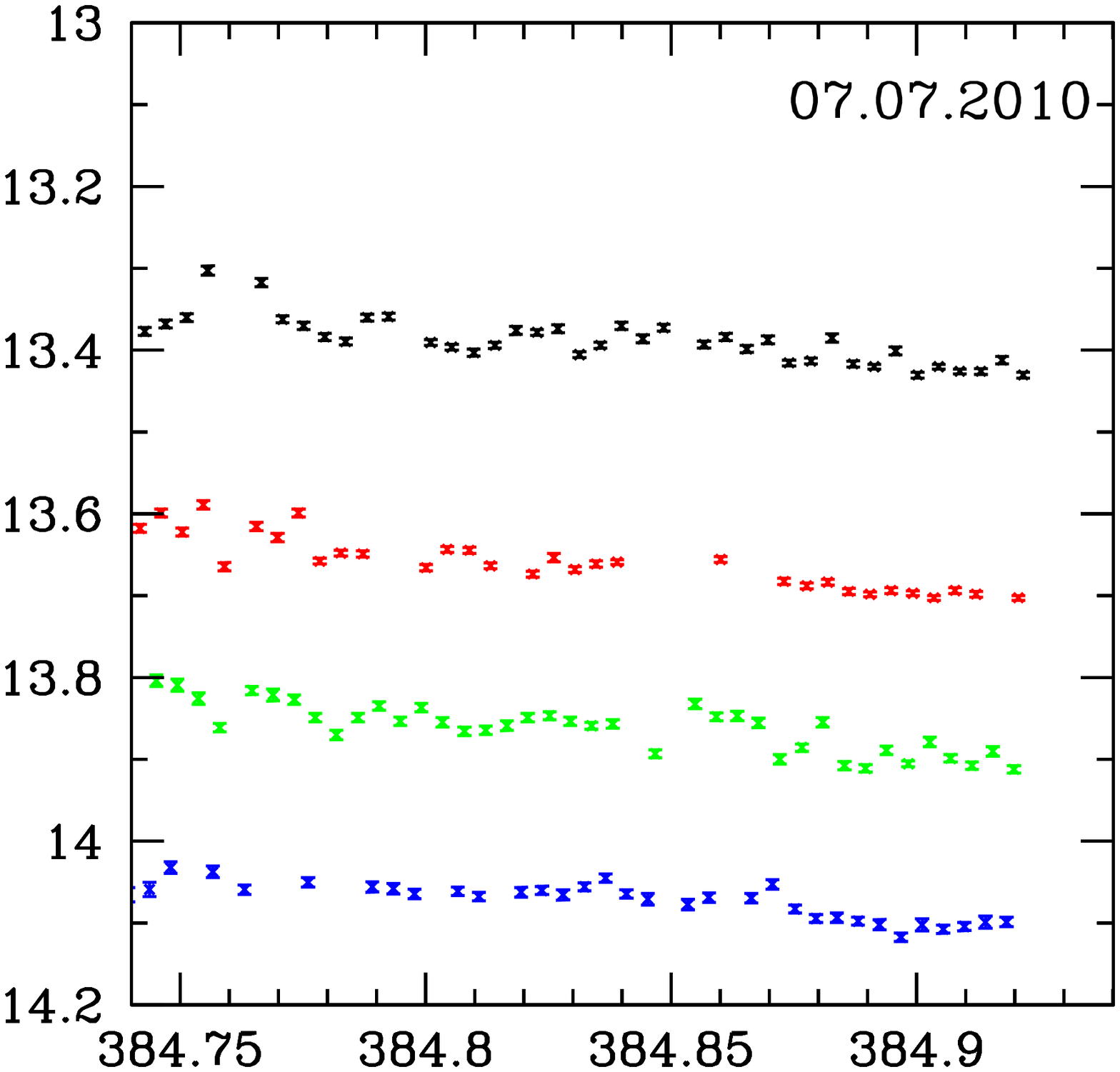}
\includegraphics[width=4cm , angle=0]{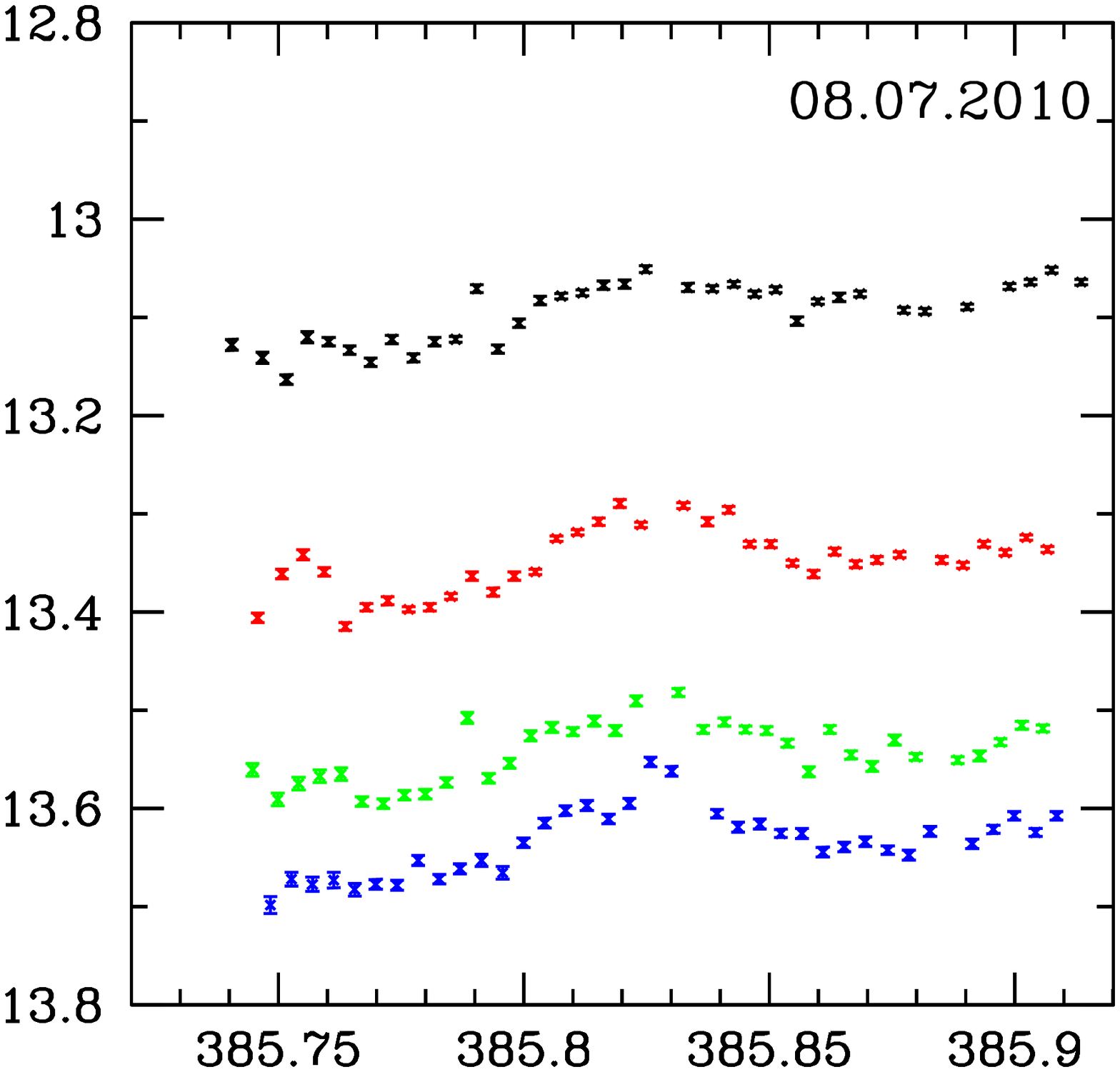}
\includegraphics[width=4cm , angle=0]{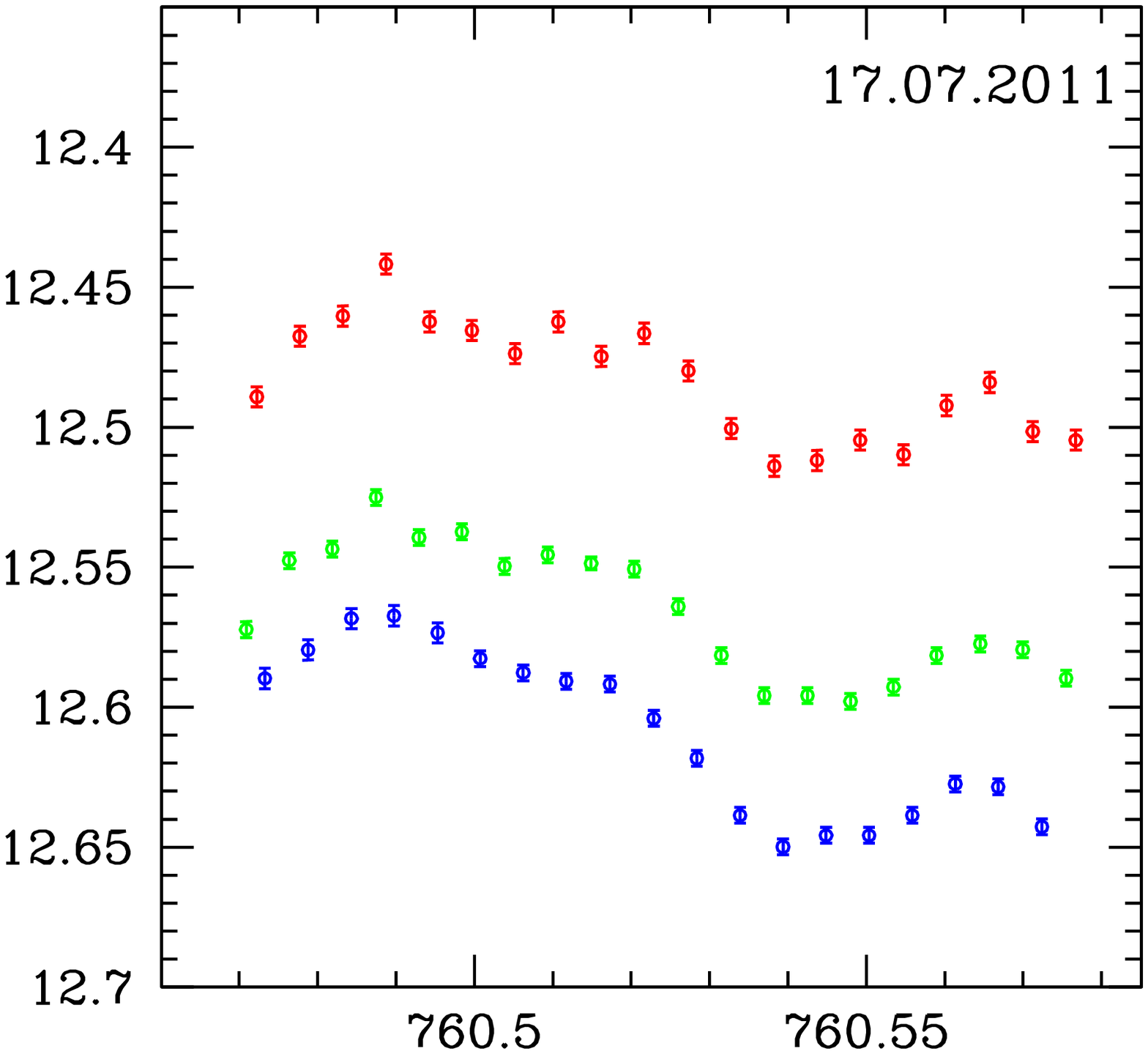}
\includegraphics[width=4cm , angle=0]{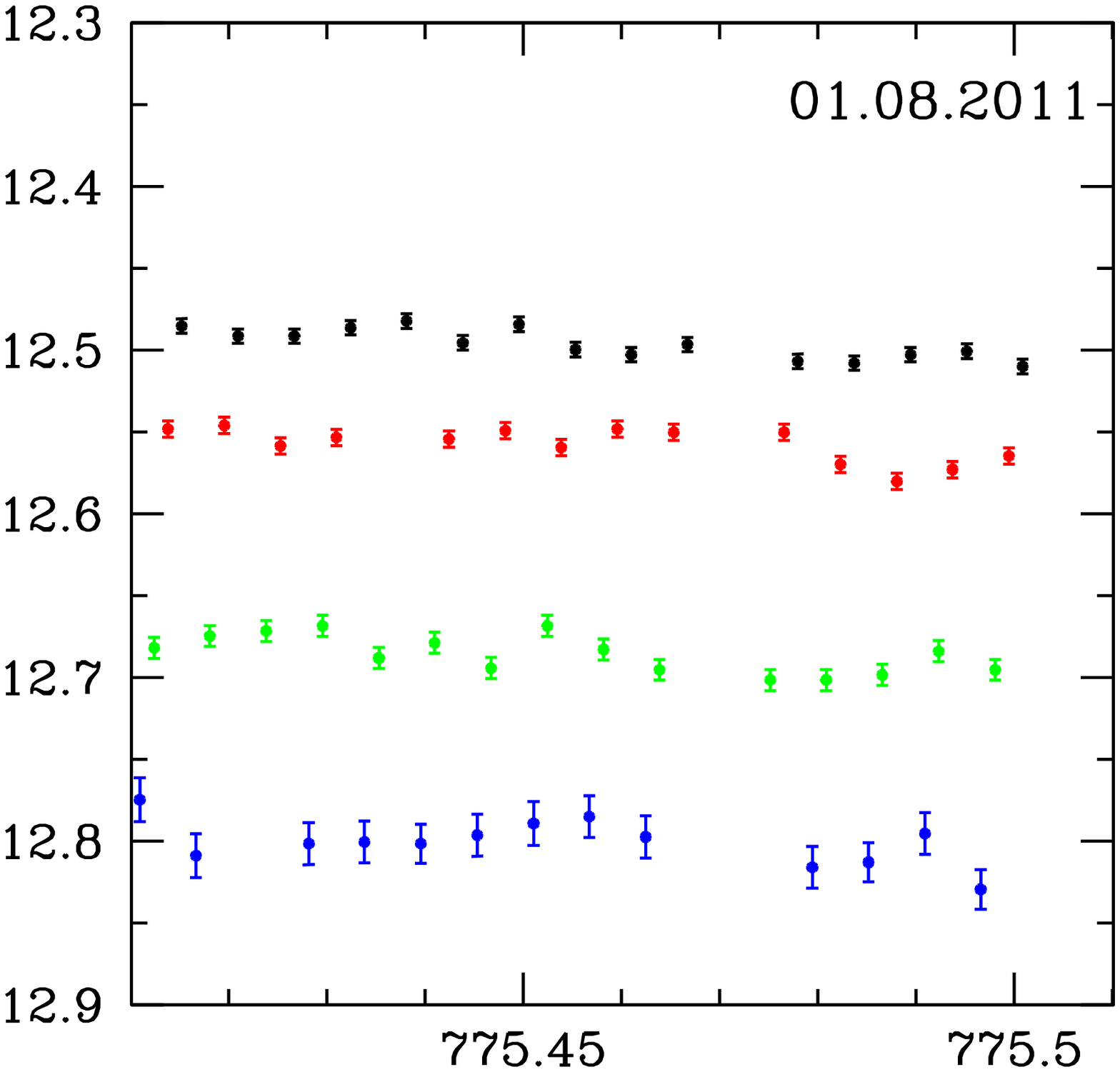}
\includegraphics[width=4cm , angle=0]{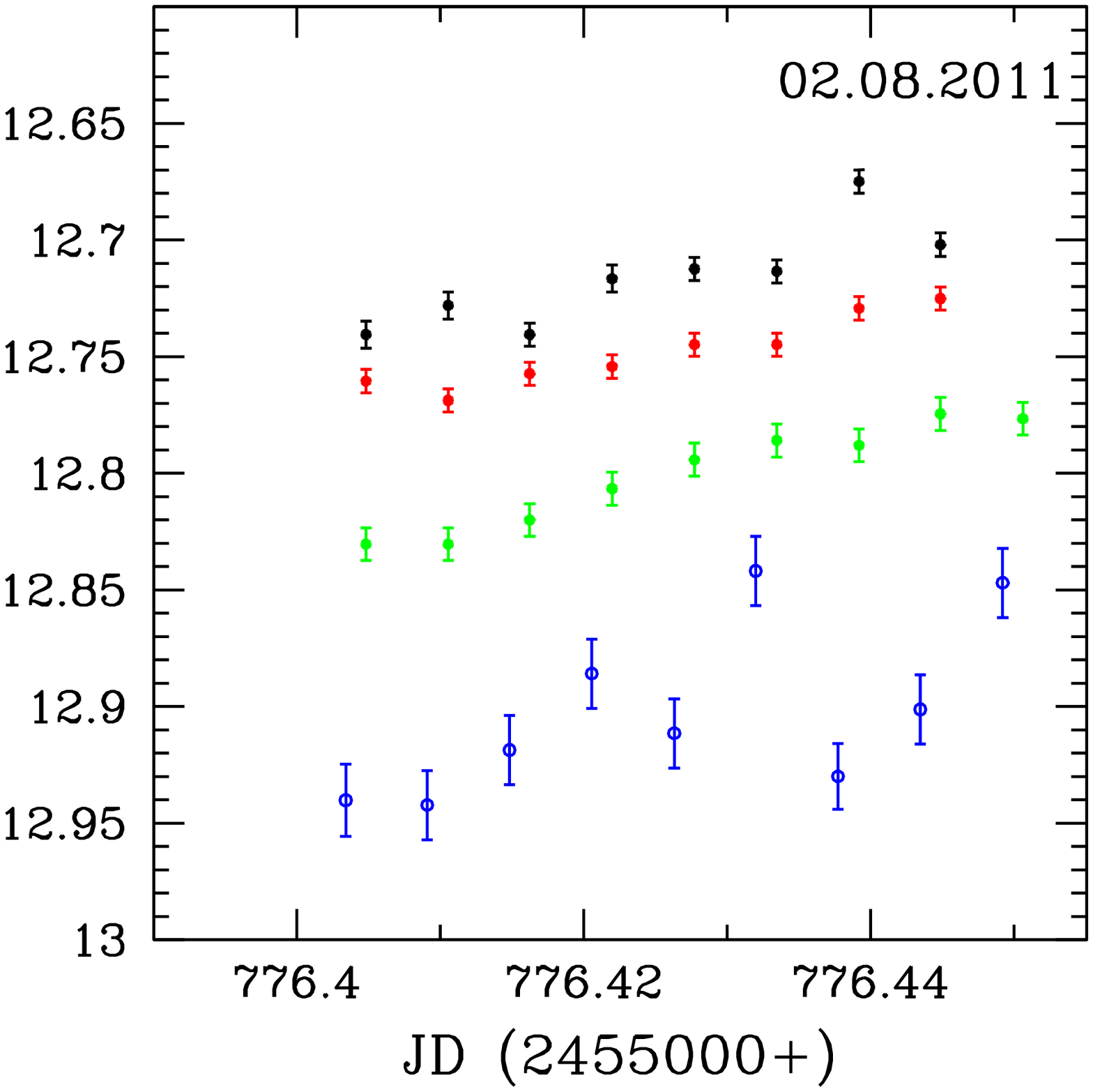}
\includegraphics[width=4cm , angle=0]{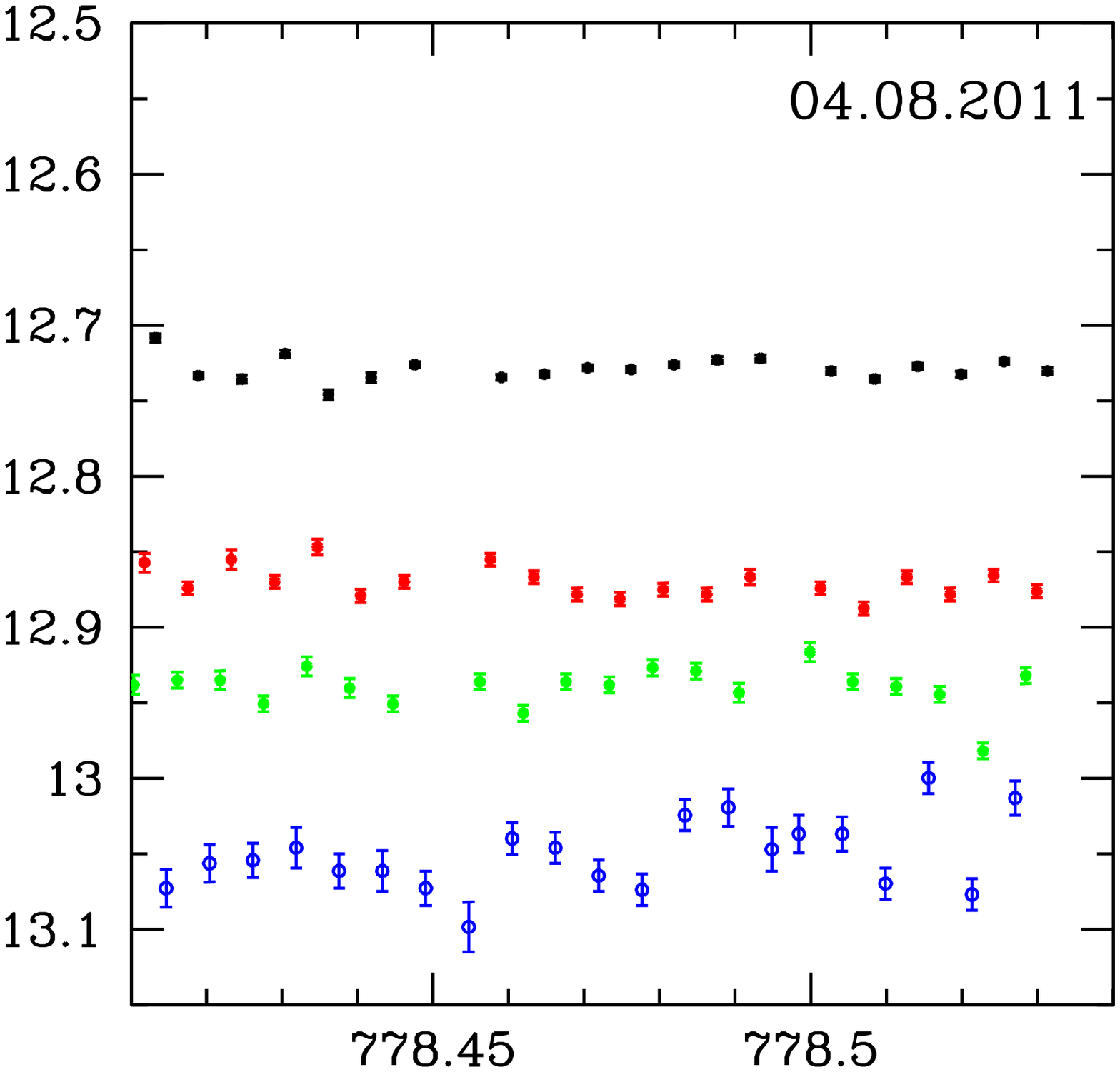}
\includegraphics[width=4cm , angle=0]{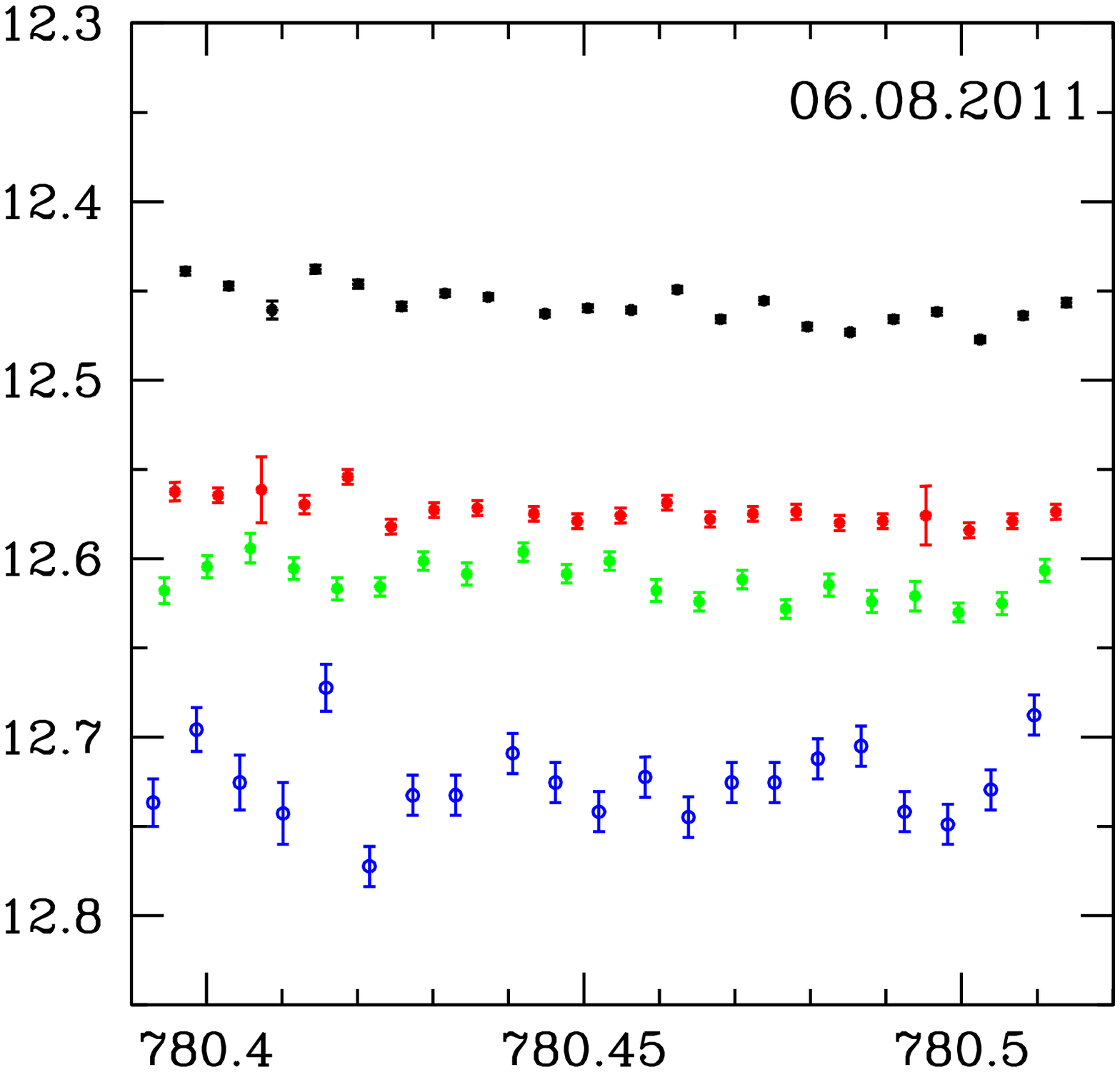}

\caption{IDV light curves of BL Lacertae during 2010 and early 2011 in the  B (blue), V(green), R(red) and I(black) bands.  The
X-axis is JD (2455000+), and the Y-axis is the calibrated magnitudes in each of the panels. The B, V and I 
bands are shifted by
arbitrary offsets with respect to R-band light curve. Observations from observatory A are represented by squares; those from
C are represented by triangles; D  by filled circles; E by open circles and F by starred symbols.}
 \end{figure*}

\begin{figure*}
   \centering
\includegraphics[width=4cm , angle=0]{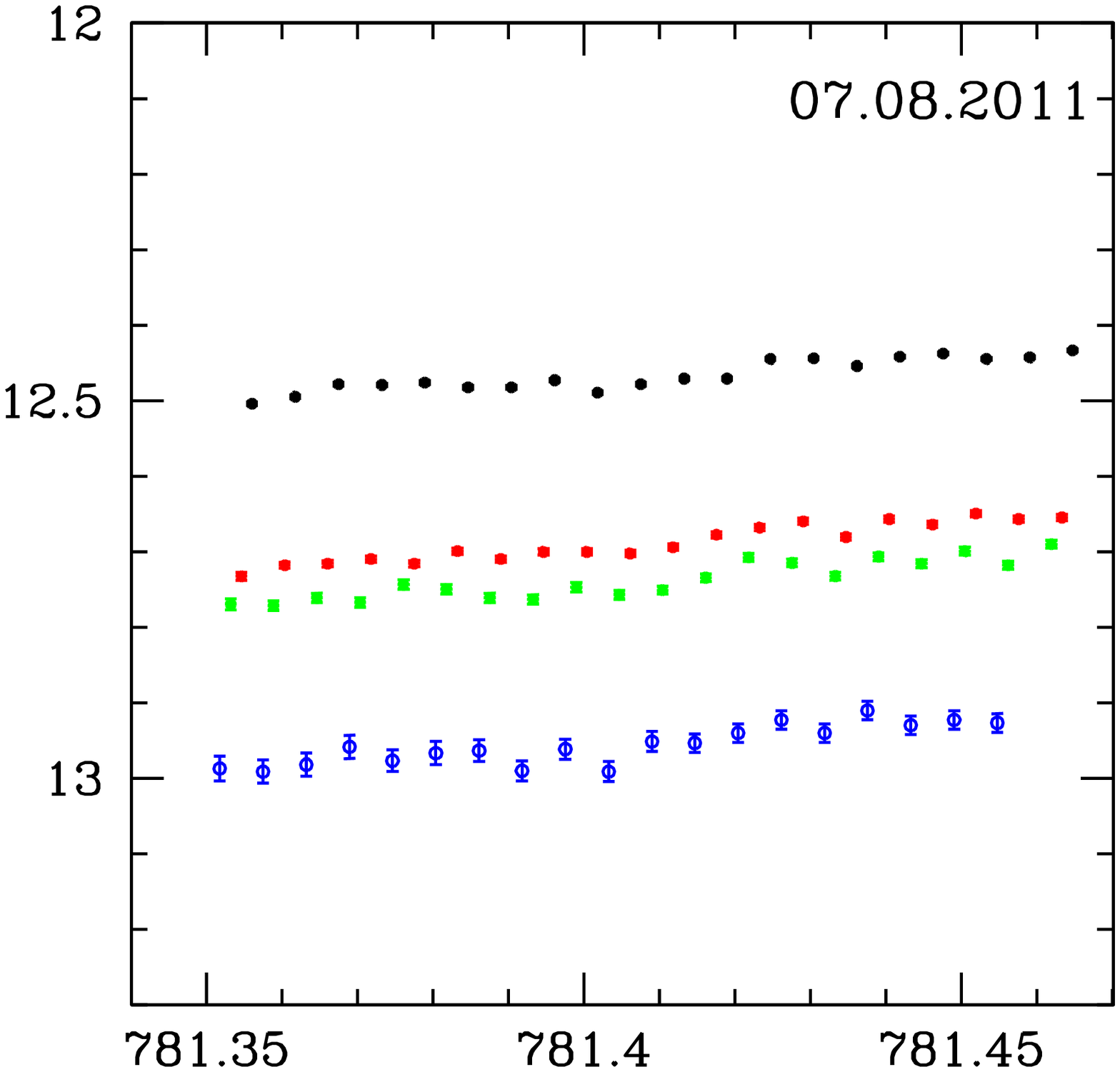}
\includegraphics[width=4cm , angle=0]{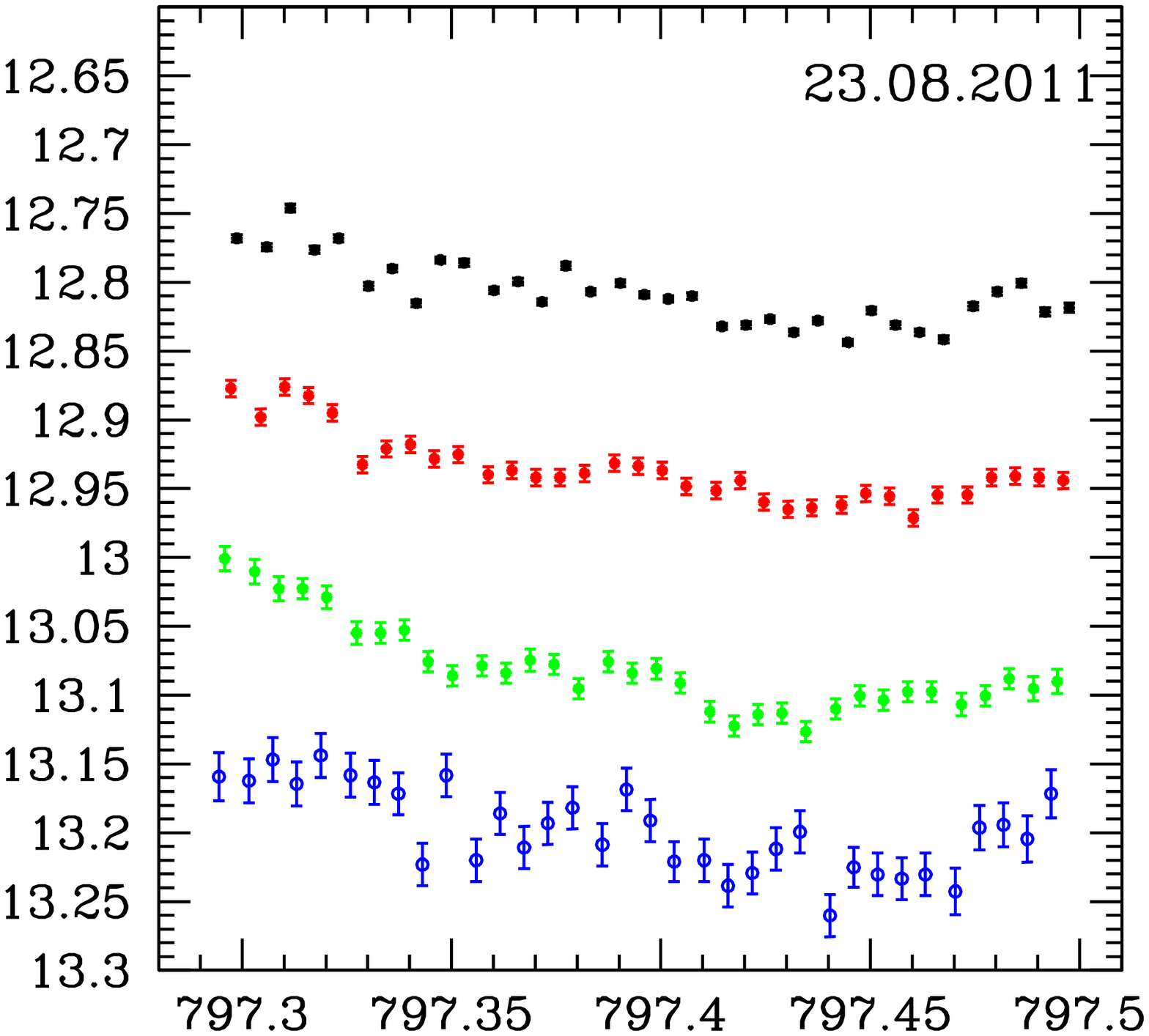}
\includegraphics[width=4cm , angle=0]{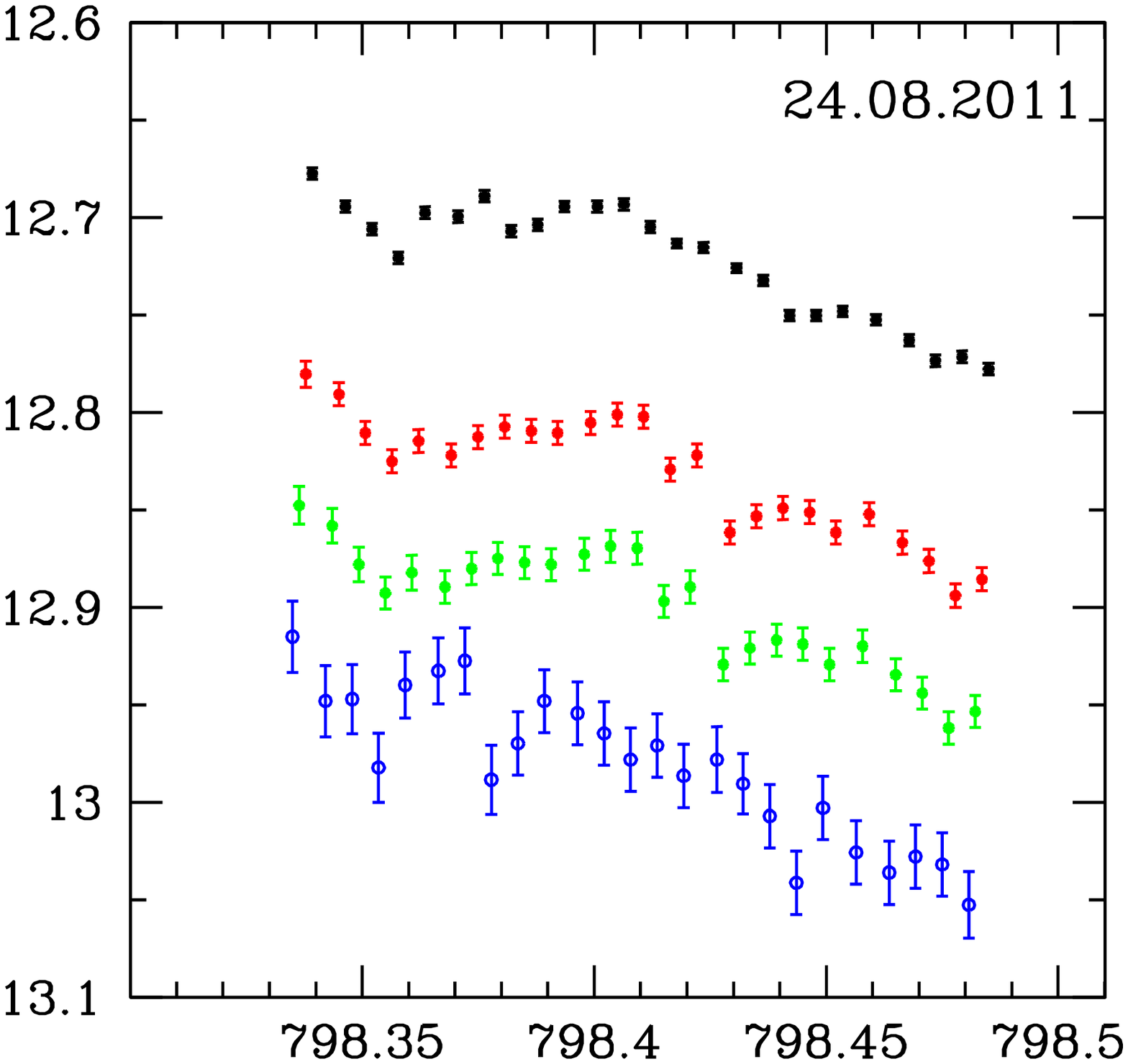}
\includegraphics[width=4cm , angle=0]{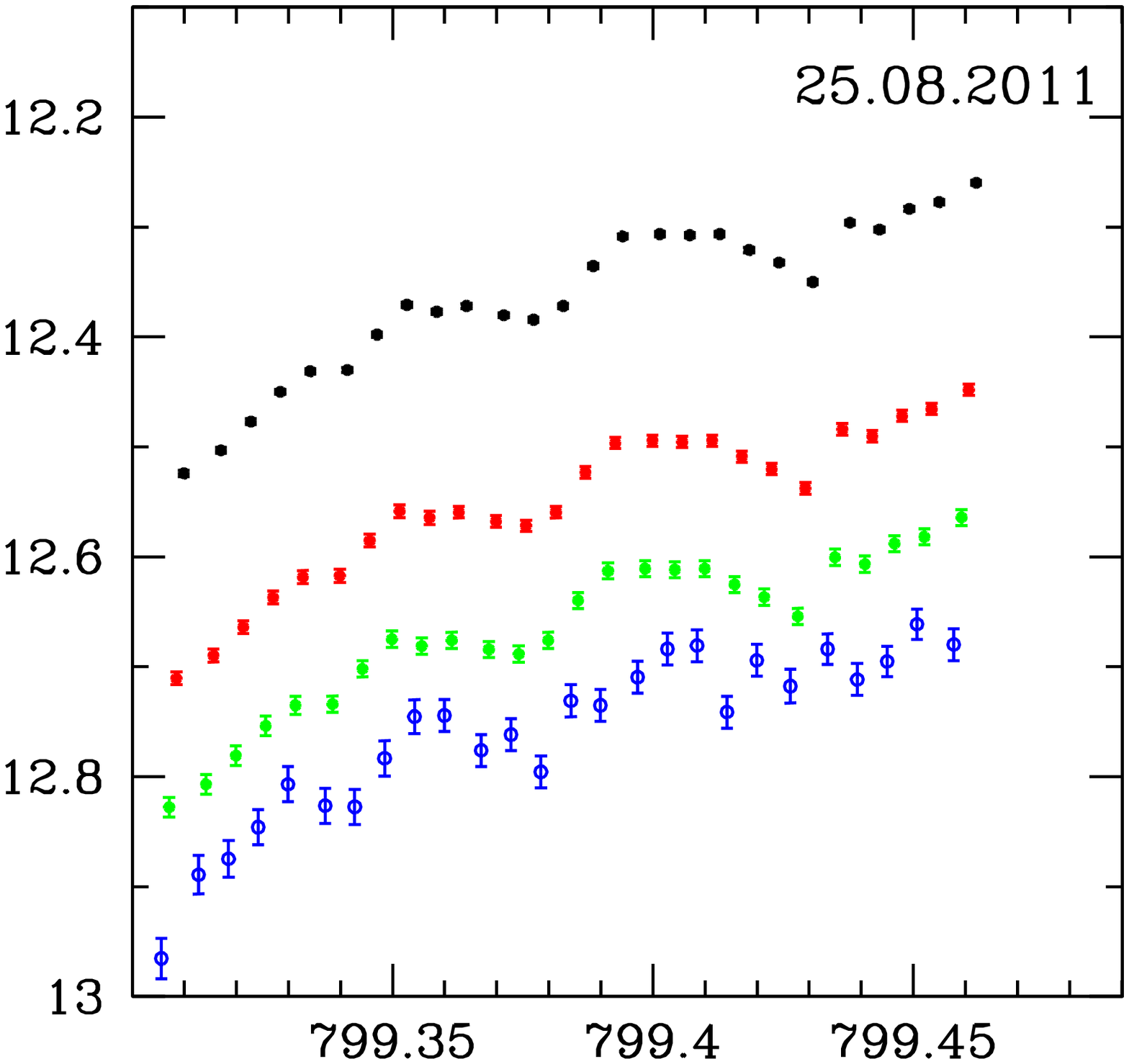}
\includegraphics[width=4cm , angle=0]{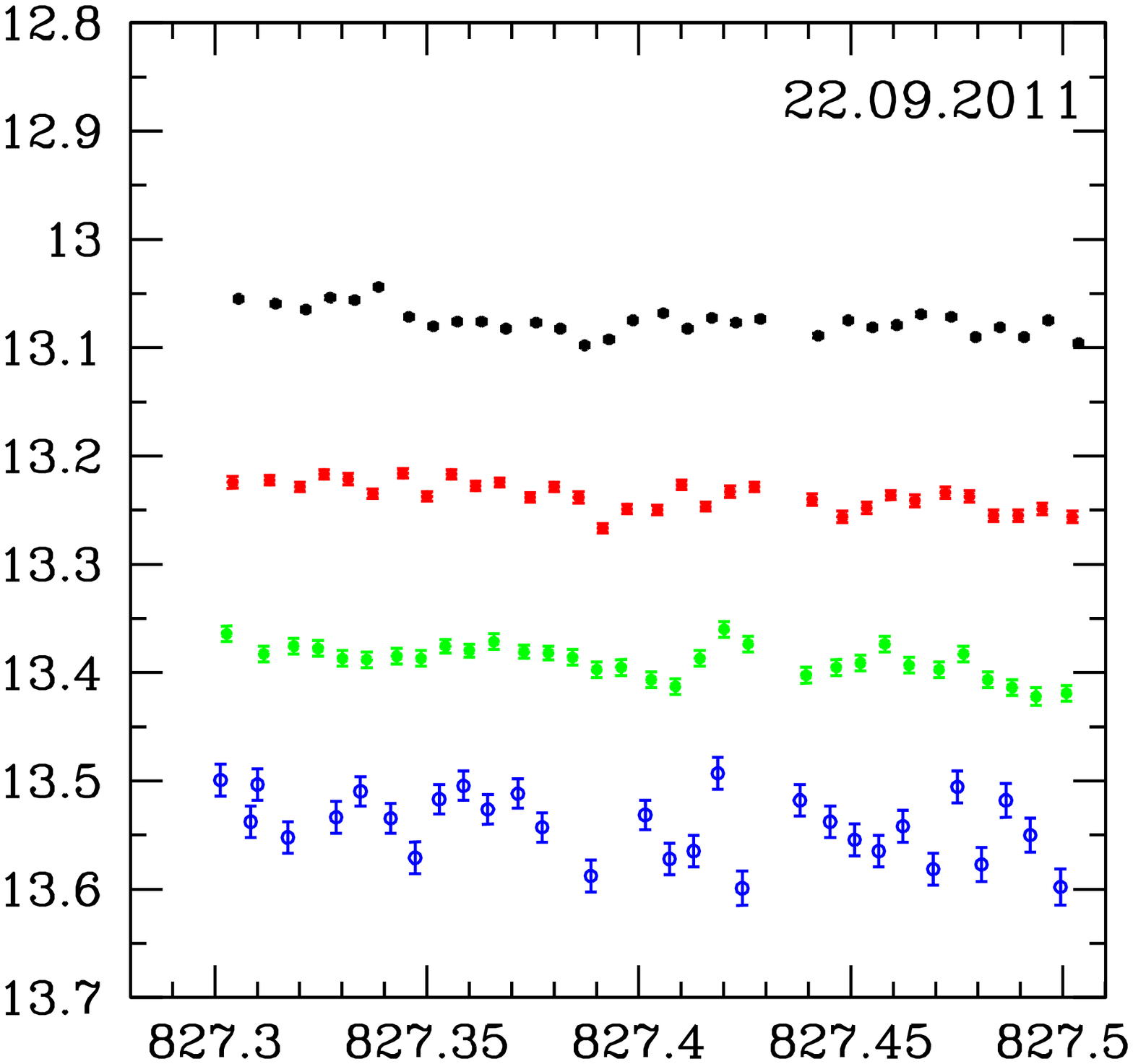}
\includegraphics[width=4cm , angle=0]{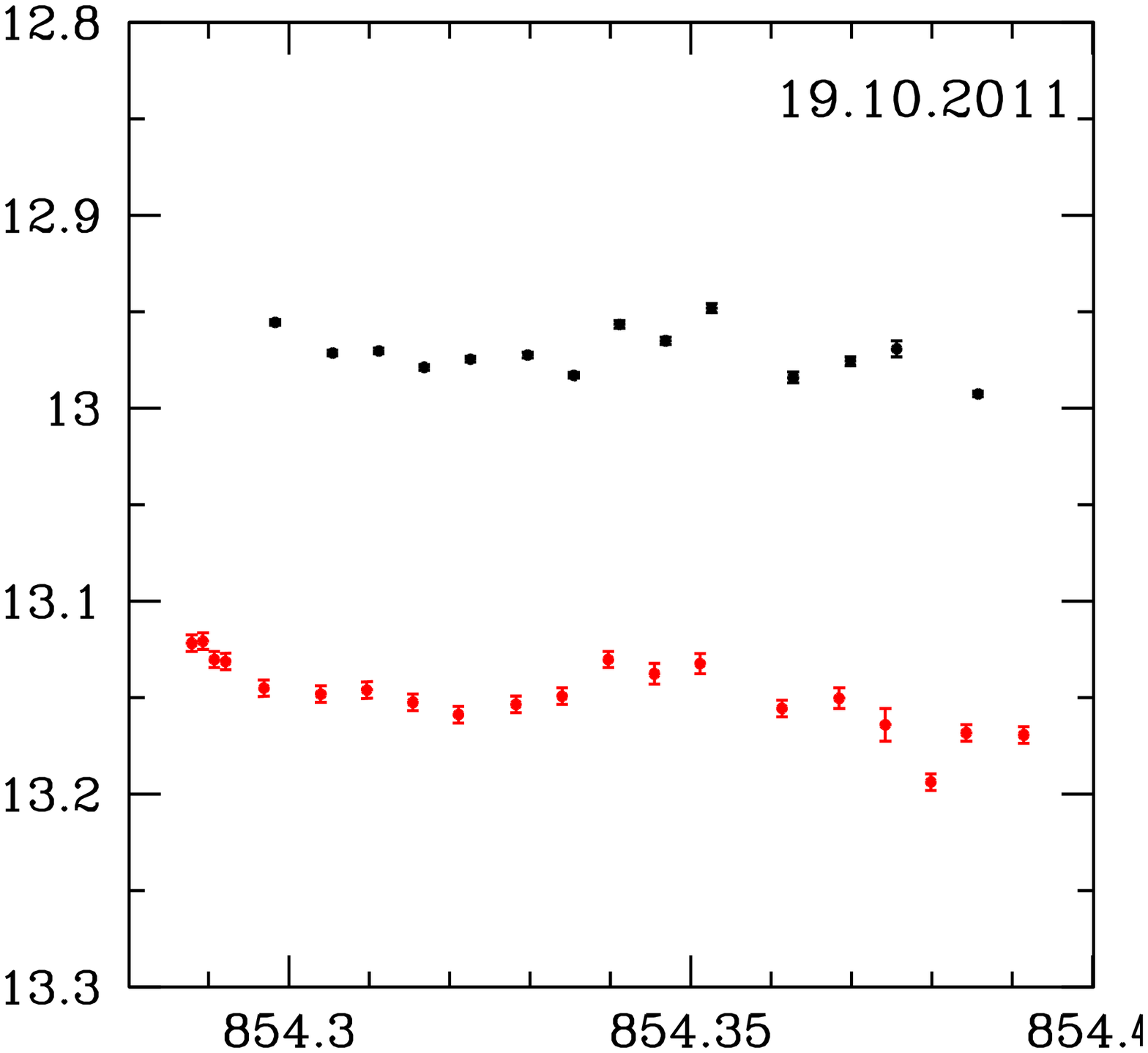}
\includegraphics[width=4cm , angle=0]{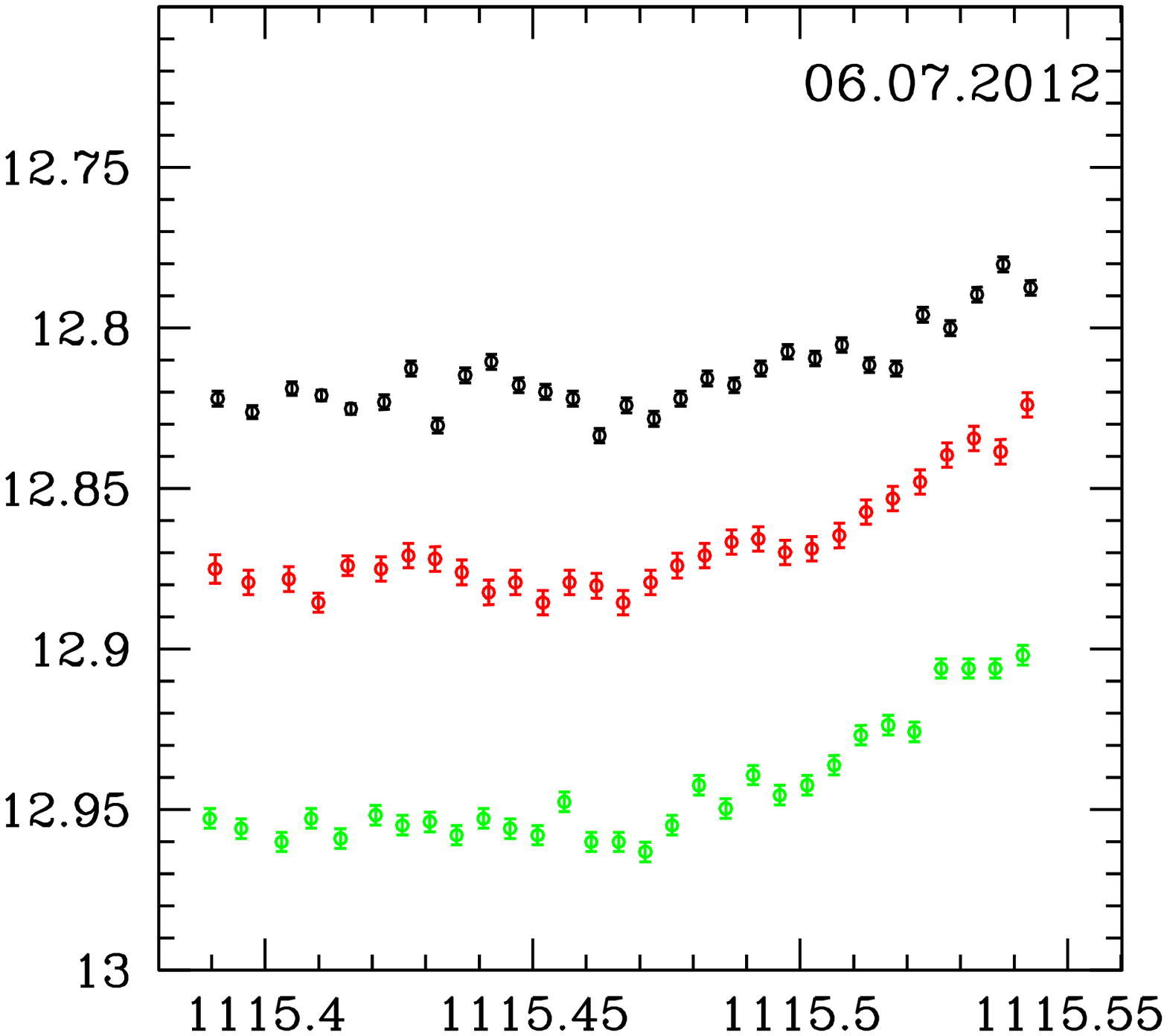}
\includegraphics[width=4cm , angle=0]{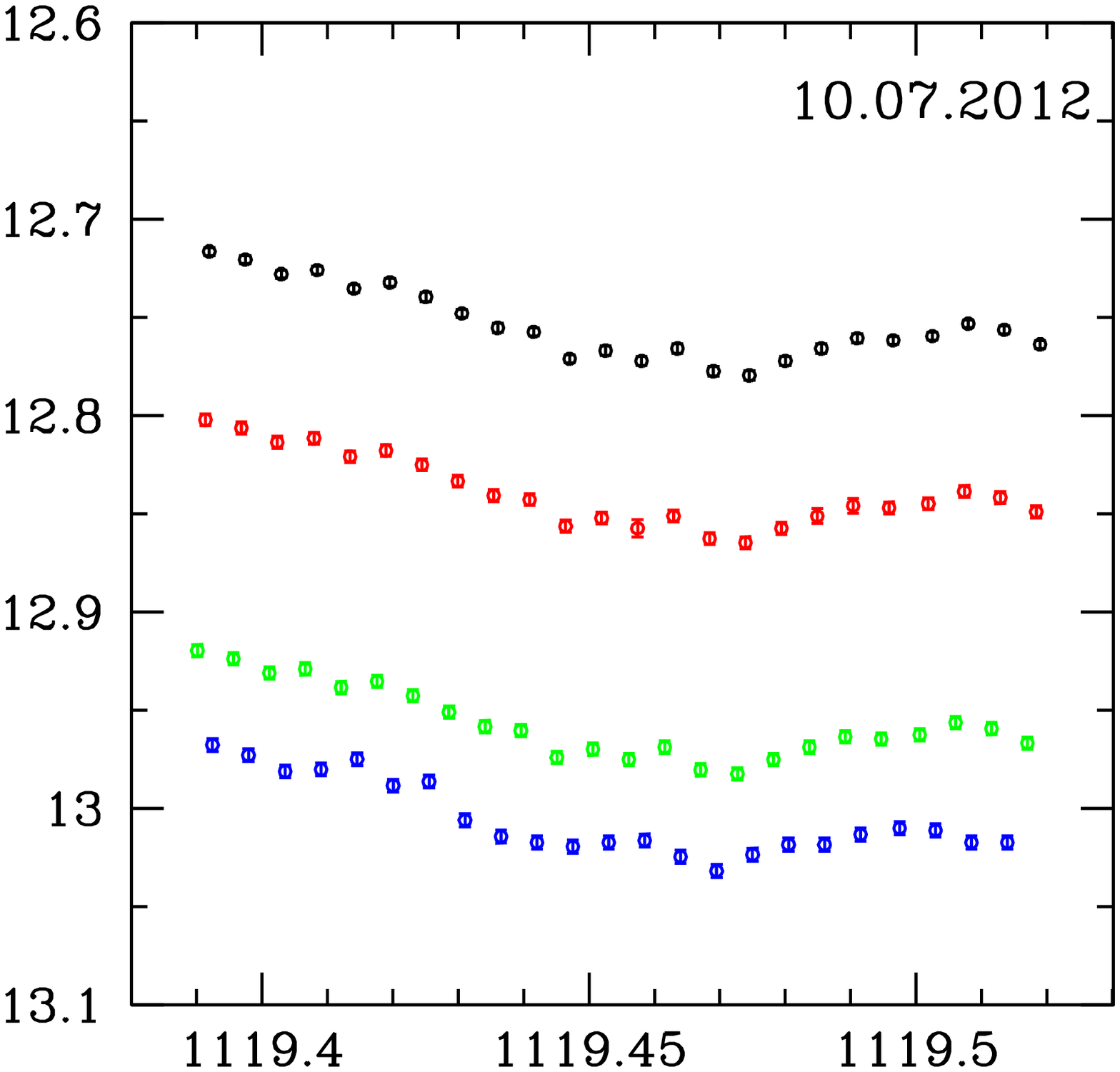}
\includegraphics[width=4cm , angle=0]{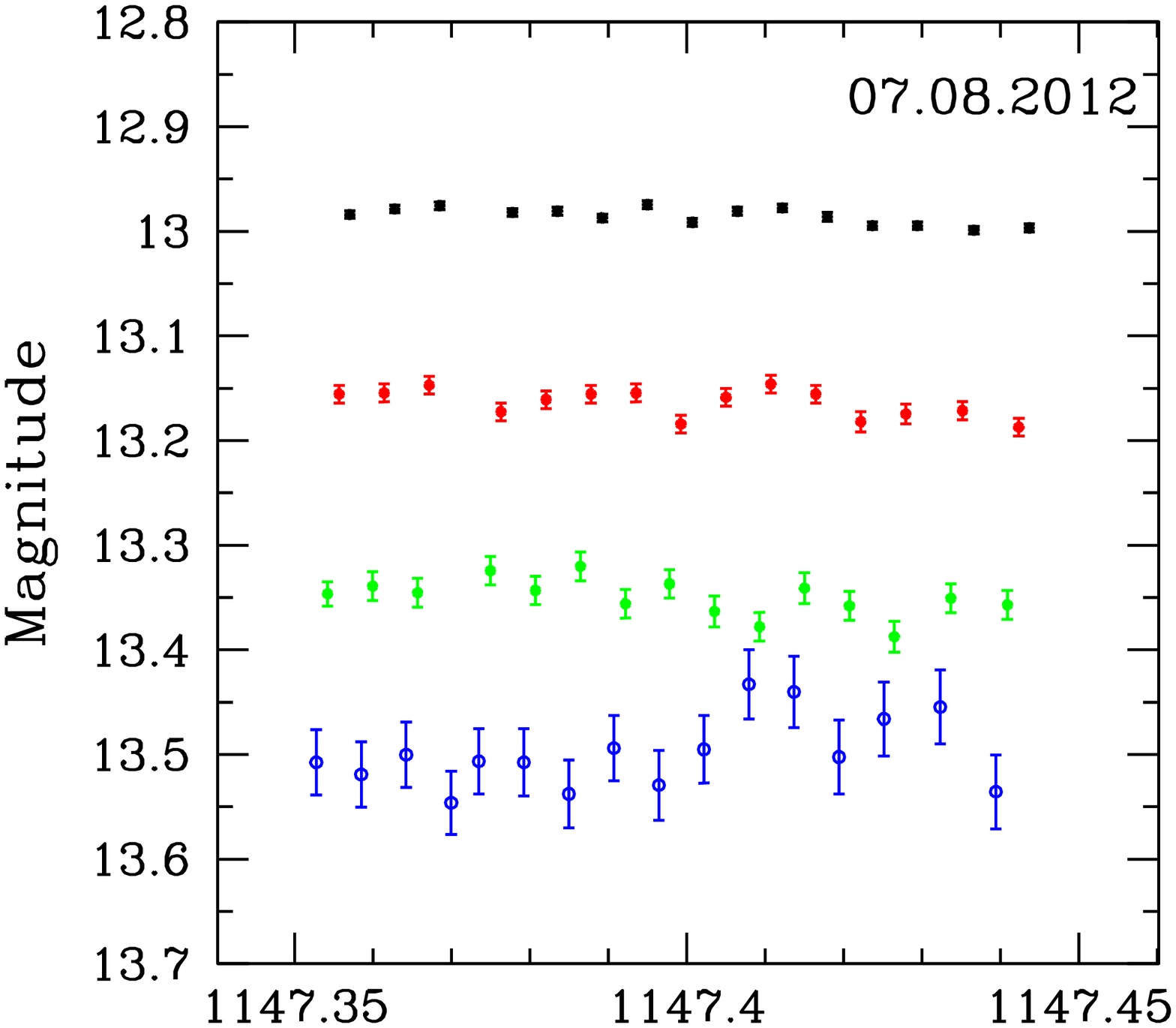}
\includegraphics[width=4cm , angle=0]{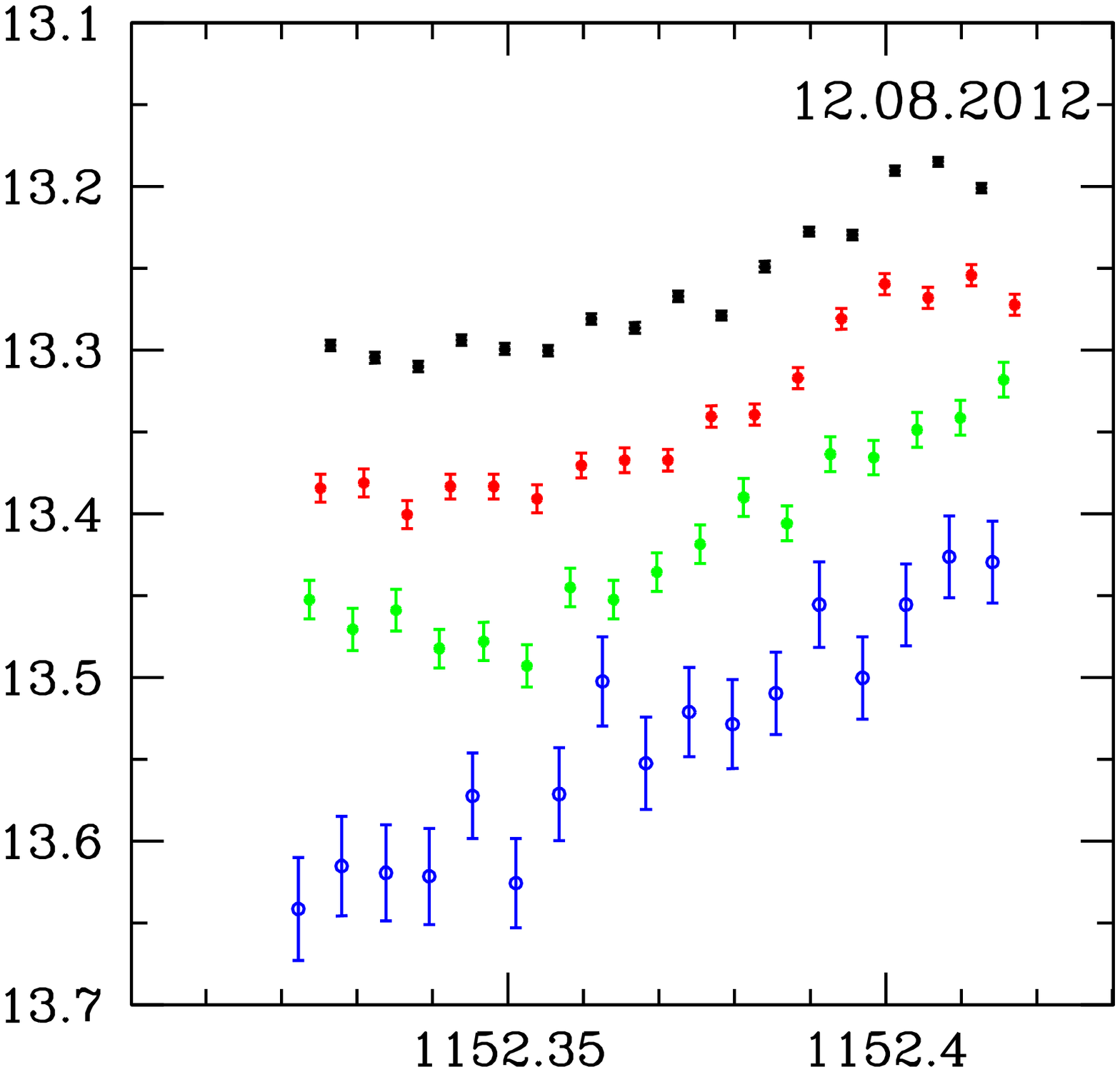}
\includegraphics[width=4cm , angle=0]{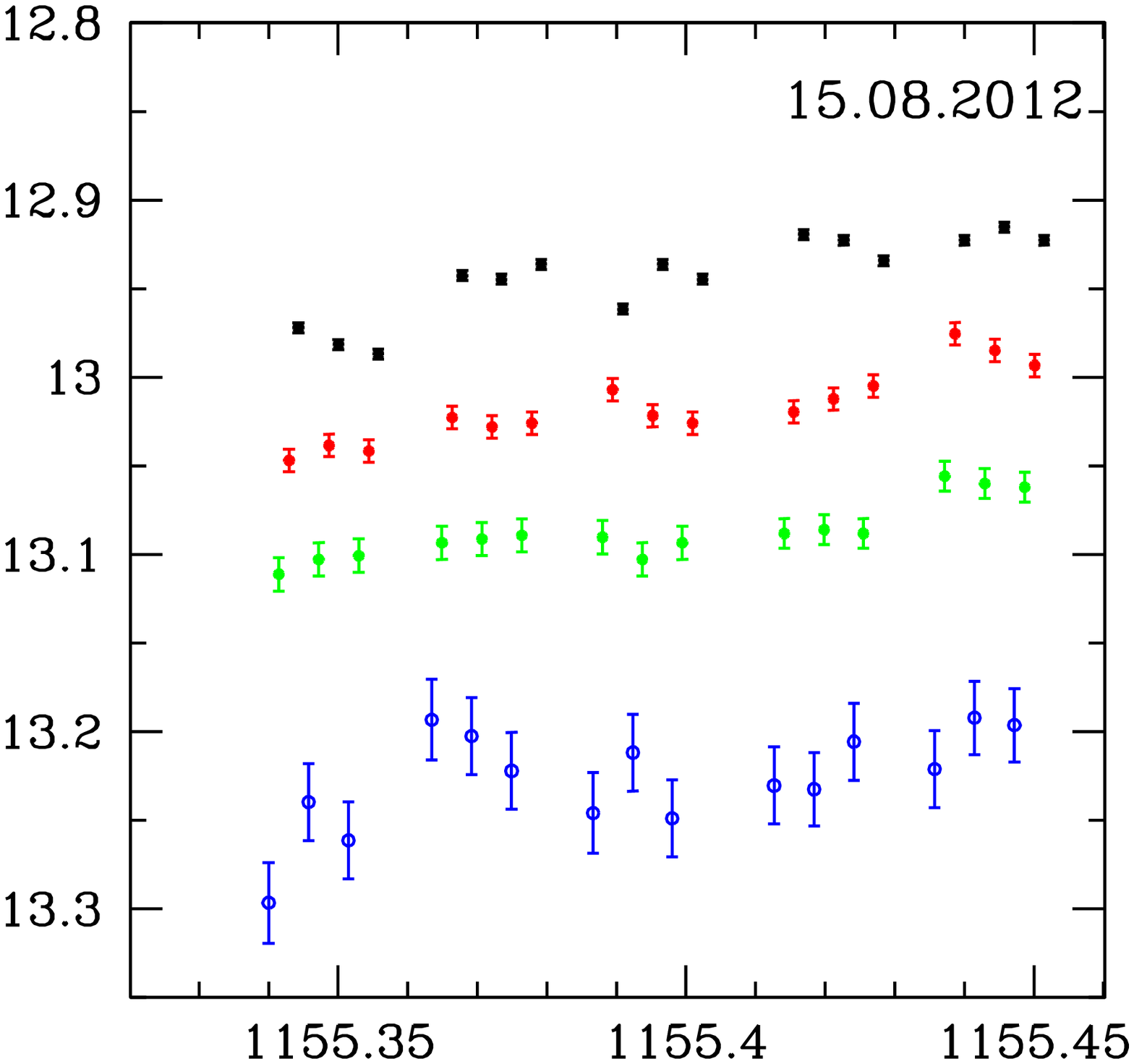}
\includegraphics[width=4cm , angle=0]{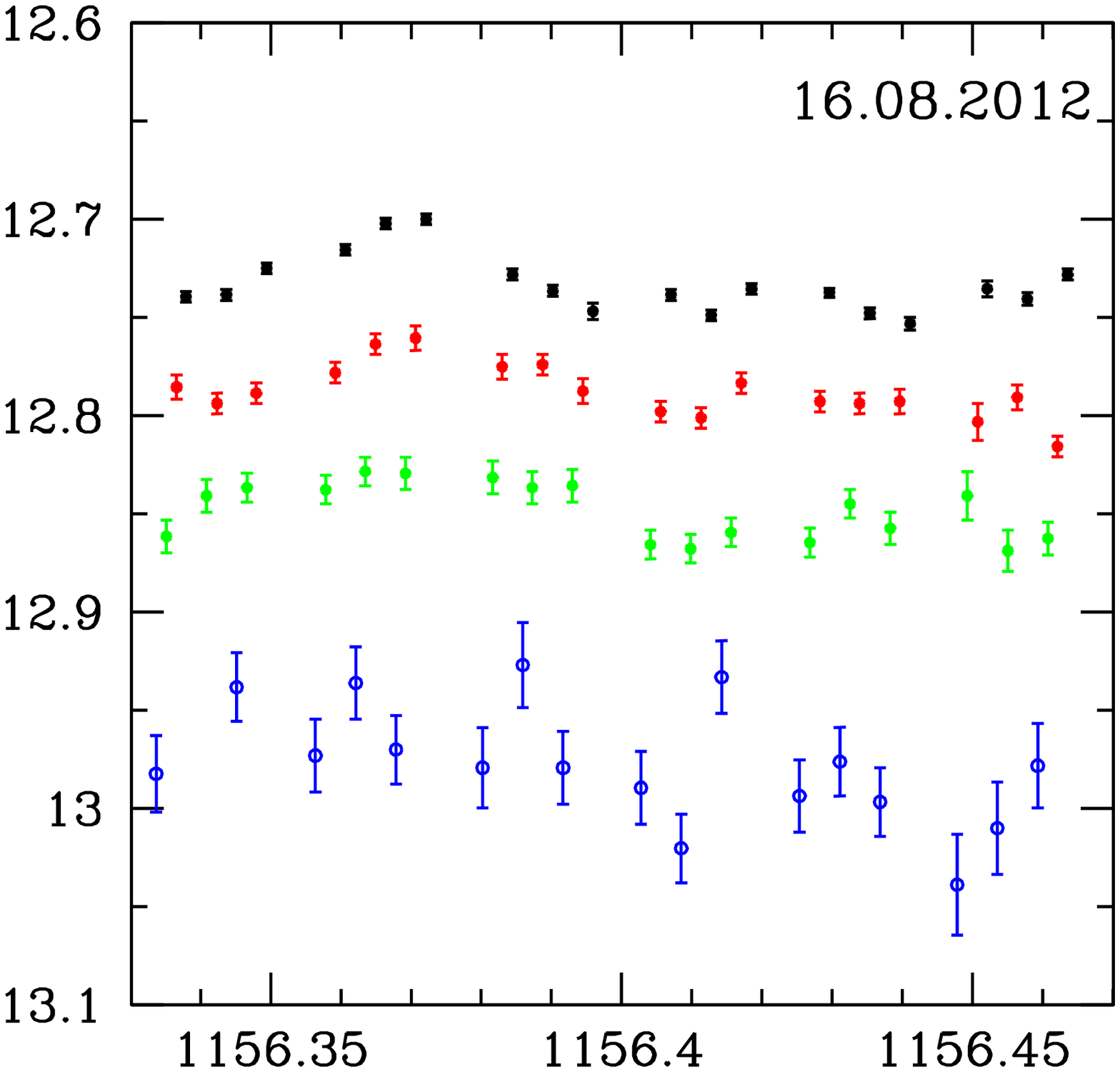}
\includegraphics[width=4cm , angle=0]{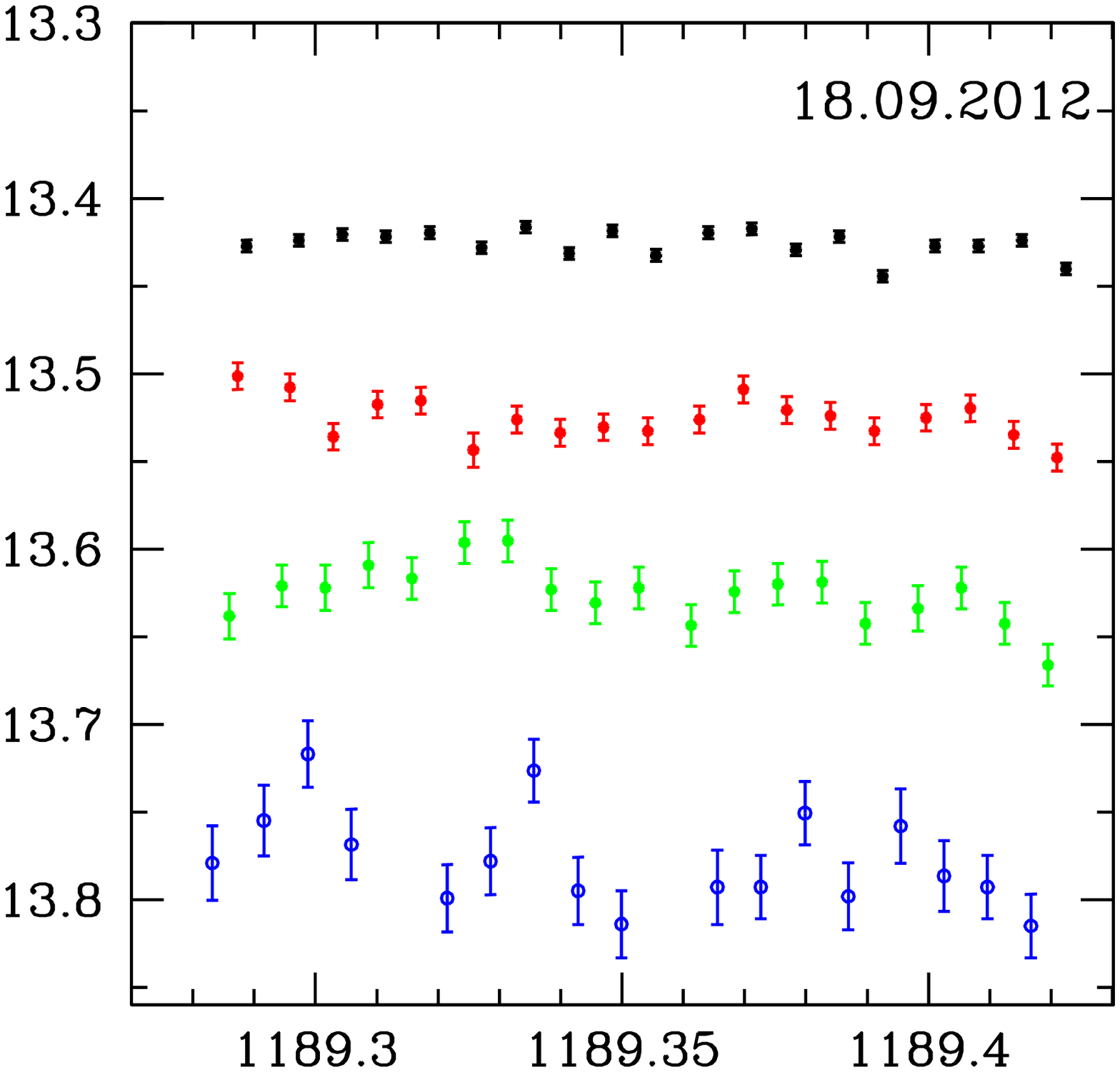}
\includegraphics[width=4cm , angle=0]{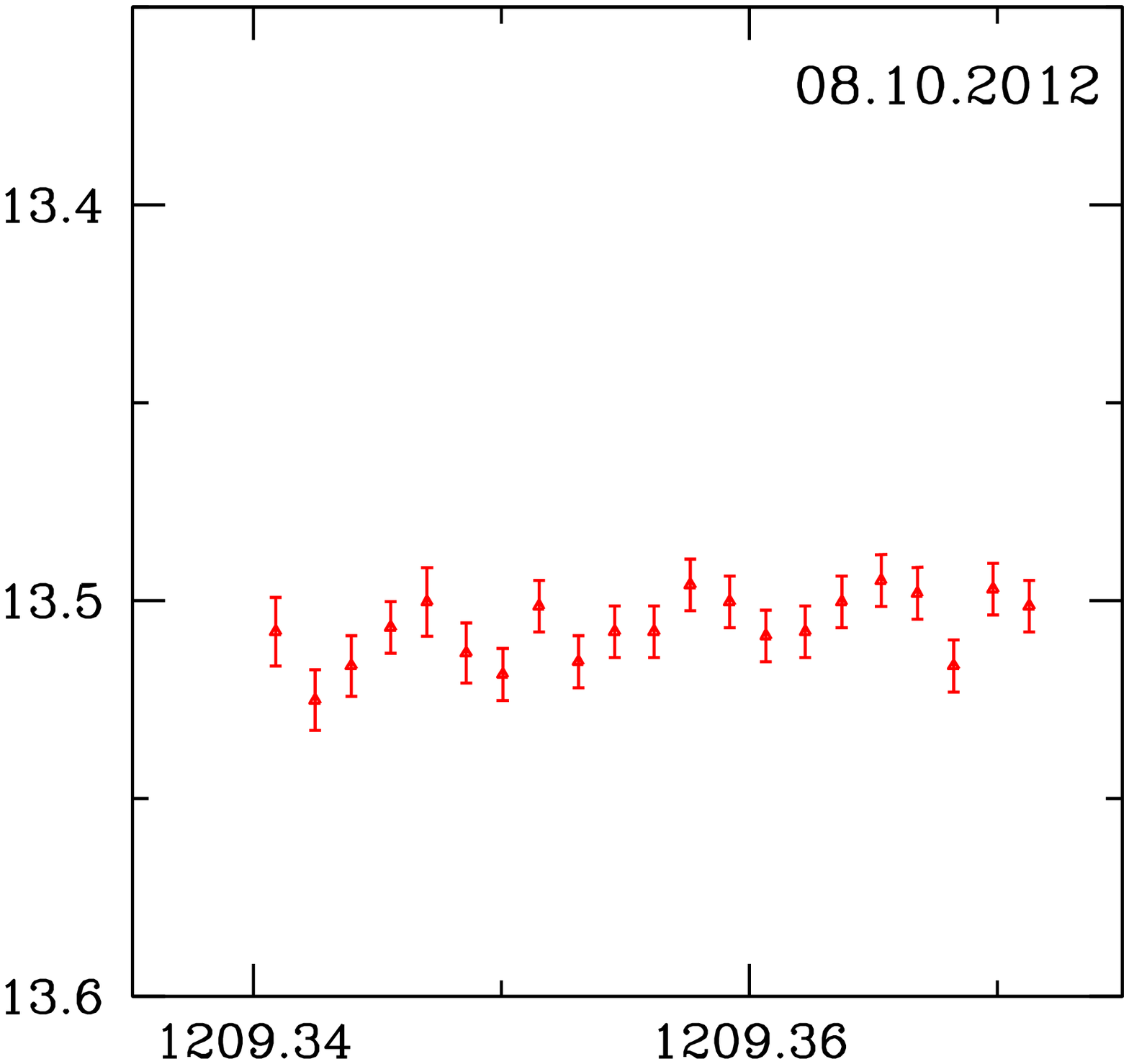}
\includegraphics[width=4cm , angle=0]{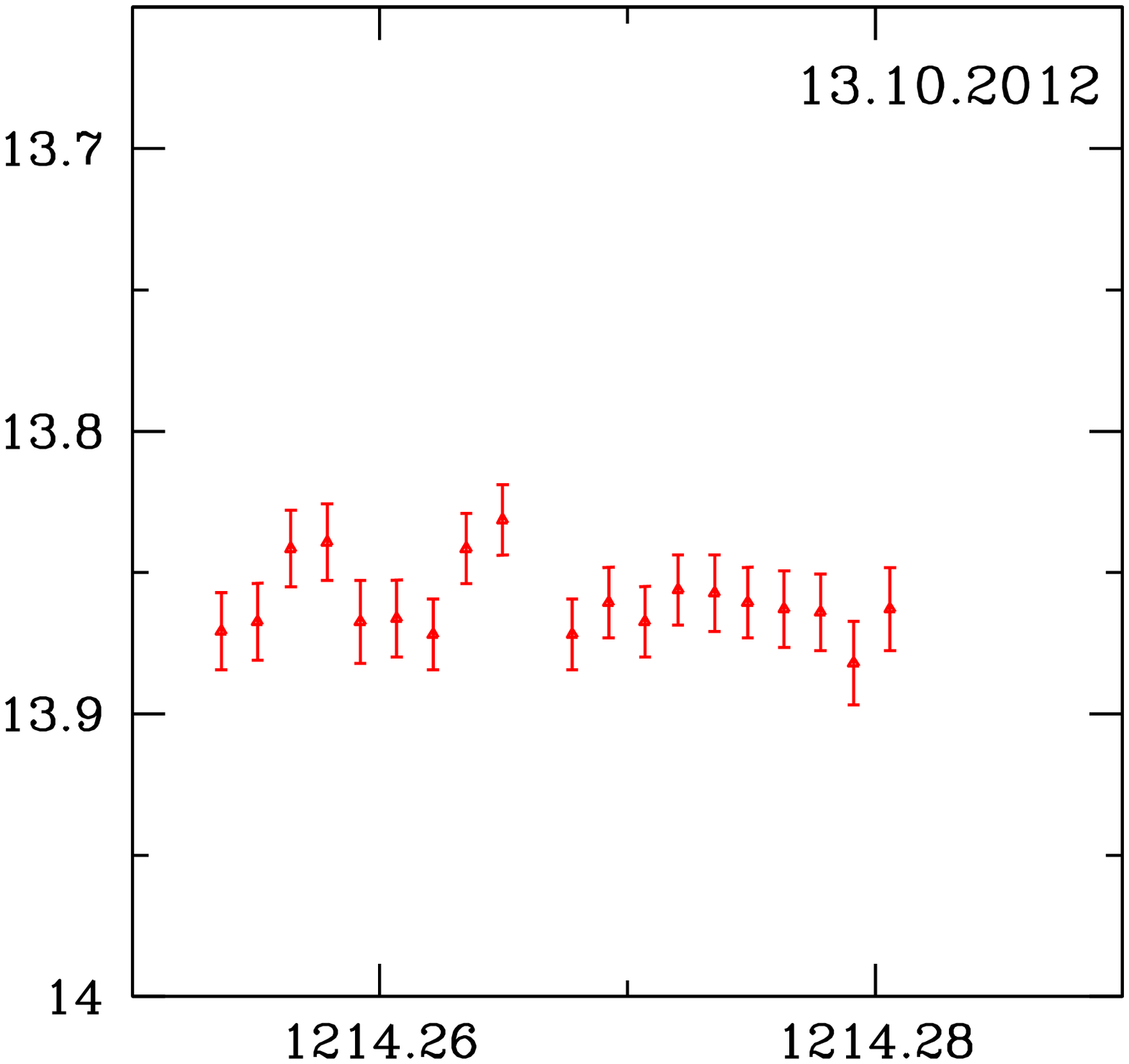}
\includegraphics[width=4cm , angle=0]{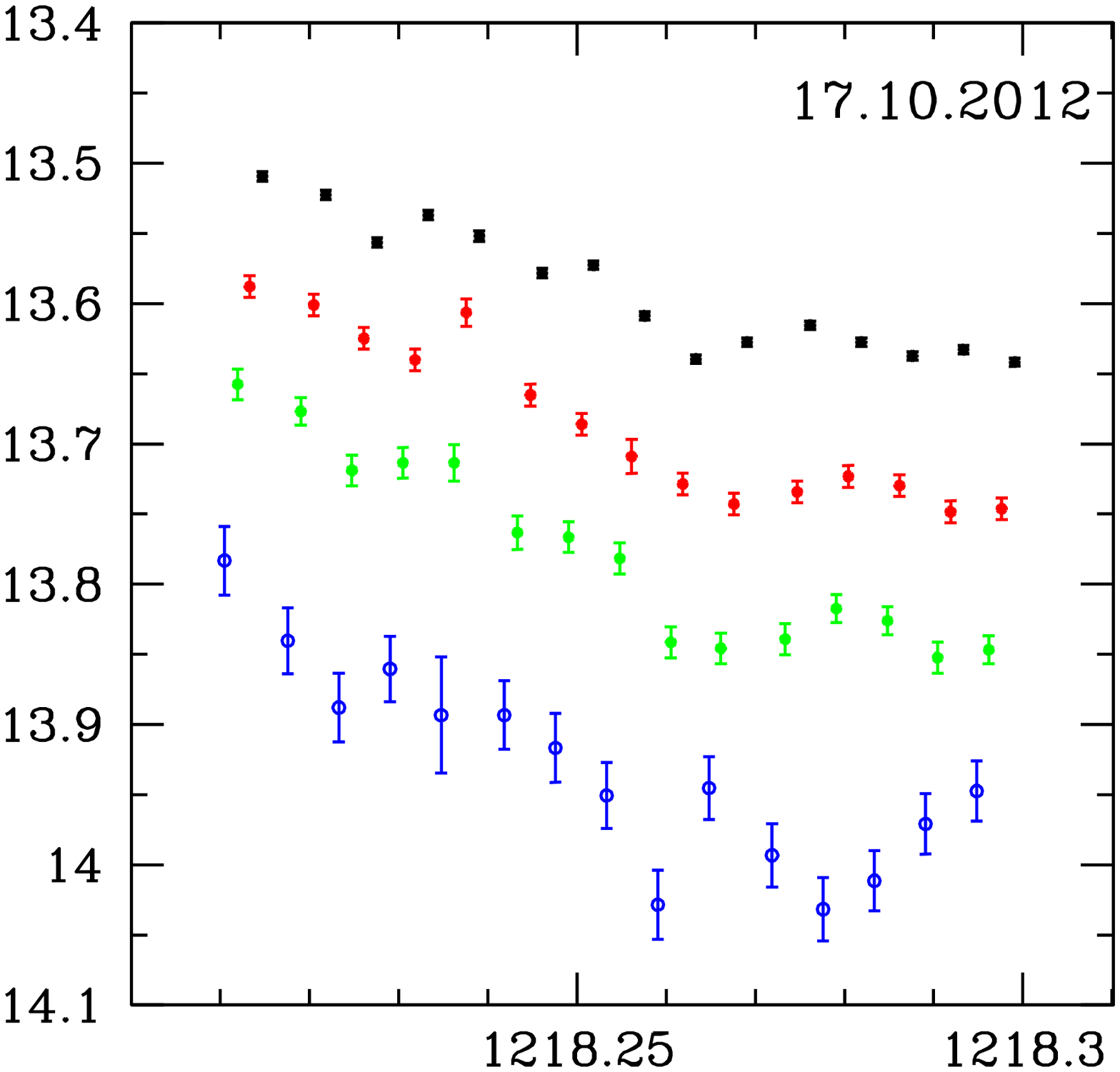}
\includegraphics[width=4cm , angle=0]{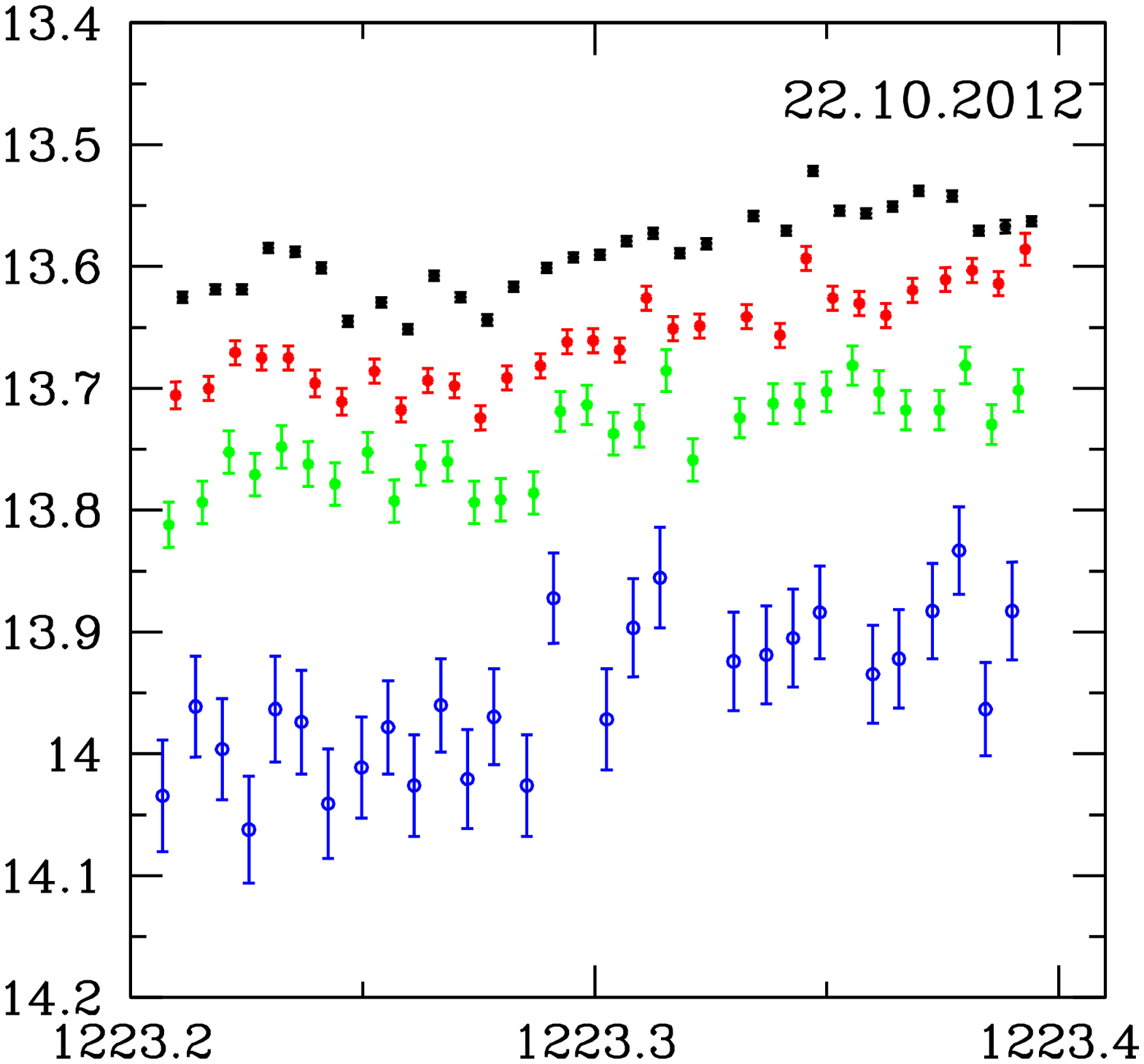}
\includegraphics[width=4cm , angle=0]{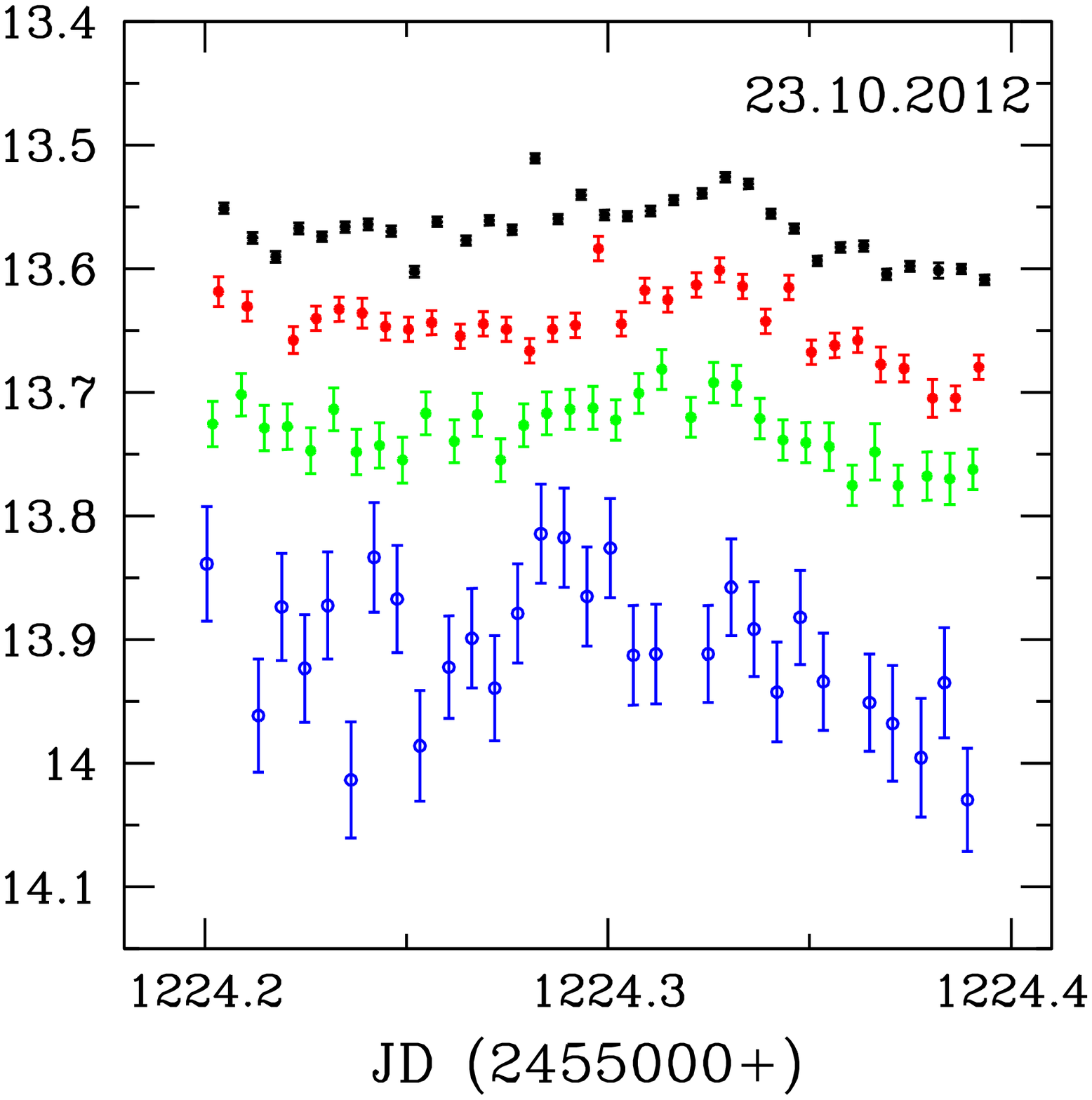}
\includegraphics[width=4cm , angle=0]{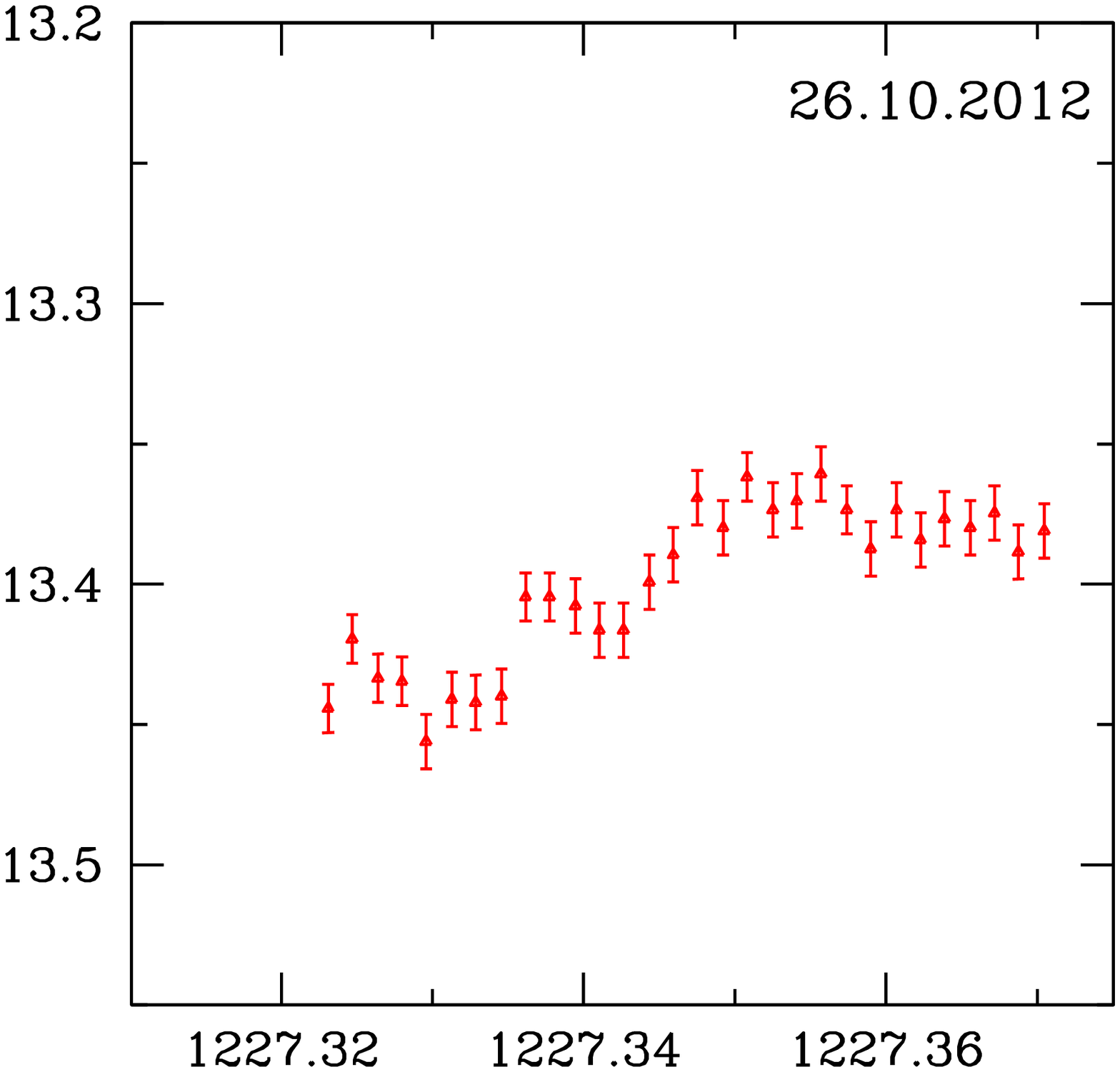}

\caption{As in Fig.\ 1 for August 2011 through October 2012.}
\end{figure*}

\subsection{Inter-band cross-correlations}

We computed the DCFs to determine the cross-correlations and time delays between the B and I, V and I, and R and I bands. 
The time delays are expected between emission in different energy bands, as the flare usually begin at higher frequencies
and then propagates to lower frequencies in the inhomogeneous jet model. Injected high energy electrons
emit synchrotron radiation first at higher frequencies and then cool, emitting at progressively lower frequencies, resulting
in time lag between high and low frequencies.
The DCFs between the light curves in all bands show close correlations among the various bands in the nights
where genuine variability is present. For the nights in which no genuine variability is present, we normally found
much weaker correlations between the bands ($<$0.4). As the peak of the DCFs are broad,  we fit the DCFs with
 Gaussian functions to determine the possible time delays; 
however, lags indicated by the DCFs are all consistent with zero. This is not surprising due to the closeness of the various optical 
bands we measured in frequency space, so if any lags are present they appear to be less than the resolution of our light curves.
An example of the DCF is shown in the right panel of fig.\ 4.

\begin{figure*}
\centering
\includegraphics[width=8cm , angle=0]{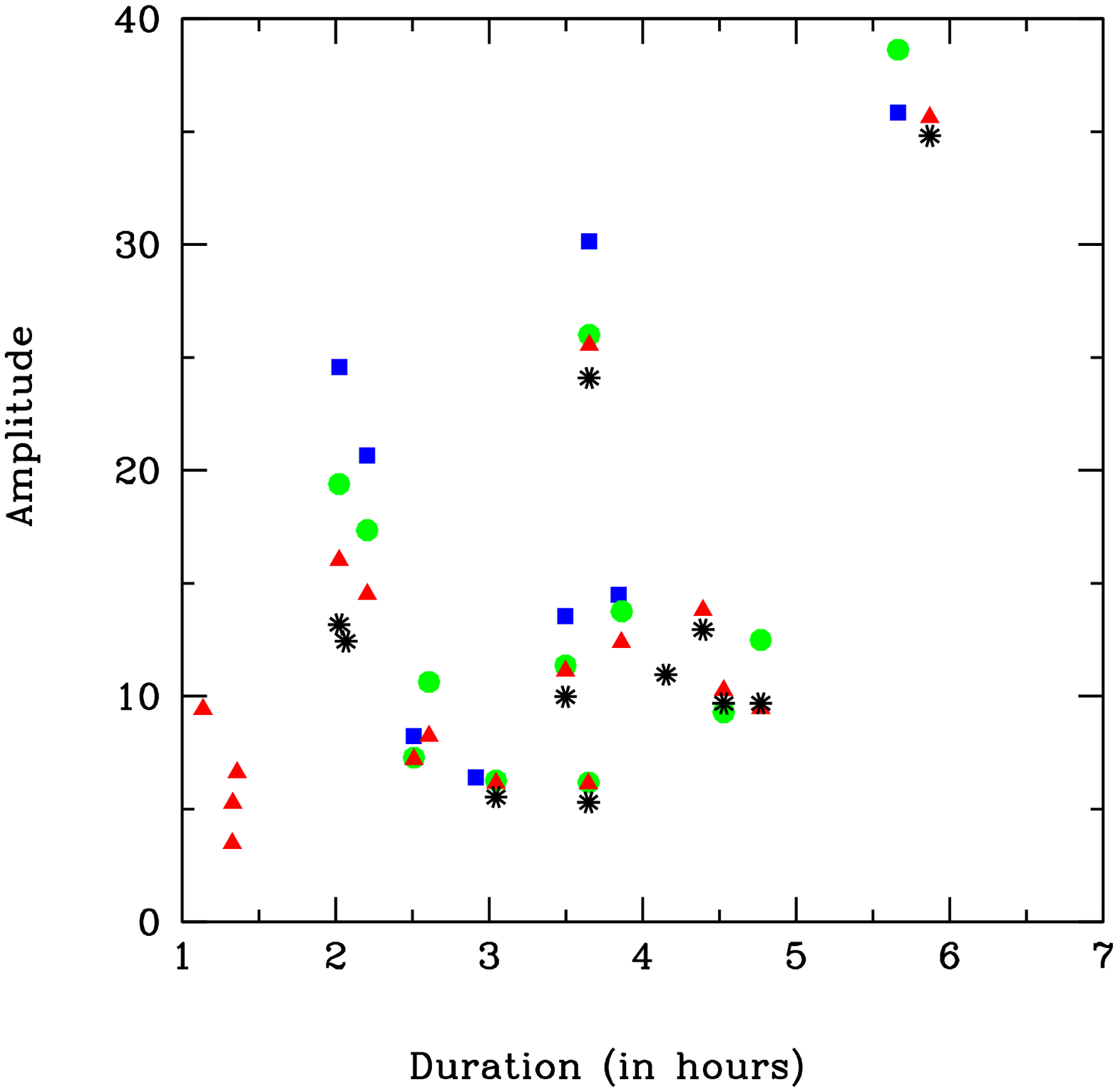}
\includegraphics[width=8cm , angle=0]{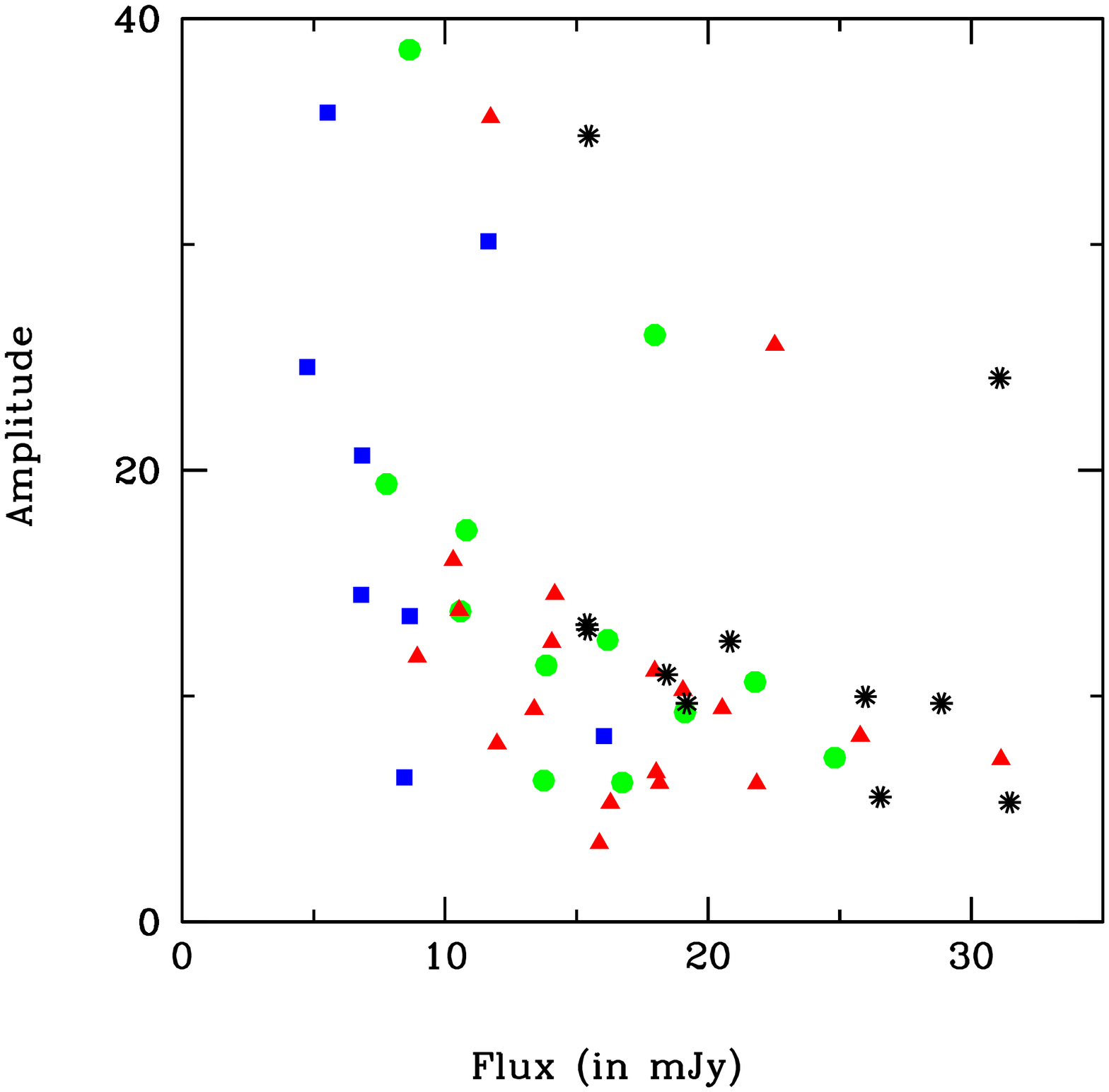}

\caption{Dependence of amplitude of variability on duration (left panel) and flux (right panel) of the observations. 
Here, B band is represented by squares (blue), V by solid circles (green), R by triangles (red) and I by starred symbols (black) .}
 \end{figure*}

\begin{figure*}
\centering
\includegraphics[width=8cm , angle=0]{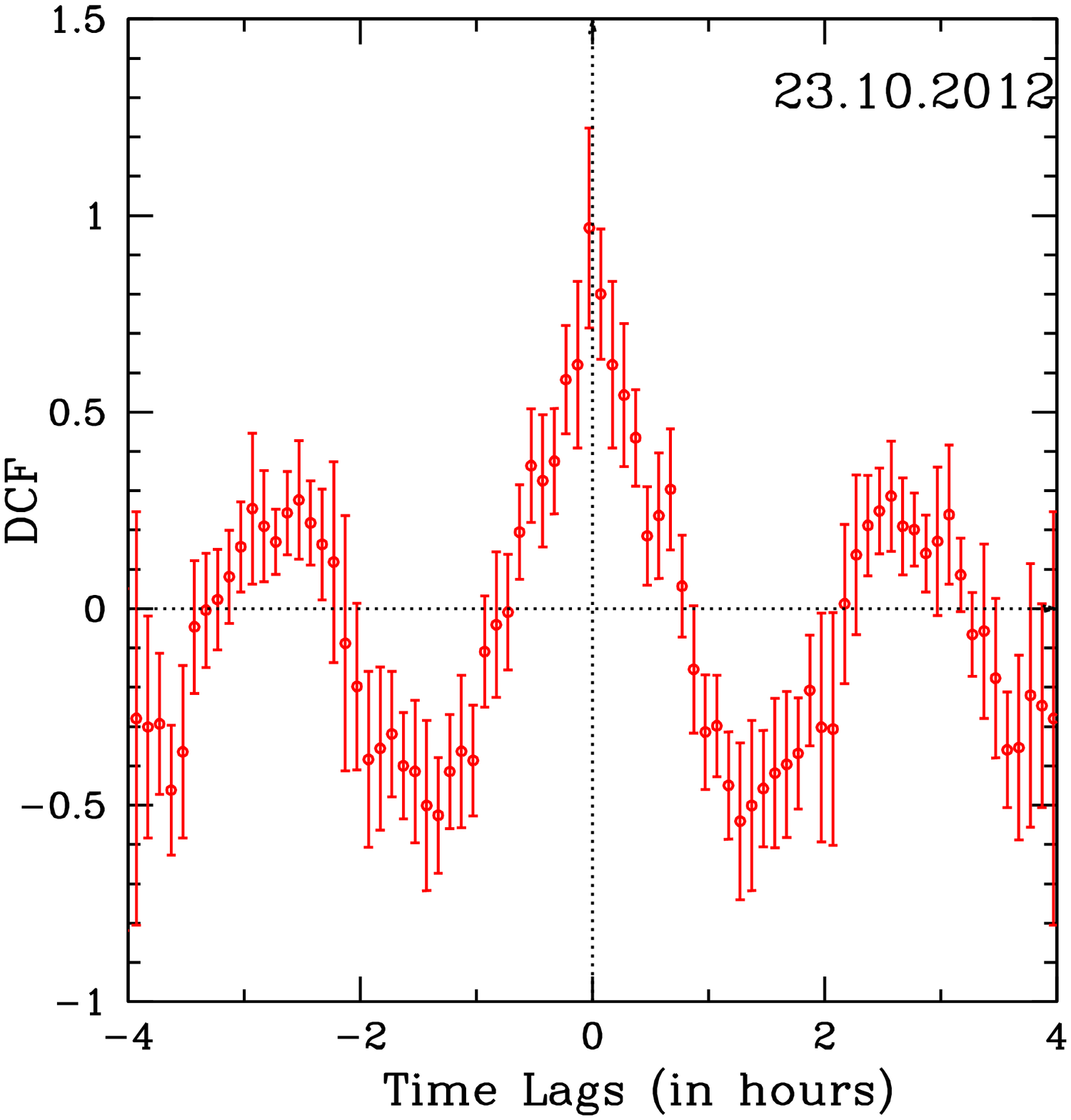}
\includegraphics[width=8cm , angle=0]{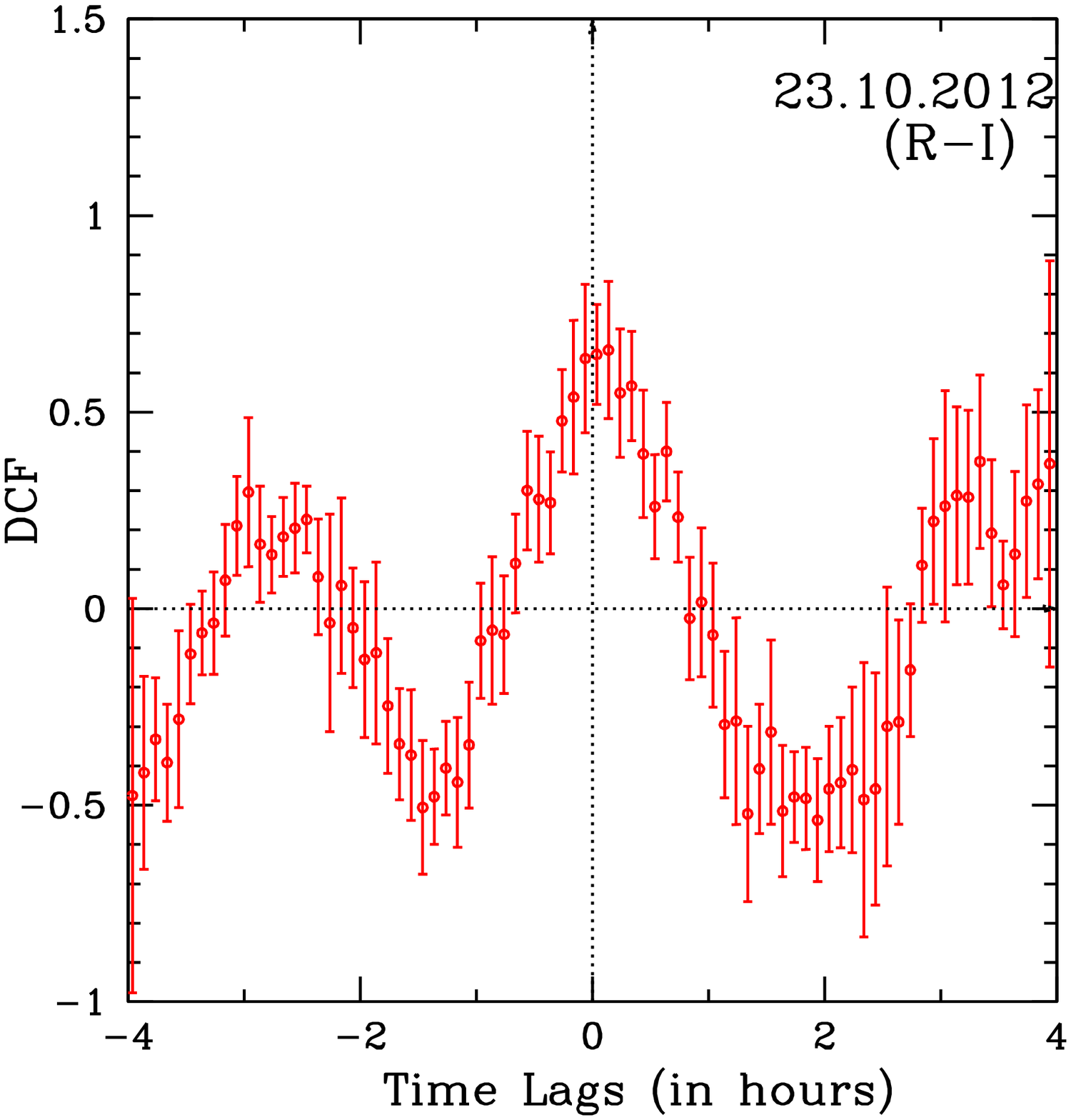}

\caption{One example of Auto-correlation (R band) is shown in the left figure and DCF (R versus I) is shown in the right figure.}
 \end{figure*}

\subsection{Colour Indices}

We next investigated the existence of spectral variations by studying the behaviour of colour variations with respect 
to the brightness of BL Lacertae. The colour indices (CIs) are calculated by combining almost simultaneous (within 8 minutes)
 B, V, R and I magnitudes to yield  CIs = B$-$V, V$-$R, R$-$I and B$-$I.  For a particular night, we studied the 
colour variability only when both light curves were identified as variable. Our data have the 
advantage of being almost simultaneous in various optical bands.  It is important to recall that the underlying host galaxy,
effect of the accretion disk component and the Gravitational microlensing can also 
lead to apparent, but unreal, colour variations (Hawkins 2002). But, since Gravitational microlensing is important
on weeks to months time-scales and during our observations, BL Lacertae was in flaring state (Raiteri et al. 2013) where the Doppler
boosting flux from the relativistic jet almost invariably swamps out the contribution of the accretion disk component, so we can
rule out the contribution of these components. Also, the data we have used in calculating the colour indices are host 
galaxy subtracted so we conclude that  our results indicates  
variability of the non-thermal continuum radiation. 

We studied the variations of colour indices with respect to brightness in 13 of these observations. We fit all the 
colour--magnitude diagrams with a linear model of the form CI = $m \times $ mag $+$~c (where $m$ is the slope in the 
fit,  V magnitude is taken as mag and $c$ is its respective intercept). The Pearson correlation coefficient (r), 
its p-value (null hypothesis probability; we consider a confirmed colour index correlation with the V magnitude when p $<$0.01) 
along with the slopes and intercepts are presented in Table 3. The significant positive correlations between the 
colour index and magnitude along with a change of slope $>$3 $\sigma$ indicates that the source exhibits a bluer-when-brighter trend. 

In five of the nights, which are marked by asterisks in Table 3,  we found significant variations in colour indices at the  3-$\sigma$ level.
In these observations, colour indices correlate with the source brightness and the overall correlation is positive, 
which indicates hardening of the spectrum as the source brightens. So, in these observations, BL Lacertae exhibits bluer-when-brighter
trend with different regression slopes. No significant negative correlations are found for the source.
The bluer-when-brighter tendency in BL Lacertae has been seen by several groups on long-term as well as short-term time-scales
(e.g.\ Papadakis et al.\ 2003; Villata et al.\ 2004; Stalin et al.\ 2006; Larionov et al.\ 2010; 
Gu et al.\ 2006; Gaur et al.\ 2012; Wierzcholska et al.\ 2015).
Villata et al.\ (2004) characterized the intra-day flares to be strong bluer-when-brighter chromatic events with a
slope of $\sim$0.4. In five observations, we found a bluer-when-brighter trend with a slope varying between 0.18--0.28 (in Table 3). 

In the other eight observations, we did not find significant linear correlations between colour indices and magnitudes (Table 3). 
In some of them significant colour variations are seen but are not well fitted by linear functions. 
During the observations on 6 July 2010 (Fig.\ 1, fourth panel), we found a strong flare with magnitude variation of $\sim$0.3
in all the optical  bands and the behaviour of the colour magnitude diagram varies  according to the different flux states. 
But, on intra-day timescales it is difficult to judge the variations of the colour indices with respect to different brightness states
 as the flare is on hours like timescales.  
Also, in other observations, i.e, 8 July 2010, 24 and 25 August 2011,  we saw small sub-flares superimposed on the long-term trend. 
In these cases, the colour indices vary significantly within the individual observations, sometimes showing different branches in the 
colour magnitude diagrams according to the flux states. Spectral steepening during the flux rise can be explained by the
presence of two components, one variable with a flatter slope which dominates during the flaring states and  another one that is
more stable and contributes to the long term achromatic emission (Villata et al.\ 2004). So, it could be possible that during these 
observations, the superposition of many distinct new variable components lead to the overall weaking of the colour-magnitude 
correlations. 
Bonning et al.\ (2012) studied a sample of FSRQs and BL Lacertaes and found that FSRQs follow redder-when-brighter trends while BL Lacertaes
show no such trends. They found complicated behaviour of the blazars on colour-magnitude diagrams: hysteresis tracks, and
acromatic flares which depart from the trend suggesting different jet components becoming important at different times.
Therefore, our colour variability results show that the intra-night 
flares between 2010 and early 2012 are chromatic but do not always follow simple bluer-when-brighter trends.

\begin{table*} 
\center
\caption{ Results of Intra-day Variability of BL Lacertae}
\setlength{\tabcolsep}{0.03in}
\begin{tabular}{lccccccr@{\hskip 0.3in}lcccccc} \cline{1-7} \cline{9-15}
Date            &Telescope &Band       &$F_{enh}$    &$F_{c}$(0.001)   &$Amp\%$   &Variable  & & Date            &Telescope &Band       &$F_{enh}$    &$F_{c}$(0.001)   &$Amp\%$   &Variable   \\ \cline{1-7} \cline{9-15}
10.06.2010           &A    &R        &1.018  &2.386   &-   & NV & & 24.08.2011           &D    &B        &4.035 &2.849    &13.55   &Var\\
11.06.2010           &A    &R        &0.985  &2.008   &-   & NV & &                      &    &V         &7.389 &2.849    &11.36   &Var   \\
12.06.2010           &A    &R         &1.034 &2.386   &-  &NV & &                      &    &R         &16.187&2.849    &11.12 &Var   \\
14.06.2010           &A    &R         &1.054 &2.076     &-   &NV & &                     &    &I         &13.665&2.849    &9.98 &Var  \\
                     &C    &R         &0.934 &2.033     &-    &NV & & 25.08.2011           &D   &B         &11.247&2.790   &30.14 &Var \\
18.06.2010           &E    &R         &51.328&1.940      &6.63 &Var  & &                      &    &V         &56.7659&2.790   &26.30 &Var  \\
19.06.2010           &E    &R         &1.861 &1.940      &-       &NV  & &                     &    &R         &63.145&2.790   &25.55 &Var\\
20.06.2010           &E    &R         &21.725&1.930      &3.48&Var   & &                     &    &I         &17.878&2.790   &24.10 &Var\\
21.06.2010           &E    &R         &1.227 &1.972     &-        &NV & & 22.09.2011           &D   &B         &1.354 &2.639   &-        &NV\\
22.06.2010           &E    &R         &29.890 &1.94     &5.28 &Var & &                     &    &V         &1.565 &2.517   &-        &NV\\
04.07.2010           &F    &B         &2.164 &2.281    &-        &NV  & &                     &    &R         &1.329 &2.517  &-  &NV\\
                     &    &V          &2.155 &2.481    &-        &NV & &                     &    &I         &0.930 &2.517   &- &NV\\
                     &    &R          &0.146 &2.236    &-        &NV & & 19.10.2011           &D   &R         &0.574 &3.239   &- &NV \\
                     &    &I          &0.938 &2.281    &-        &NV & &                      &    &I         &0.841 &4.142   &-  &NV \\
05.07.2010           &F    &B         &1.527 &2.305    &-        &NV & & 06.07.2012           &E   &V        &25.590&2.596   &6.18 &Var \\
                     &    &V          &0.618 &2.281    &-        &NV & &                     &    &R         &21.632&2.596   &6.14  &Var \\
                     &    &R          &0.109 &2.258    &-        &NV & &                     &    &I         &9.829  &2.596   &5.30 &Var \\
                     &    &I          &1.052 &2.281    &-        &NV & & 10.07.2012           &E   &B         &16.174&2.983   &6.40 &Var\\
06.07.2010           &F    &B         &32.4002&1.995    &35.85 &Var & &                     &    &V         &57.750&2.913   &6.27 &Var \\
                     &    &V         &36.577 &1.995    &38.64 &Var & &                     &    &R         &28.700&2.913   &6.15  &Var\\
                     &    &R          &45.931&2.047    &35.64        &Var & &                     &    &I         &11.878&2.913   &5.54 &Var\\
                     &    &I          &55.757&1.995    &34.82 &Var & & 07.08.2012           &D   &B         &2.110 &3.753   &-        &NV \\
07.07.2010           &F    &B         &2.307 &2.481    &-        &NV & &                     &    &V         &0.558 &3.932   &-        &NV\\
                     &    &V          &2.284 &2.305    &-        &NV & &                     &    &R         &0.638 &3.932   &-        &NV\\
                     &    &R          &1.611 &2.258    &-        &NV & &                     &    &I         &0.746 &3.932   &- &NV\\
                     &    &I          &1.247 &2.258    &        &NV   & & 12.08.2012           &D    &B        &4.669 &3.598   &20.67 &Var  \\
08.07.2010           &F    &B         &17.271&2.305    &14.49        &Var & &                     &    &V         &18.835 &3.598  &17.35 &Var \\
                     &    &V          &3.824 &2.281    &13.76       &Var & &                     &    &R         &36.030&3.598   &14.52 &Var \\
                     &    &R          &6.864 &2.281    &12.40       &Var & &                     &    &I         &24.429&3.753   &12.43 &Var  \\
                     &    &I          &4.720 &2.331    &10.96 &Var  & & 15.08.2012           &D    &B        &0.670 &3.932   &-        &NV\\
17.07.2011           &E   &B         &35.089&3.239    &8.23  &Var & &                     &    &V         &0.417 &3.932   &-        &NV\\
                     &    &V          &90.629&3.239    &7.28 &Var   & &                     &    &R         &0.908 &3.932   &-  &NV\\
                     &    &R          &50.931&3.239    &7.20 &Var   & &                     &    &I         &0.754 &3.932   &- &NV \\
01.08.2011           &D   &B         &1.860 &3.932     &-       &NV & & 16.08.2012           &D    &B        &0.908 &3.463   &-        &NV\\
                     &    &V         &2.774 &3.932     &-       &NV & &                     &    &V         &1.726 &3.463   &-        &NV\\
                     &    &R         &2.094 &3.932     &- &NV  & &                     &    &R         &1.622 &3.463   &-        &NV\\
                     &    &I         &1.862 &3.932     &- &NV   & &                     &    &I         &0.764 &3.463   &- &NV \\
02.08.2011           &D   &B         &1.584 &6.195     &- &NV   & & 18.09.2012           &D    &B        &0.540 &3.345   &-        &NV\\
                     &    &V         &4.858 &6.195     &- &NV   & &                     &    &V         &1.395  &3.345   &-        &NV\\
                     &    &R         &5.322&7.077    &- &NV   & &                     &    &R         &1.034 &3.345   &-        &NV\\
                     &    &I         &2.540 &7.077    &- &NV   & &                     &    &I         &0.678 &3.345   &-       &NV \\
04.08.2011           &D   &B         &0.907 &2.849    &-       &NV  & & 08.10.2012           &C    &R        &3.007 &1.961    &7.89       &Var\\
                     &    &V         &0.552 &2.983    &-  &NV   & & 13.10.2012           &C    &R        &6.686 &3.239    &11.76       &Var\\
                     &    &R         &1.507 &2.983    &-  &NV   & & 17.10.2012           &D    &B        &6.863 &3.932    &24.58  &Var  \\
                     &    &I         &1.346 &2.983    &- &NV   & &                     &    &V         &27.535 &3.932    &19.40 &Var  \\
06.08.2011           &D   &B         &0.368 &2.736    &-       &NV & &                     &    &R         &42.842&3.932    &16.02  &Var   \\
                     &    &V         &0.300 &2.736    &-       &NV & &                     &    &I         &33.239&3.932    &13.18   &Var   \\
                     &    &R         &0.670 &2.790     &-       &NV & & 22.10.2012           &D    &B        &1.383   &2.686    &-       &NV  \\
                     &    &I         &0.859 &2.849    &- &NV   & &                     &    &V         &1.462 &2.555    &-       &NV\\
07.08.2011           &D   &B         &0.997 &3.239    &-       &NV & &                     &    &R         &10.608 &2.555    &13.80 &Var  \\
                     &    &V         &7.058&3.239    &10.63  &Var   & &                     &    &I         &10.043 &2.555    &12.95  &Var   \\
                     &    &R         &7.773 &3.239    &8.23  &Var    & & 23.10.2012           &D    &B        &0.606 &2.596    &       &NV\\
                     &    &I         &2.046 &3.239    &      &NV    & &                      &    &V         &3.585 &2.481    &9.29       &Var\\
23.08.2011           &D   &B         &0.730 &2.448    &-       &NV & &                     &    &R         &5.166 &2.517    &10.25  &Var   \\
                     &    &V         &4.960 &2.448    &12.49  &Var    & &                     &    &I         &3.329 &2.481    &9.68  &Var  \\
                     &    &R         &8.400  &2.448    &9.47  &Var   & & 26.10.2012           &C   &R         &4.330 &2.596    &9.42 &Var   \\
                     &    &I         &5.092 &2.448    &9.69   &Var   \\

\cline{1-7} \cline{9-15}
\end{tabular} \\
\footnotesize
\begin{tabbing}
A: \= 1.04 meter Samprnanand Telescope, ARIES, Nainital, India  \=
\hspace{3em} \=
C: \= 50/70-cm Schmidt Telescope at National Astronomical \\
D: \> 60-cm Cassegrain Telescope at Astronomical Observatory \> \> \> Observatory, Rozhen, Bulgaria   \\
\> Belogradchik, Bulgaria \> \>
E: \> 1.3-m Skinakas Observatory, Crete, Greece \\
F: \> 1.3 m McGraw-Hill Telescope, Arizona, USA\\
$F_{enh}$:   \> \hspace{1em} Enhanced F-test values \> \>
$F_{c}$(0.001): \> \hspace{3em} Critical Values of F Distribution at 0.1\%; \\
Amp:  \> \hspace{1em} Variability Amplitude \> \>
 Var / NV: \> \hspace{3em} Variable / Non-variable \\
\end{tabbing}

\end{table*}


\begin{table}
\caption{ Details of telescopes and instruments}
\textwidth=6.0in
\textheight=9.0in
\vspace*{0.2in}
\noindent
\begin{tabular}{ll} \hline
Telescope:        &1.3 m McGraw-Hill Telescope     \\\hline
Chip size:   & $4064\times4064$ pixels      \\
Pixel size:  &$15\times15$ $\mu$m           \\
Scale:       &0.315\arcsec/pixel             \\
Field:       & $21.3\arcmin\times21.3\arcmin$   \\
Gain:        &2.2-2.4 $e^-$/ADU                   \\
Read Out Noise:         &5 $e^-$ rms         \\ \hline
\end{tabular} \\
\noindent
\end{table}

\begin{figure*}
\centering
\includegraphics[width=5cm , angle=0]{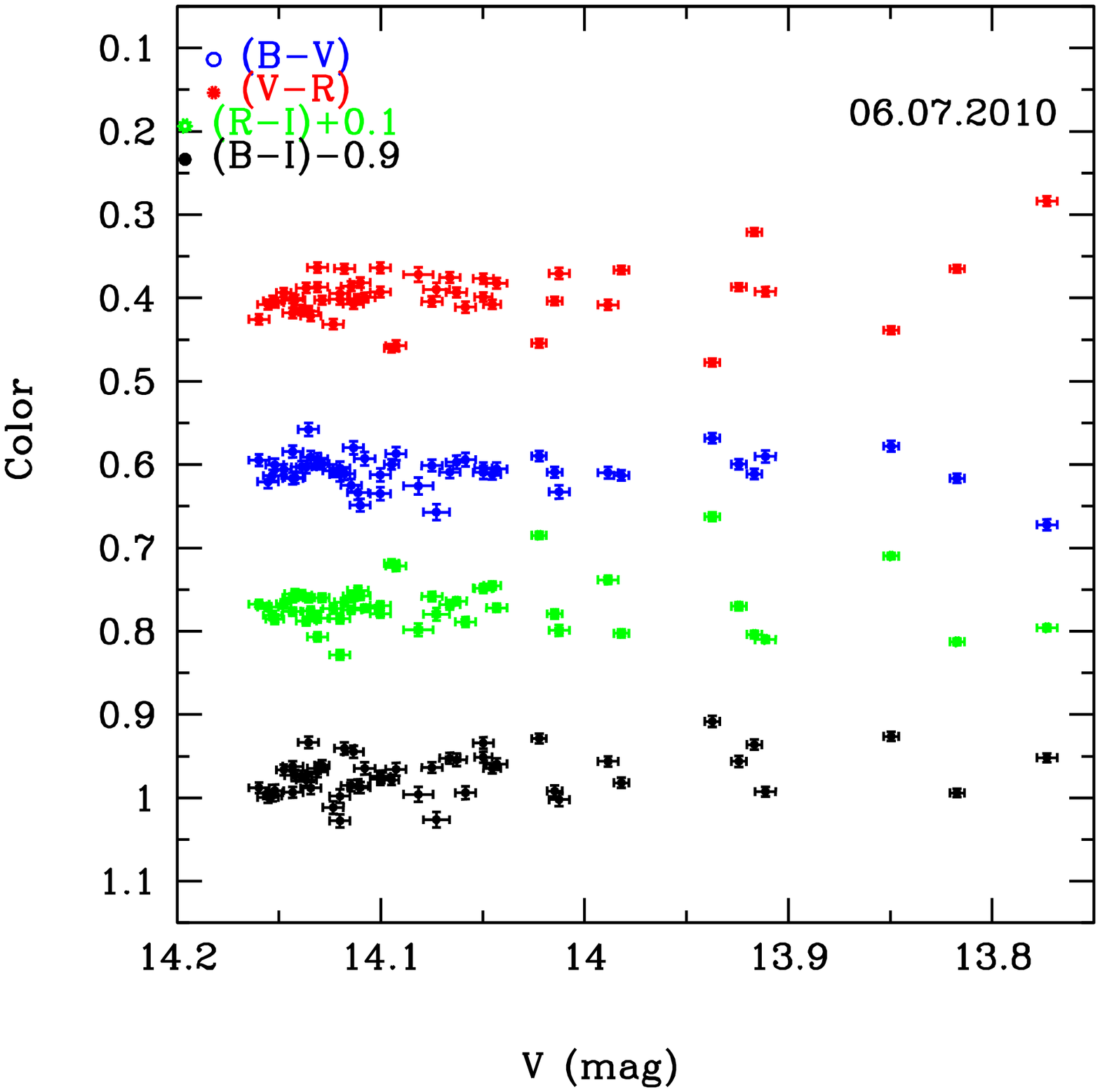}
\includegraphics[width=5cm , angle=0]{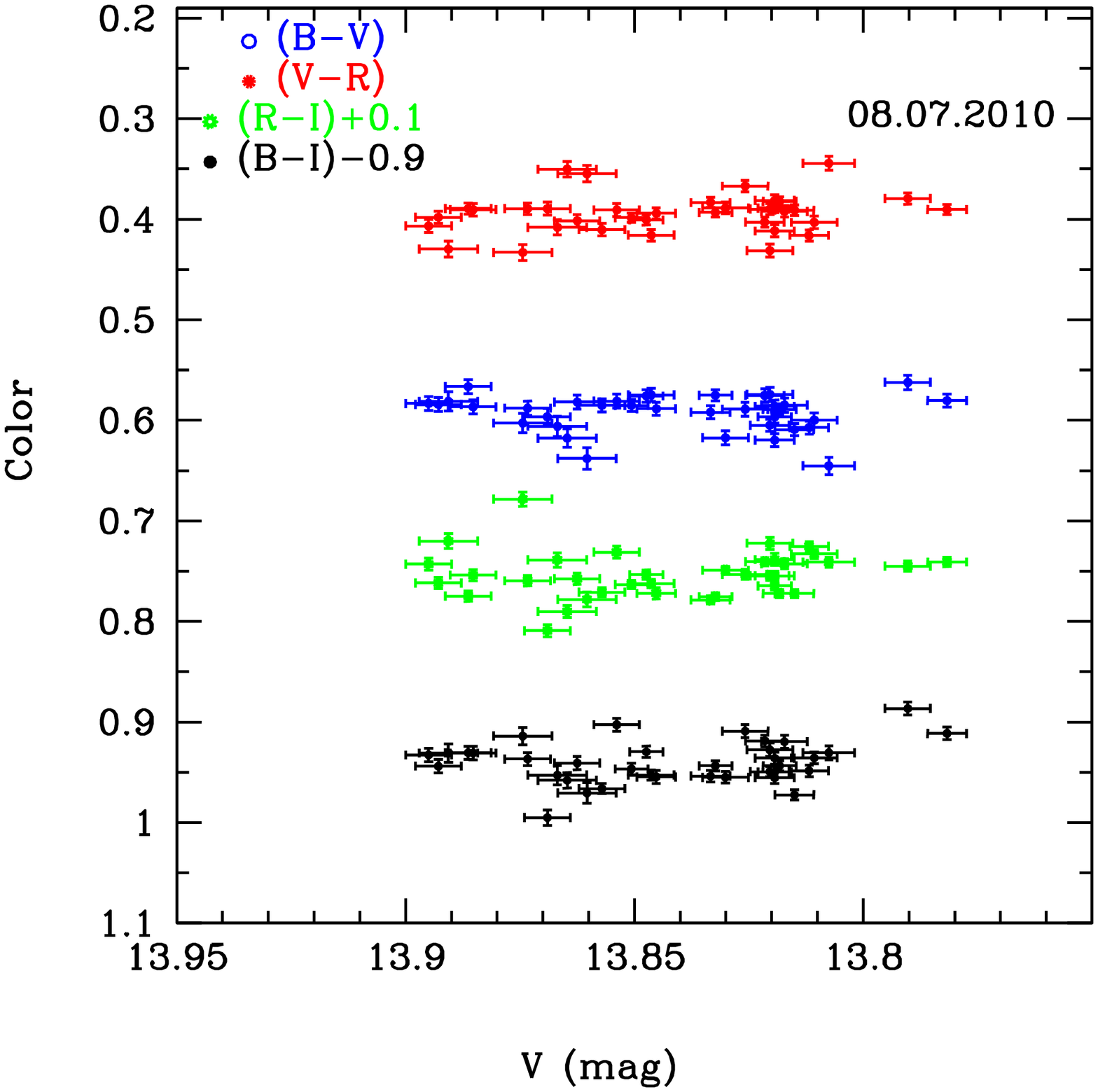}
\includegraphics[width=5cm , angle=0]{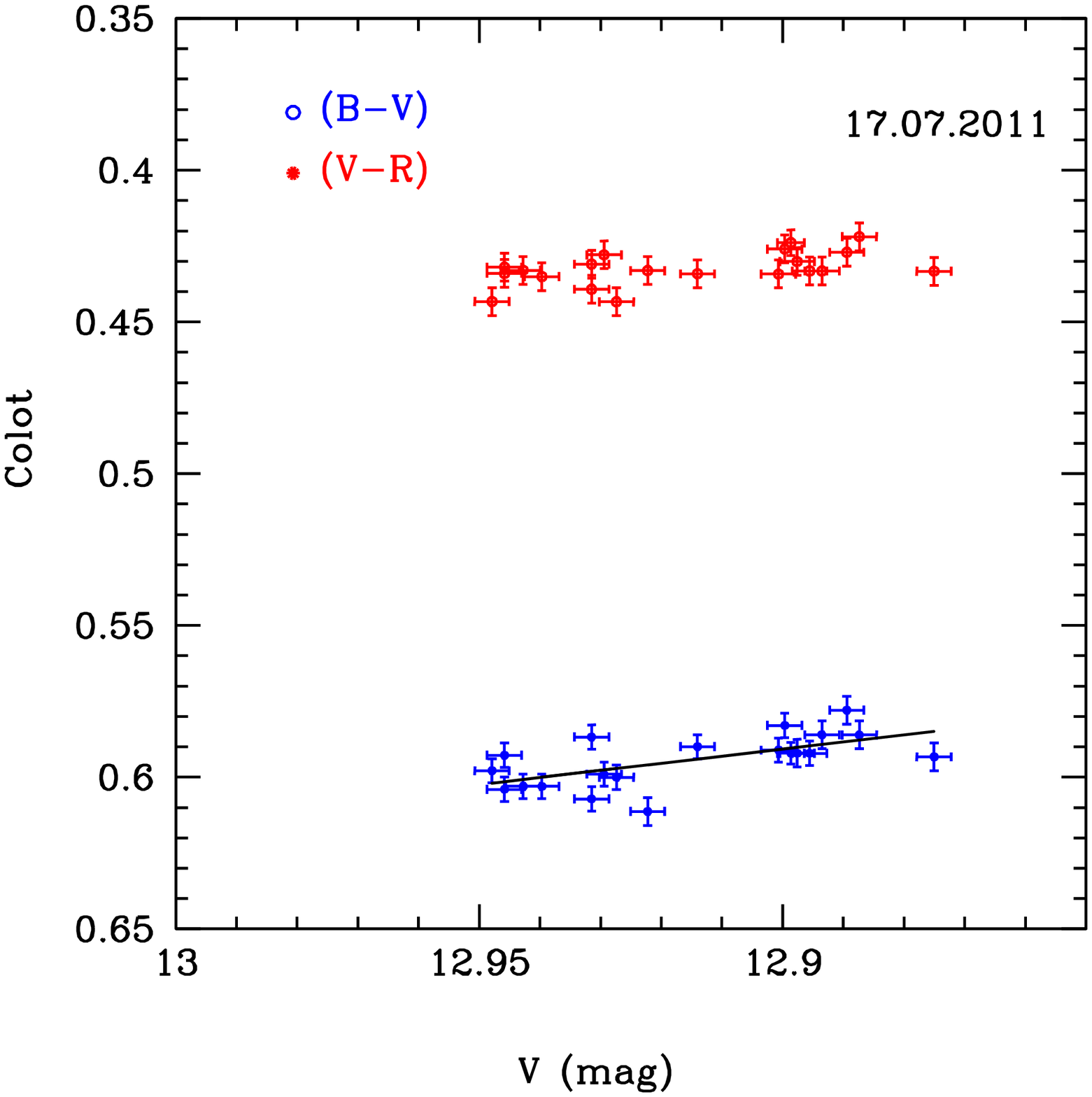}
\includegraphics[width=5cm , angle=0]{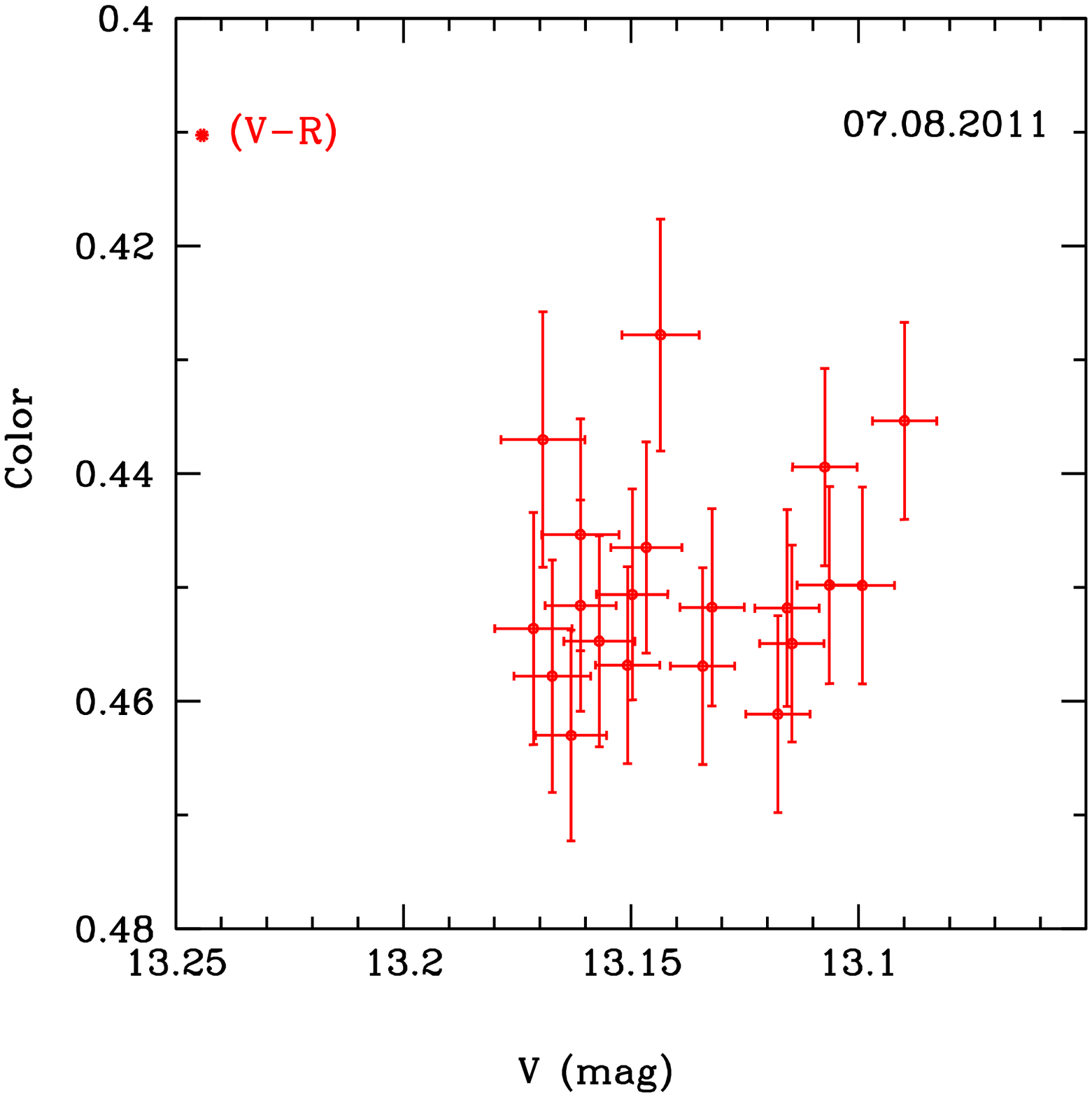}
\includegraphics[width=5cm , angle=0]{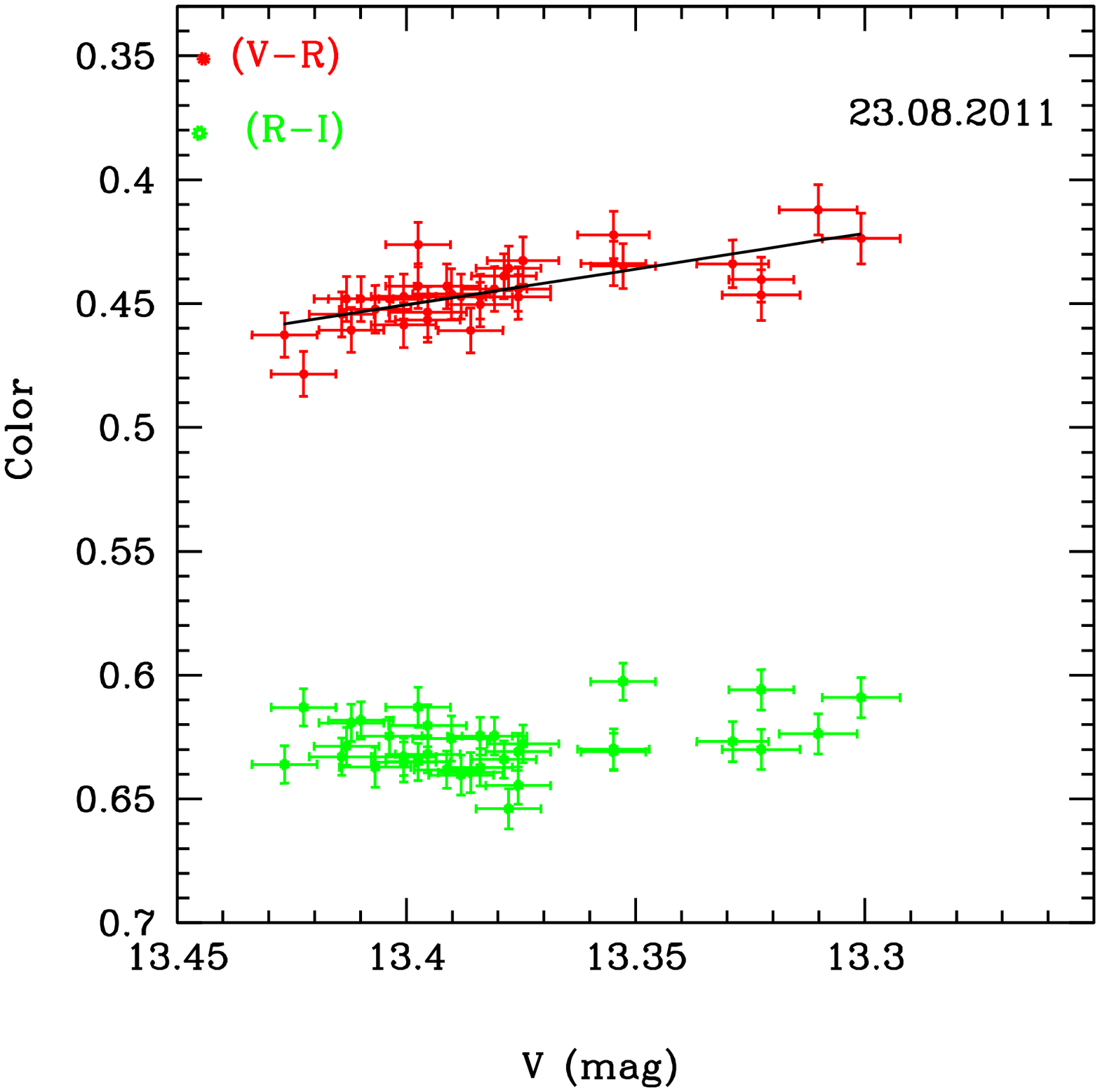}
\includegraphics[width=5cm , angle=0]{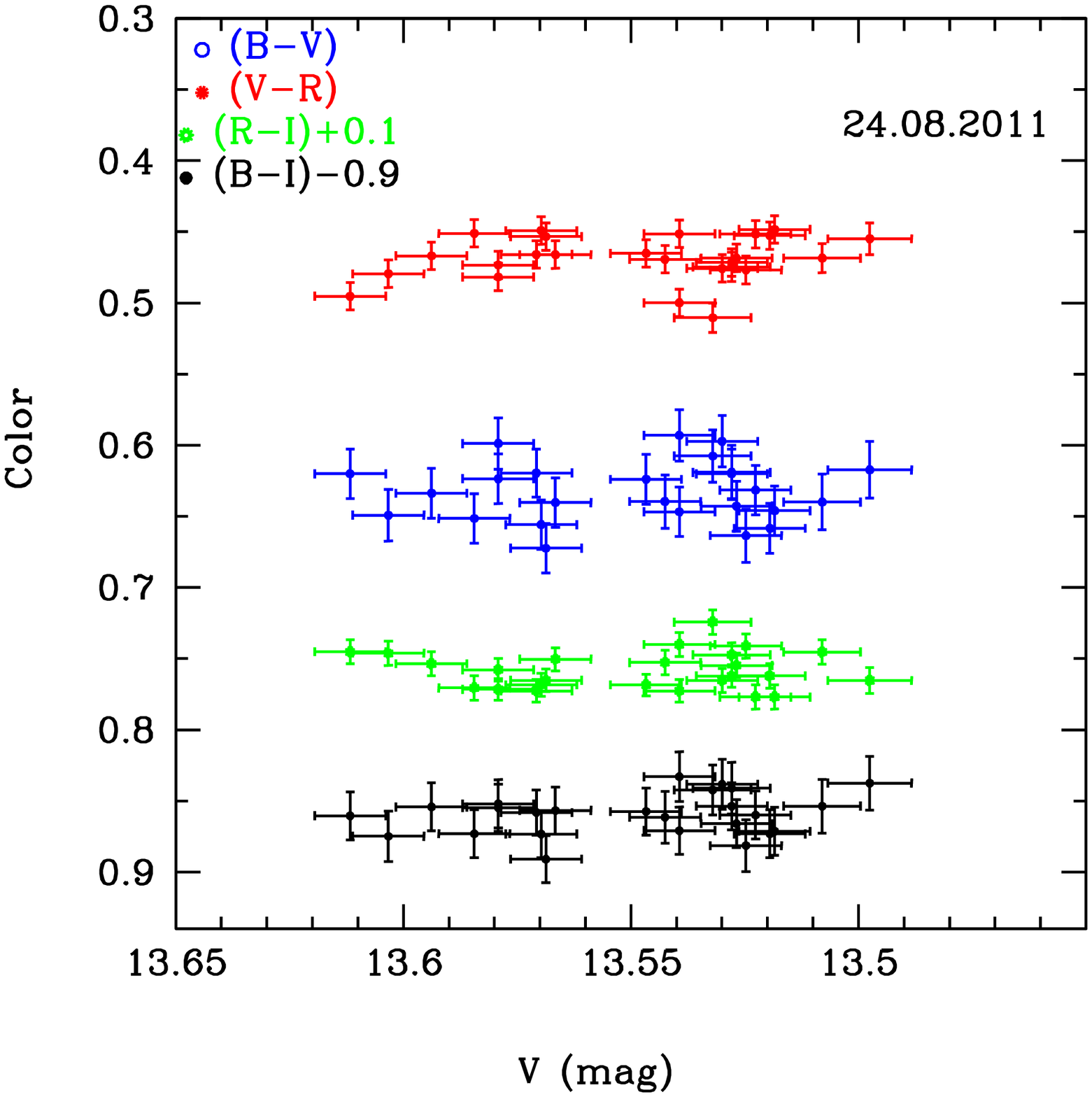}
\includegraphics[width=5cm , angle=0]{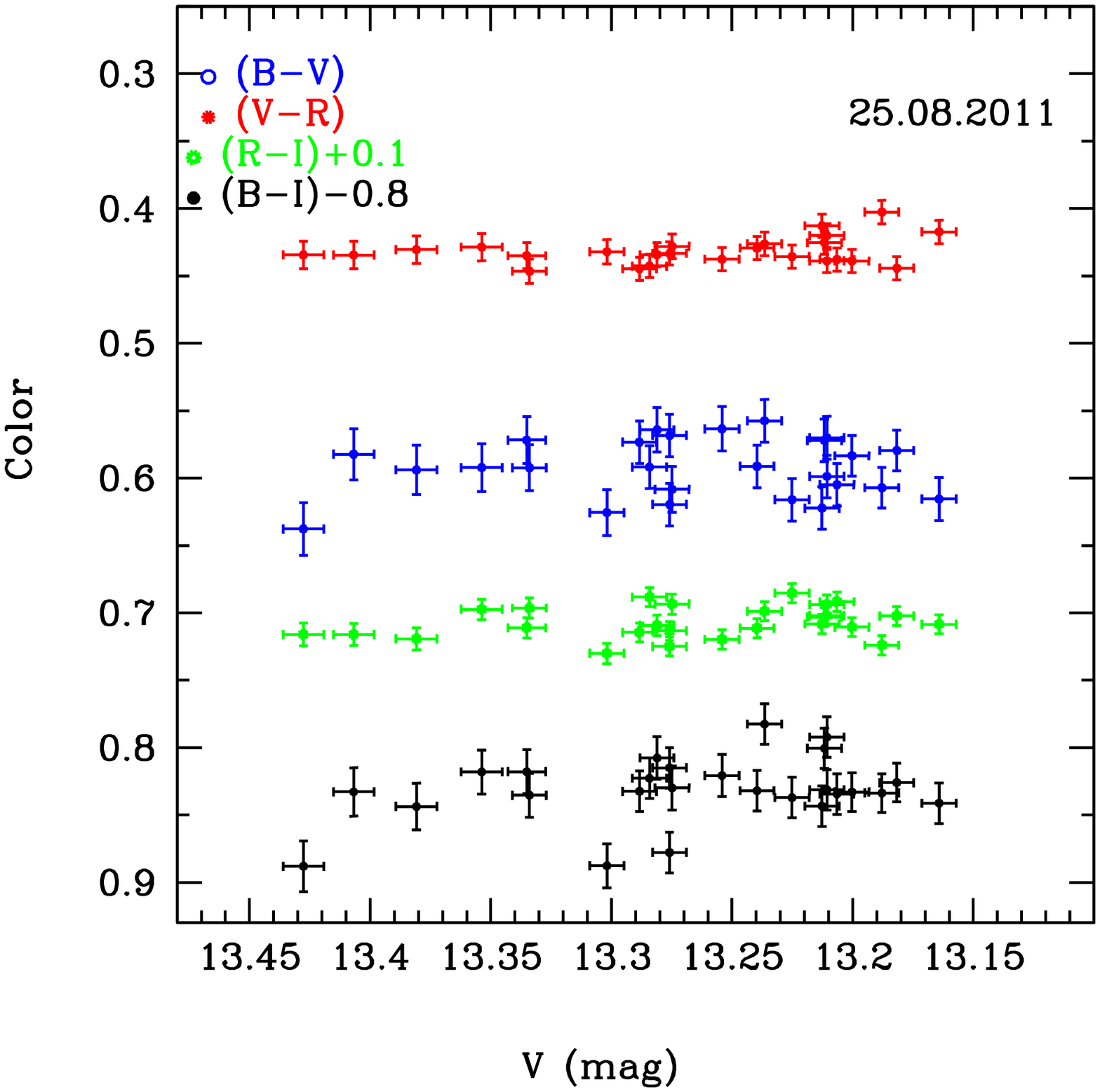}
\includegraphics[width=5cm , angle=0]{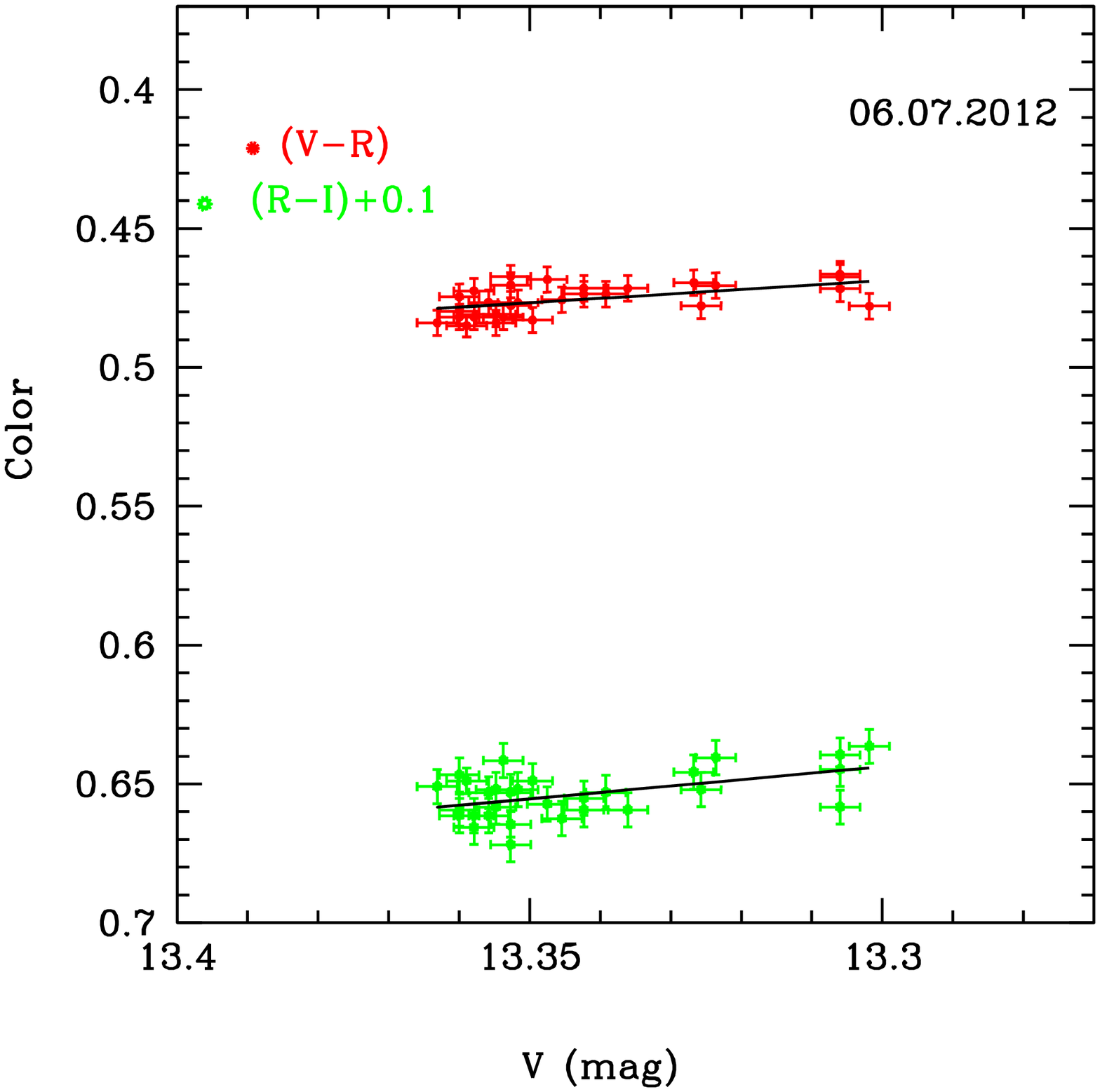}
\includegraphics[width=5cm , angle=0]{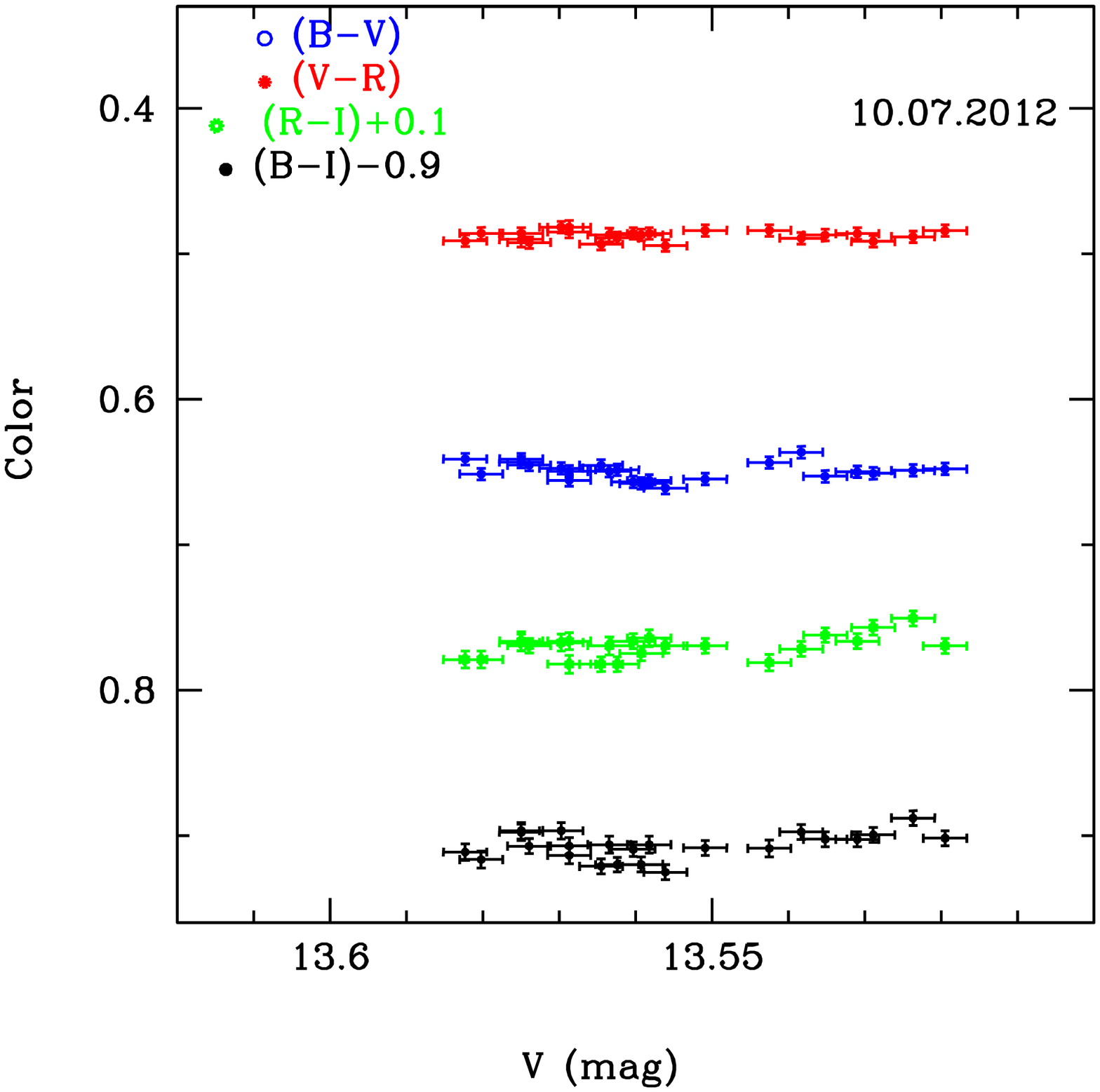}
\includegraphics[width=5cm , angle=0]{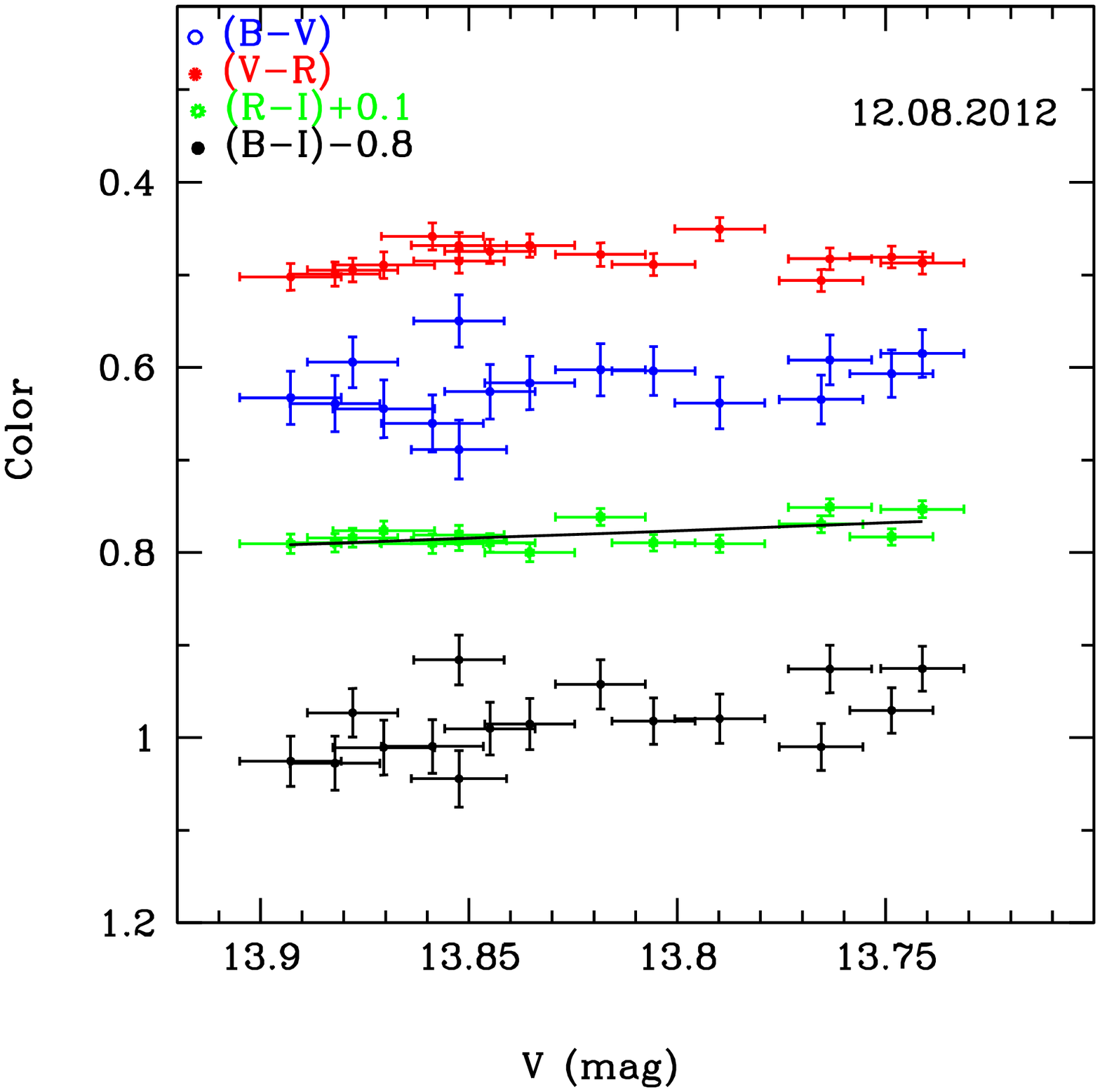}
\includegraphics[width=5cm , angle=0]{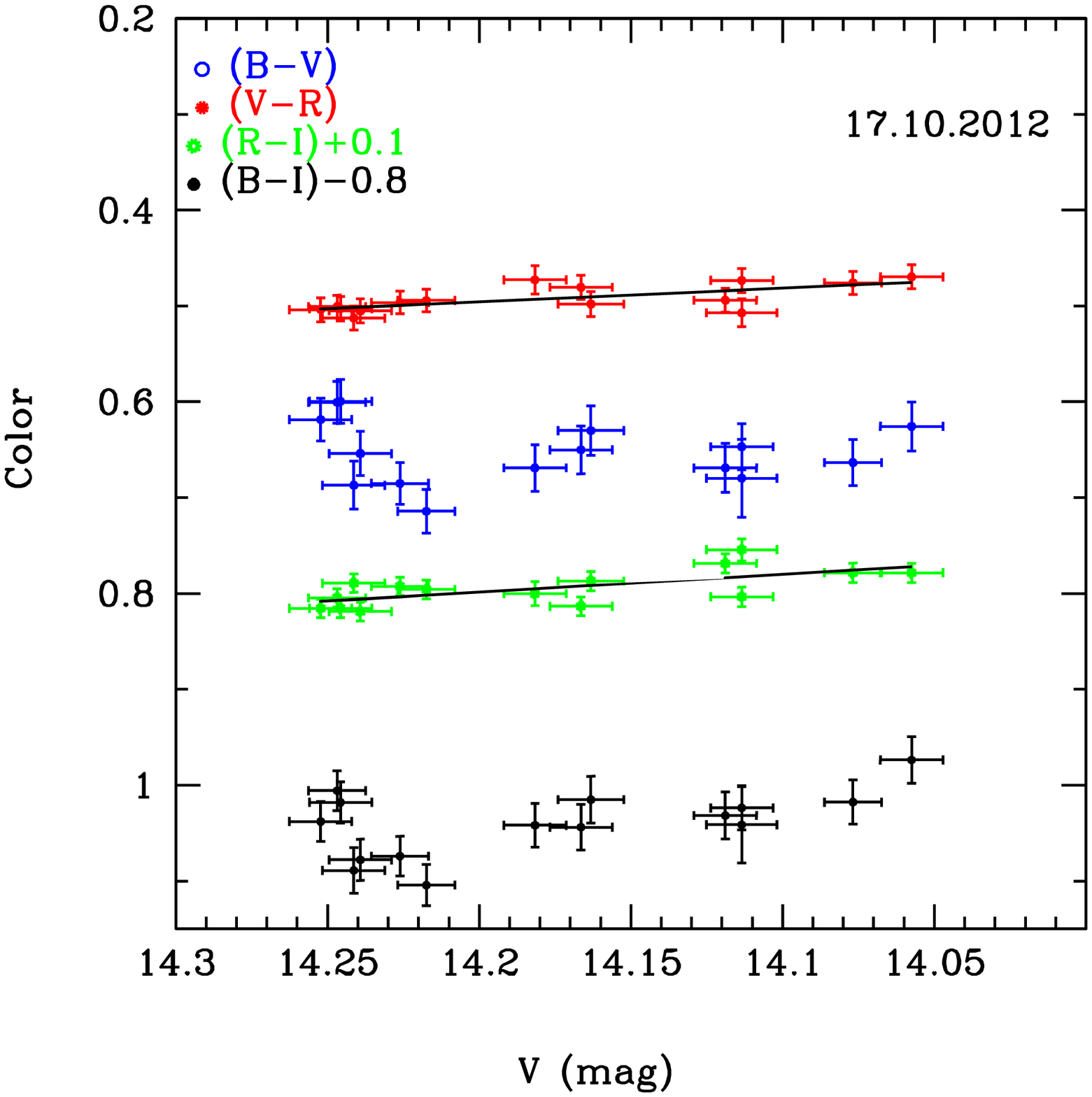}
\includegraphics[width=5cm , angle=0]{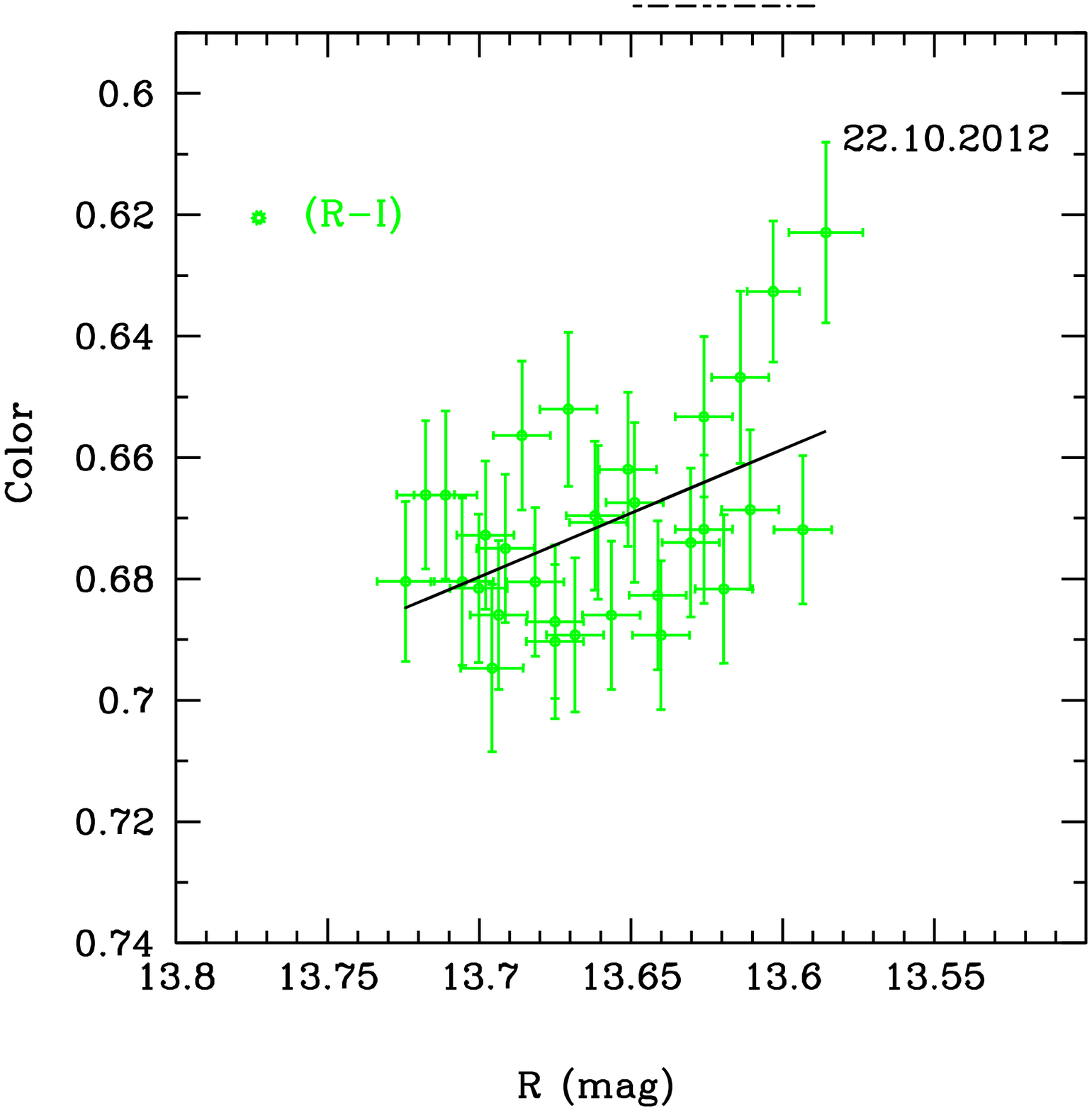}
\includegraphics[width=5cm , angle=0]{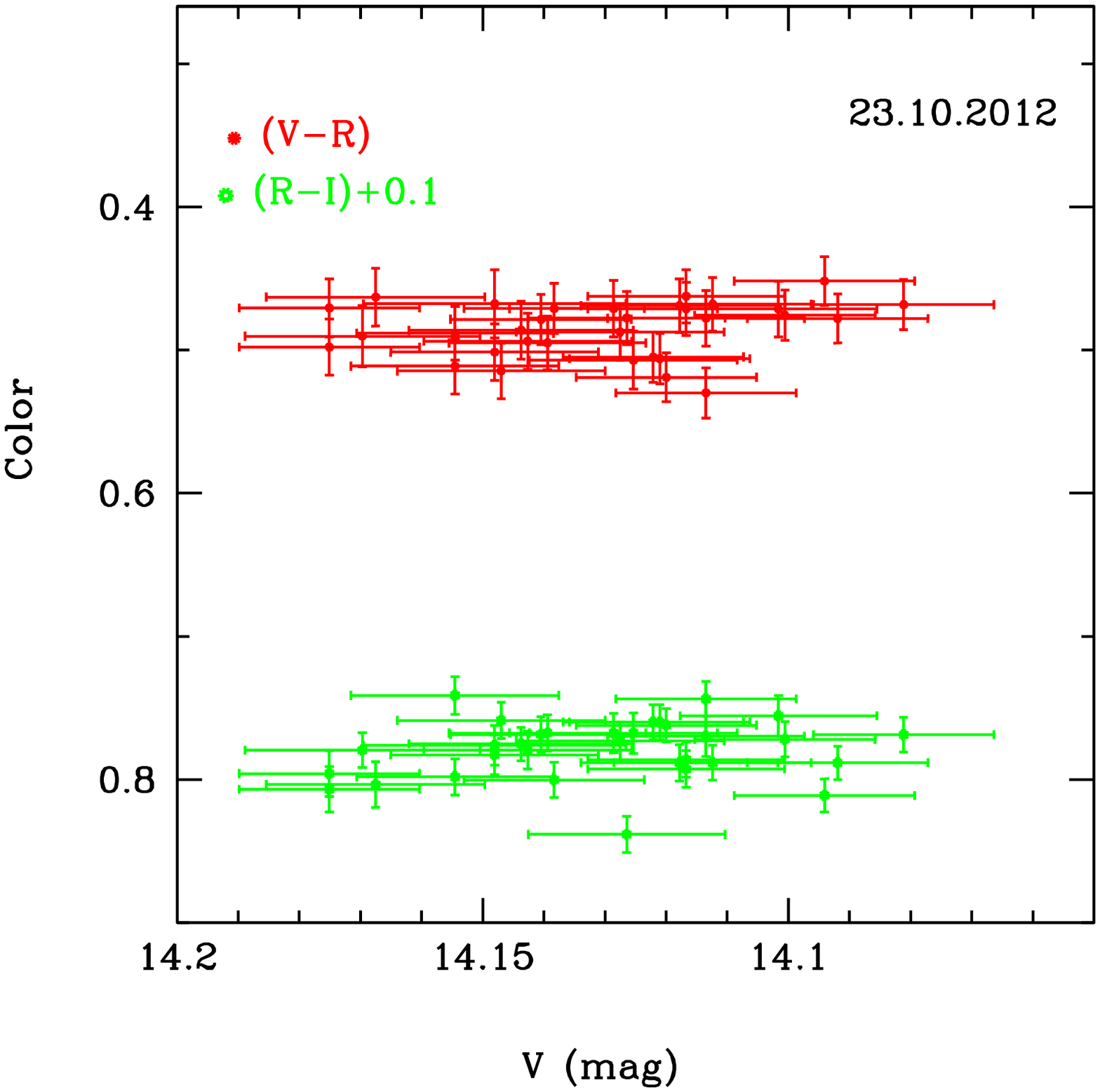}
\caption{Colour index versus magnitude diagrams for BL Lacertae. The solid line is the best fit on the observations.}
 \end{figure*}

\begin{table}
\caption{ Results of linear fits to Colour-Flux Diagrams.}
\setlength{\tabcolsep}{0.03in}
\begin{tabular}{lccccc} \hline \hline

Date of          &Band             &r   &p    &c$\pm$$\Delta$c  &m$\pm$$\Delta$m     \\            
Observation      &                 &    &        &Intercept   &Slope                 \\\hline
06.07.2010   &(B-V)                &-0.138 &0.339     & 1.038$\pm$0.446     &-0.032$\pm$0.032     \\
             &(V-R)                &0.315  &0.026     &-1.055$\pm$0.675     & 0.110$\pm$0.048      \\
             &(R-I)                &0.023  &0.873     & 0.459$\pm$0.675     &0.008$\pm$0.048       \\
             &(B-I)                &0.316  &0.026     & 0.442$\pm$0.534     &0.087$\pm$0.038     \\
08.07.2010   &(B-V)                &-0.106 &0.537     & 1.500$\pm$1.455     &-0.066$\pm$0.105      \\
             &(V-R)                &0.205  &0.230     &-1.389$\pm$1.540     &0.136$\pm$0.111       \\
             &(R-I)                &0.110  &0.545     &-0.626$\pm$1.836     &0.085$\pm$0.133       \\
             &(B-I)                &0.220  &0.198     &-0.516$\pm$1.643     &0.156$\pm$0.119       \\
17.07.2011   &(B-V)*               &0.629  &0.003     &-2.424$\pm$0.879     &0.234$\pm$0.068      \\
             &(V-R)                &0.512  &0.021     &-1.150$\pm$0.627     &0.123$\pm$0.049      \\
07.08.2011   &(V-R)                &0.193  &0.415     &-0.434$\pm$1.058     &0.067$\pm$0.081       \\
23.08.2011   &(V-R)*               &0.714  &$<$0.001  &-3.423$\pm$0.670     &0.289$\pm$0.050      \\
             &(R-I)                &0.280  &0.109     &-0.658$\pm$0.779     &0.096$\pm$0.058       \\
24.08.2011   &(B-V)                &0.057  &0.788     &0.113$\pm$1.910      &0.038$\pm$0.141      \\
             &(V-R)                &0.181  &0.388     &-0.805$\pm$1.446     &0.094$\pm$0.107       \\
             &(R-I)                &-0.022 &0.917     &0.787$\pm$1.219      &-0.009$\pm$0.090      \\
             &(B-I)                &0.264  &0.202     &0.095$\pm$1.269      &0.123$\pm$0.094       \\
25.08.2011   &(B-V)                &0.087  &0.673     &0.235$\pm$0.836      &0.027$\pm$0.063       \\
             &(V-R)                &0.345  &0.118     &-0.162$\pm$0.366     &0.045$\pm$0.028       \\
             &(R-I)                &0.251  &0.217     &0.051$\pm$0.439      &0.042$\pm$0.033       \\
             &(B-I)                &0.325  &0.105     &0.125$\pm$0.894      &0.114$\pm$0.067      \\
06.07.2012   &(V-R)*               &0.530  &0.003     &-1.643$\pm$0.641     &0.159$\pm$0.048       \\
             &(R-I)*               &0.506  &0.004     &-2.429$\pm$0.994     &0.231$\pm$0.074       \\
10.07.2012   &(B-V)                &-0.084 &0.703     &1.019$\pm$0.956      &-0.027$\pm$0.071       \\
             &(V-R)                &0.061  &0.781     & 0.332$\pm$0.550     &0.011$\pm$0.040       \\
             &(R-I)               &0.487  &0.019     &-2.201$\pm$1.125     &0.212$\pm$0.083       \\
             &(B-I)                &0.395  &0.062     &-0.850$\pm$1.347     & 0.196$\pm$0.099       \\
12.08.2012   &(B-V)                &0.310  &0.243     &-2.208$\pm$2.320     &0.205$\pm$0.168       \\
             &(V-R)                &0.101  &0.709     &0.052$\pm$1.131      &0.031$\pm$0.082       \\
             &(R-I)                &0.581  &0.011     &-1.618$\pm$0.861     &0.166$\pm$0.062      \\
             &(B-I)                &0.517  &0.040     &-3.774$\pm$2.458     &0.402$\pm$0.178     \\
17.10.2012   &(B-V)                &-0.103 &0.716     &1.370$\pm$1.927      &-0.051$\pm$0.136       \\
             &(V-R)*               &0.674  &0.006     &-1.526$\pm$0.614     &0.142$\pm$0.043     \\
             &(R-I)*               &0.668  &0.006     &-1.945$\pm$0.816     &0.186$\pm$0.058      \\
             &(B-I)                &0.542  &0.037     &-2.102$\pm$1.694     &0.278$\pm$0.120      \\
22.10.2012   &(R-I)*               &0.482  &0.006     &-2.203$\pm$0.970     & 0.210$\pm$0.071      \\
23.10.2012   &(V-R)                &0.217  &0.233     &-1.919$\pm$1.973     & 0.170$\pm$0.140       \\
             &(R-I)                &0.163  &0.373    &-1.262$\pm$2.147      & 0.137$\pm$0.152       \\ \hline

\end{tabular} \\
$ $*: Significant variations are found in these observations.  \\
r \& p: Pearson Correlation Coefficient and its probability values respectively. \\
\end{table}

\section{Discussion}

We performed photometric monitoring of  BL Lacertae during the period 2010--2012 for a total of 38 nights 
in the B, V, R and I bands in order to study its flux and spectral variability. In 19 of those nights, we found 
genuine IDV. The light curves often show gradual rises and decays,  sometimes with smaller 
sub-flares superimposed. No evidence for periodicity or other characteristic time scales was found.  
 We find the duty cycle of the source during this period to be $\sim$44\%.
In the earlier studies, it has been found that LBLs display stronger IDV than HBLs (high frequency peaked blazars)
and the DC has been estimated to be $\sim$70\% for LBLs and $\sim$30--50\% for HBLs (Heidt \& Wagner 1998; Romero et al. 2002;
Gopal-Krishna et al. 2003). Gopal-Krishna et al. (2011) studied a large sample of blazars and found that if variability
amplitude (Amp) $>$3\% is considered, the DC is 22\% for HBLs and 50\% for LBLs. We found DC of $\sim$44\% for BL Lacertae 
(which is a well known LBL) is in accordance with the previous studies.

In the literature, there are various models which explains intra-day variability of blazars.  Intrinsic ones focus on the
 evolution of the electron energy density distribution of the relativistic paricles leading to a variable synchrotron
emission, with shocks accelerating turbulent particles in the plasma jet which then cools by synchrotron emission (e.g., 
Marscher, Gear \& Travis 1992; Marscher 2014; Calafut \& Wiita 2015). 
Extrinsic ones involve  geometrical effects like swinging jets where the path of the relativistic moving blobs along the jet
deviated slightly from the line of sight, leading to a variable Doppler factor (e.g., Gopal-Krishna \& Wiita 1992). The long term 
periodic and acromatic BL Lacertae variability may be mostly explained by the geometrical scenarios where viewing angle
 variation can be due to the rotation of an inhomogeneous helical jet which causes variable Doppler boosting of the
 corresponding radiation (Villata et al 2002; Larionov et al. 2010 and references therein). As we are considering the 
faster intra-night flux
variations that are associated with the colour variations, they are more likely to be associated with models
involving shock propagating in a turbulent plasma jet. 

When variability is clearly detected, its amplitude is usually greater at higher frequencies, which is consistent with previous studies
(Papadakis et al.\ 2003; Hu et al.\ 2006) and can be well explained by electrons that are accelerated at the shock front and then 
lose energy as they move away from the front. Higher-energy electrons lose energy faster through the production of synchrotron radiation, 
 and are produced in a thin layer behind the shock front. In contrast, the lower-frequency emission is spread out over a larger volume 
behind the shock front (Marscher \& Gear 1985). This leads to time lags of the peak of the light curves toward lower frequencies and
amplitude of variability higher at higher frequencies which is clearly visible in the multi-frequency blazars light curves. 
Due to the closeness of the various optical bands, the starting time of a flare should be almost the same and hence 
on short time-scales, it is difficult to detect the time lags between the optical bands.
Although the  amplitude of variability is an inherent property of the source,  we had to examine whether
it has any dependence on the duration of observation, and found a
significant positive correlations between the observed amplitude of variability of the light curves and the duration of the observations.
As noted by Gupta \& Joshi (2005),  on intra-day
timescales, the probability of seeing a significant intra-day variability generally increases if the source is continuously observed for
long durations. In our observations, duration of monitoring varies between 1.5--5.8 hours. Also, we found that the 
amplitude of variability decreases as the source flux increases which can be explained as the source flux increases, the irregularities
in the turbulent jet (Marscher 2014) decrease and the jet flow becomes more uniform leading  to a decrease in amplitude of variability. 

We searched for the possible correlations between colour versus magnitude and found significant positive correlations
between them in five of the observations out of total 13. So, BL Lacertae showed significant bluer-when-brighter trends
on these night with different regression slopes (in Table 3). This behaviour is very well known for BL Lacertaes 
and can be interpreted as resulting from rapid, impulsive injection/acceleration of relativistic electrons, followed by
subsequent radiative cooling (e.g., B{\"o}ttcher \& Chiang 2002). However, other observations do not show significant
 linear correlations and show complicated behaviour on colour-magnitude diagrams i.e different slopes according to the
different flux states or nearly zero slopes between colour-magnitude (Table 3).
Of course it is possible that superposition of different spectral slopes from many variable components (standing shocks in
different parts of the jet) could lead to the overall weakening of the colour-magnitude correlations (Bonning et al.\ 2012).

Hence, the behaviour of the colour--magnitude diagrams  provides us with indirect information on the amplitude 
difference and time-lags between these bands as Dai et al.\ (2011) performed simulations to confirm that both the 
amplitude differences and time delays between variations at different wavelengths result in 
a hardening of the spectrum during the flare rise. They showed that if there is a difference in amplitude 
in two light curves, it leads to the evolution of the object along a diagonal path in the colour-magnitude diagram.  
If there is a time-lag along with the amplitude change, counter-clockwise loop patterns on the colour index--magnitude 
diagram arise (Dai et al.\ 2011). However, any such lags are probably shorter than the sampling time of 
our observations, so we are not able to detect them through the DCFs we computed. 

\section{Conclusions}
 
Our conclusions are summarized as follows: 

\begin{itemize}
\item During our observations in 2010-2012 BL Lacertae was highly variable in B, V, R and I bands. 
The variations were well correlated in all four bands and were very smooth, with gradual rises/decays.
In one of the observations, on  6 October 2010, we found a pronounced flare-like event, and the highest variability amplitude 
is found in the V band at $38$ per cent.
\item In the cases with significant variability, the amplitude of variability is highest in the highest energy band. 
\item The amplitude of variability correlates positively with the duration of the observation and
 decreases as the flux of the source increases. 
\item We searched for time delays between the B, V, R and I bands in our observations, but we did not find any
significant lags. This implies that the variations are almost simultaneous in all of the bands and any time lags, if present, 
are less than our data sampling interval of $\sim$8 minutes. 
\item The flux variations are associated with spectral variations on intra-day time-scales. In  5 of the 13 observations, the
optical spectrum showed the overall bluer-when-brighter trend which could well represent  highly variable jet emission.
\item The colour vs.\ magnitude diagrams show different behaviours which could represent the contribution of different variable 
components during the flaring states.
\item We conclude that the acceleration and cooling timescales
are very short for these optical variations and hence dense optical observations with even shorter cadence and higher sensitivity are required
to better characterize them.  
\end{itemize}

We thank the referee for useful and constructive comments.
This research was partially supported by Scientific Research Fund of the Bulgarian Ministry of Education and Sciences under 
grant DO 02-137 (BIn 13/09). The Skinakas Observatory is a collaborative project of the University of Crete, the Foundation 
for Research and Technology -- Hellas, and the Max-Planck-Institut f\"ur Extraterrestrische Physik. 
H.G. is sponsored by the Chinese Academy of Sciences Visiting Fellowship for Researchers from Developing Countries
 (grant No. 2014FFJB0005), and supported by the NSFC Research Fund for International Young Scientists (grant No. 11450110398).
ACG is partially supported by the Chinese Academy of Sciences Visiting Fellowship for Researchers
from Developing Countries (grant no. 2014FFJA0004).
MB acknowledges support by the South African Department of Science and Technology through the National Research Foundation 
under NRF SARChI Chair grant no. 64789. MFG acknowledges support from the National Science Foundation
of China (grant 11473054) and the Science and Technology Commission of Shanghai Municipality (14ZR1447100).

{}

\end{document}